# TopoLa: a novel embedding framework for understanding complex networks

Kai Zheng[1], Qilong Feng[1], Yaohang Li[2], Qichang Zhao[1], Jinhui Xu[3,*], Jianxin Wang[1,*]

[1]School of Computer Science and Engineering, Central South University, Changsha, 410083, China

[2]Department of Computer Science, Old Dominion University, Norfolk, VA 23529, United States.

[3]Department of Computer Science and Engineering, State University of New York at Buffalo, Buffalo, NY 14260, United States

*To whom correspondence should be addressed:

Jianxin Wang. Email: jxwang@mail.csu.edu.cn
Jinhui Xu. Email: jinhui@buffalo.edu

**Abstract.** Complex networks, which are the abstractions of many real-world systems, present a persistent challenge across disciplines for people to decipher their underlying information. Recently, hyperbolic geometry of latent spaces has gained traction in network analysis, due to its ability to preserve certain local intrinsic properties of the nodes. In this study, we explore the problem from a much broader perspective: understanding the impact of nodes' global topological structures on latent space placements. Our investigations reveal a direct correlation between the topological structure of nodes and their positioning within the latent space. Building on this deep and strong connection between node distance and network topology, we propose a novel embedding framework called Topology-encoded Latent Hyperbolic Geometry (TopoLa) for analyzing complex networks. With the encoded topological information in the latent space, TopoLa is capable of enhancing both conventional and low-rank networks, using the singular value gap to clarify the mathematical principles behind this enhancement. Meanwhile, we show that the equipped TopoLa distance can also help augment pivotal deep learning models encompassing knowledge distillation and contrastive learning.

## Introduction

Analyzing complex networks is an often-encountered challenging task in many disciplines. As pointed out in previous studies[1-2], to decipher those complex networks that are reflective of real-world entities, it is crucial to understand their structures as they underpin collective behaviors and reveal the main features of relevant processes. These inherent structures closely align with network geometry, which prompted the proposal of various geometric models to fathom and represent network properties[1-5]. Among these models, hyperbolic geometry of latent space (called latent hyperbolic geometry) has emerged as a promising tool for simulating real-world networks[4], explaining efficient routing mechanisms[6], and identifying community structures[7]. Given the link between network paths that follow hyperbolic geodesics in the latent space and real-world shortest paths, positions within the latent space serve as potent indicators of inter-node relationships[4,8-10].

So far, researchers have devised a series of methods to ascertain the distances between nodes in their corresponding latent space. These methods roughly fall into two categories based on how the distance is measured: energy distance measure and spatial distance measure. Energy distance measure, founded on the maximum entropy principle, is determined jointly by energy (spatial distance) and chemical potential (a function of the expected degree)[11]. The number of common neighbors has been shown to correlate with energy distance in the latent space[11]. Spatial distance measure is typically classified into three principal categories: generative models, data-driven models, and hybrid models[12]. Generative models often rely on techniques such as Monte Carlo sampling and maximum likelihood estimation to facilitate statistical inference[13]. Data-driven methods derive coordinates in the latent space either by utilizing unsupervised learning techniques for nonlinear dimensionality reduction or by leveraging network community structures[14,15]. The hybrid models synthesize the strengths of both, ensuring accurate results with efficient computation[16]. Although spatial distance measure excels in pinpointing latent coordinates, they overlook the influence of chemical potential on node relationship metrics. Existing energy distance measure relies on the local connectivity of nodes, and thus neglects the great impact of nodes' global connectivity and topological structural information on latent space placements. Hence, a thorough investigation is needed for us to fully understand the relationship between nodes' placement in the latent space and their local and global features (such as degrees and local and global connectivity) in the network.

In this study, we show, for the first time, that topological similarity (calculated by using global connectivity information) between each pair of nodes can be encoded in the latent hyperbolic geometry space so that the pairwise distances of nodes in the space are tightly correlated to their topological similarity in the network. Building on this deep and strong connection between node distance and network topology, we introduce a new embedding framework called Topology-encoded Latent Hyperbolic Geometry (TopoLa), which includes latent space embedding, energy distance measure, and spatial specificity (such as triangle inequality). From a network perspective, the energy distance (called TopoLa distance) measures the global topological similarity of the two involving nodes by counting the weighted number of even-hop paths (i.e., paths of even lengths or even number of hops) that connect them (Fig.1A). Thus, different from the existing methods which consider only 2-hop paths, the TopoLa distance accounts for the effects of the nodes' global connectivity information on latent space positioning. Given our observation that nodes exhibit a tendency to form connections with neighbors of those nodes with analogous topological structure, we propose a network enhancement approach rooted in node topology, Network Reconstruction via Global Connectivity-Based Topological Similarity (NR). To mitigate the computational burden of matrix inversion in NR, we also propose a Fast Network Reconstruction via Global Connectivity-Based Topological Similarity (fastNR), tailored for large-scale network computations. Empirical validations emphasize the multifaceted performance of NR: reconstructing the topological structure of nodes; sharpening community boundaries among distinct biological cell types; reducing noise in Hi-C contact maps from the human genome; enhancing the precision of fine-grained species identification; refining movie recommendations; and improving precision of multi-label web content classification. Furthermore, we show that the TopoLa distance can be used to enhance deep learning models such as knowledge distillation and contrastive learning. Specifically, we show that this distance allows us to explore sample relationships by network geometry, whether inheriting relationships from teacher models in Knowledge Distillation or measuring positive and negative sample relationships in Contrastive Learning. Experiments showed that these models exhibit improved performance with TopoLa distance compared to using distances and vector angles in the standard Euclidean space. In summary, our work considerably enhances our ability in network analysis and introduces a sample relationship measure for knowledge distillation and contrastive learning.

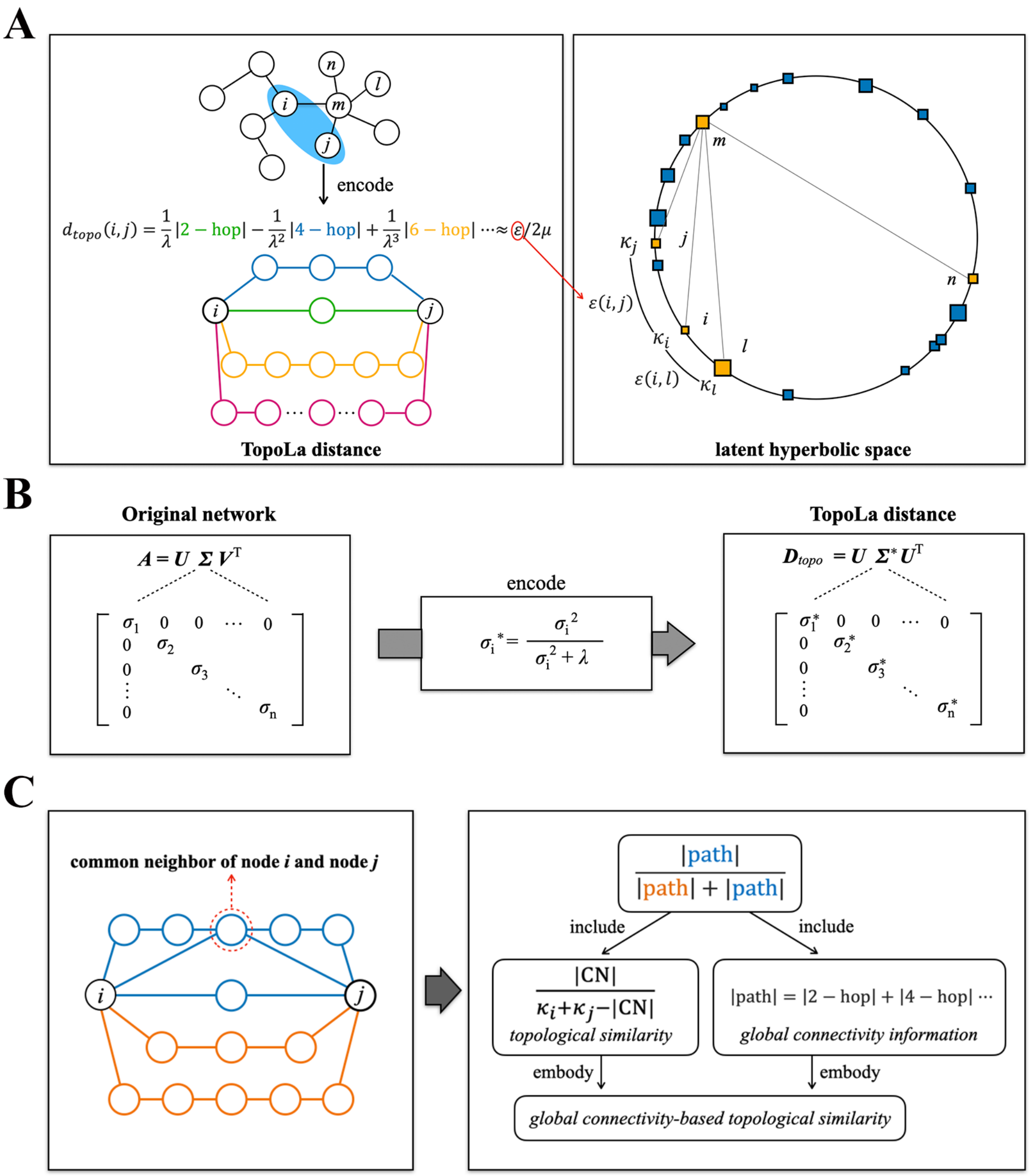

**Fig.1 Overview of the topology-encoded latent hyperbolic geometry (TopoLa)**. (**A**) TopoLa distance $d_{topo}$ is determined by counting the weighted number of even-hop paths connecting the corresponding pair of nodes. (**B**) For efficient computation of the inverse operation, the original networks $\boldsymbol{A}$ are decomposed into singular values and vectors. This decomposition facilitates the use of singular value $\sigma_i$ from $\boldsymbol{A}$ to calculate the eigenvalue $\sigma_i^*$ of the TopoLa distance matrix $\boldsymbol{D}_{topo}$. (**C**) All even-hop paths are categorized into two types: blue paths containing common neighbors and orange paths without them. The proportion of blue paths among all paths reflects the topological similarity between the two nodes. For instance, if the number of common neighbors $|\mathrm{CN}|$ between the two nodes equals their respective degrees $\kappa_i$ and $\kappa_j$, then the count of orange paths is zero. Incorporating the statistics of all even-hop paths provides global connectivity information. Hence, this similarity measure is termed as 'global connectivity-based topological similarity'.

# Results

## Overview of TopoLa

The topology-encoded latent hyperbolic geometry (TopoLa) is a novel embedding framework that uses node topological information to refine inter-node energy distance measures in the latent space. Grounded in the maximum entropy principle, the energy distance in latent space represents the ratio between energy and chemical potential[11,17]. In networks, the connectivity between nodes $i$ and $j$ can be characterized by all the paths connecting them. The quantity of 2-hop paths correlates positively with the complement of energy distance[11]. However, the current method of measuring energy distance using 2-hop paths lacks precision (Supplementary text 3.1, Theorem 1). This imprecision primarily stems from an emphasis on local connectivity and a failure to consider the impact of high-degree nodes on the network's geometry[18]. Thus, the energy distance should be farther between high-degree nodes compared to those with low-degree, given an equal number of common neighbors. To address these limitations, we need to consider not only the information of node degrees and local connectivity, but more importantly the global connectivity information between each pair of nodes. Consequently, we propose the TopoLa distance $\boldsymbol{D}_{topo}$:

$$\boldsymbol{D}_{topo} = \frac{1}{\lambda}\boldsymbol{A}\boldsymbol{A}^T - \frac{1}{\lambda^2}\boldsymbol{A}\boldsymbol{A}^T\boldsymbol{A}\boldsymbol{A}^T + \frac{1}{\lambda^3}\boldsymbol{A}\boldsymbol{A}^T\boldsymbol{A}\boldsymbol{A}^T\boldsymbol{A}\boldsymbol{A}^T - \cdots \quad (1)$$

where $\boldsymbol{A}$ is the adjacency matrix of the original network. The TopoLa framework consists of latent space embedding, energy distance measure, and spatial specificity (such as the triangle inequality). $\boldsymbol{D}_{topo}$ represents the TopoLa distance expressed in matrix form, while the TopoLa distance between two nodes is denoted by $d_{topo}$. Our findings reveal that $d_{topo}$ can effectively capture global connectivity information, enabling the measurement of topological similarity between two nodes. We introduce the concept as 'global connectivity-based topological similarity' (Fig.1C). $d_{topo}$ establishes a clear correlation between the topological structure of nodes and their spatial positioning in latent space, notably clustering nodes with similar topology (Supplementary text 3.1, Theorem 3). Using this framework, we have developed two network enhancement algorithms: Network Reconstruction via Global Connectivity-Based Topological Similarity (NR) and Fast Network Reconstruction via Global Connectivity-Based Topological Similarity (fastNR). These algorithms are designed for network analysis as detailed below.

The prevailing thought suggests that as the distance between nodes in latent space increases, the likelihood of connection diminishes[19-21]. It implies that nodes in closer proximity are more likely to be interconnected[12,22]. Counterintuitively, experimental evidence has shown that nodes tend to form connections with neighbors of those nodes that have similar topological structures (Supplementary text 4). Based on this phenomenon, we propose a network enhancement technique called Network Reconstruction via Global Connectivity-Based Topological Similarity (NR), aimed at reshaping the topological structure of nodes from the perspective of latent space energy distance for better robustness against noises, errors, or data missing. NR takes a network as input to generate an enhanced network $\boldsymbol{A}^*$ without supervision or prior knowledge. The formula is as follows:

$$\boldsymbol{A}^* = \boldsymbol{D}_{topo}\boldsymbol{A}. \qquad (2)$$

We demonstrated, from the perspective of singular value gaps, that the same noise has a lesser impact on $\boldsymbol{A}^*$ compared to $\boldsymbol{A}$ (Supplementary text 3.2, Theorem 5). To mitigate the computational burden of matrix inversion in NR, we propose a Fast Network Reconstruction via Global Connectivity-Based Topological Similarity algorithm (fastNR). Our analysis reveals that the low-rank approximation of any matrix can benefit from the disturbance resilience provided by the NR algorithm (Supplementary text 3.3, Theorem 6). By using randQB_fp[23], we derived a low-rank approximation of matrix $\boldsymbol{A}$ that preserves the top $k$ singular values, denoted as $\boldsymbol{A}_k = \boldsymbol{U}_k\boldsymbol{\Sigma}_k\boldsymbol{V}_k^T$. So, the formula for enhancing the input network in fastNR is defined as follows:

$$\boldsymbol{A}_k^* = \boldsymbol{U}_k\boldsymbol{\Sigma}_k^T\boldsymbol{\Sigma}_k(\boldsymbol{\Sigma}_k^T\boldsymbol{\Sigma}_k + \lambda\boldsymbol{I})^{-1}\boldsymbol{\Sigma}_k\boldsymbol{V}_k^T \quad (3)$$

We proved that, for large-scale matrices, fastNR significantly reduces time complexity compared to NR (Supplementary text 8). Additionally, our results affirm that fastNR achieves comparable accuracy to NR across a broad spectrum of applications (Supplementary text 8).

**Comparing the two energy distance measures**

Energy distance, initially proposed by Krioukov, enables better network properties identification[11]. However, a precise method for quantifying energy distance has yet to be established. Nevertheless, Krioukov demonstrated that $|\mathrm{CN}|$, the number of common neighbors, provides an approximation of the complement of the energy distance.

Previous research has demonstrated that a higher value of the logarithm of thermodynamic activity facilitates more precise measurements of energy distance[11]. We further proved that $\alpha'$, representing the logarithm of thermodynamic activity for $d_{topo}$, exceeds $\alpha$ associated with $|\mathrm{CN}|$ (Supplementary text 3.1, Theorem 1). The precision of energy distance measures is determined by integrating graphons, based on the principle of maximum entropy[11]. Hence, it is difficult to directly compare the precision of two energy distance measures using real-world examples. We employed $\alpha' = 15$ and $\alpha = 5$ as an example to visually illustrate how enhancements in the logarithm of thermodynamic activity affect measurement precision (Fig.2A). As shown in the figure, $C_n(\alpha', r)$ is closer to $1 - r$ compared to $C_n(\alpha, r)$, indicating that $C_n(\alpha', r)$ provides a more precise measurement of $r$, where $C_n(\cdot)$ denotes the integration of all 2-hop paths between the two involving nodes11. $C_n(\alpha', r)$ represents a function of the integration of common neighbors and the energy distance $r$, where $\alpha'$ is a fixed value. Therefore, $d_{topo}$ offers superior precision compared to $|\mathrm{CN}|$.

We demonstrated the difference in precision between $|\mathrm{CN}|$ and $d_{topo}$ using the logarithm of the thermodynamic activity. Next, we use a simple example to analyze the difference in their physical properties. Specifically, we generated an undirected graph with 500 nodes and 10,000 edges. The topological similarities between nodes are $\frac{|\mathrm{CN}|}{\kappa_i + \kappa_j - |\mathrm{CN}|}$. As shown in Fig. 2B, the topological similarities in this network mostly range from 5% to 15%.

Firstly, we explored the relationship between the union of neighbors ($\kappa_i + \kappa_j - |\mathrm{CN}|$) and different measures. Specifically, we analyzed node pairs with topological similarities around 5%, 10%, and 15%, within a range of ±0.5%. As shown in Fig.2B and 2C, when using $|\mathrm{CN}|$ as a measure, the same value can correspond to multiple topological similarities. For instance, $|\mathrm{CN}| = 14$ can correspond to node pairs with topological similarities of 10% and 15%. In contrast, $d_{topo}$ more accurately distinguishes between node pairs with different topological similarities. Additionally, $d_{topo}$ exhibits a negative correlation with the union of neighbors. Previous research has shown that high-degree nodes negatively impact the detection of geometry induced by latent hyperbolic spaces[18]. Therefore, by assigning lower measurement values to high-degree node pairs, $d_{topo}$ can effectively align with the properties of latent hyperbolic geometry.

Secondly, we analyzed the relationship between topological similarities and measures (Fig.2E and 2F). Our findings indicate that a single $|\mathrm{CN}|$ value corresponds to a wide range of topological similarities (wide horizontal slices), while $d_{topo}$ provides a more detailed delineation of topological similarity.

These experiments demonstrate the physical properties of $d_{topo}$ in terms of both degree and topological similarity, confirming the conclusions of Theorem 1.

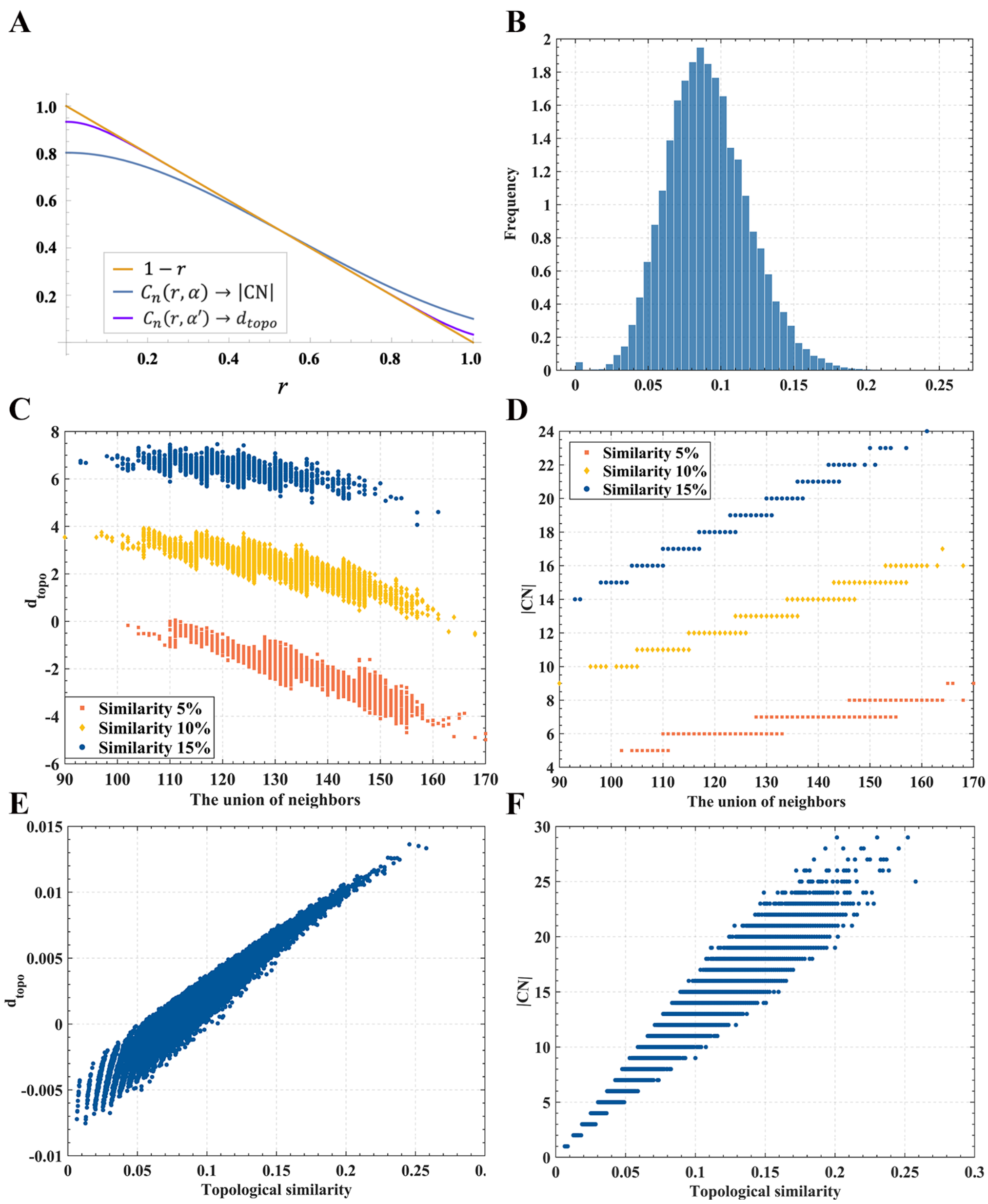

**Fig.2 The properties of energy distance measures.** (**A**) The relationships between energy distance and its measures. Utilizing the maximum entropy principle, $|\mathrm{CN}|$ and $d_{topo}$ can be expressed as functions of energy distance $r$, where $C_n(\cdot)$ denotes the integration of all 2-hop paths between the two involving nodes[11]. The distinction between $|\mathrm{CN}|$ and $d_{topo}$ lies in weighting of 2-hop paths, with $d_{topo}$ exhibiting a higher logarithm of thermodynamic activity and thus closer to the energy distance (approaching the line of $1-r$). (**B**) The topological similarity distribution between nodes in a randomly generated undirected graph ($\mathbb{R}^{500\times500}$). (**C**) The relationship between the union of neighbors and $d_{topo}$ for the sampled node pairs, with corresponding topological similarities around 5%, 10%, and 15%. (**D**) The relationship between the union of neighbors and $|\mathrm{CN}|$ for the sampled node pairs, with corresponding topological similarities around 5%, 10%, and 15%. (**E**) The relationship between the topological similarity and $d_{topo}$. (**F**) The relationship between the topological similarity and $|\mathrm{CN}|$.

## Using NR for network enhancement

The ability of NR to capture nodes' global connectivity, topological structure, and degree enables us to use it to enhance a variety of complex networks.

**Link prediction in complex networks.** Many real-world applications require the use of complex networks to describe social, biological, and information systems[24]. Link prediction models represent a critical challenge in network analysis, as they aim to capture the distribution of links within a network and predict potential links' presence[25]. Real-world networks inevitably possess structural defects, including errors, biases, and redundancies, due to the presence of noise and missing data during the construction of these networks using observed data[26,27]. To validate the utility of the TopoLa distance in real-world networks, we conducted link prediction experiments on nine real-world networks in this section. These networks represent relationships between entities ranging from US airlines to metabolites in E. coli (Supplementary text 1.1).

The reason $\boldsymbol{D}_{topo}$ can be used to enhance networks is that, when deriving the convergence conditions for Random Walk with Restart[28] ($||\boldsymbol{P}_0 - \boldsymbol{C}\boldsymbol{P}_0||_F^2 < \varepsilon$, Supplementary text 3.1, Theorem 2), we find that when $\boldsymbol{C} = \boldsymbol{D}_{topo}$, it not only ensures convergence but also dilutes the impact of the network's "ill-conditioned" on the closed-form solution of Random Walk with Restart (Supplementary text 4). Most related work approximates the closed-form solution to accelerate computations on large graphs, aiming to provide faster calculations while ensuring the approximate solutions fall within a theoretical error range[29,30]. Currently, there is little research focused on directly improving the closed-form solution itself. Therefore, proposing an enhanced closed-form solution to improve RWR performance is necessary. Consequently, we propose TRWR, a new closed-form solution for Random Walk with Restart (RWR). TRWR is defined as follows:

$$\mathrm{TRWR} = (1-\alpha)(\boldsymbol{I}-\alpha\boldsymbol{W})^{-1}\boldsymbol{D}_{topo}\boldsymbol{A}$$

$$= (1-\alpha)(\boldsymbol{I}-\alpha\boldsymbol{W})^{-1}\boldsymbol{A}^{*} \quad (4)$$

It is clear that TRWR is equivalent to initially enhancing the network using NR ($\boldsymbol{D}_{topo}\boldsymbol{A}$), followed by conducting an RWR. This provides evidence that $\boldsymbol{D}_{topo}$ can enhance networks.

To evaluate the performance of NR, we compared two closed-form solutions: $(1-\alpha)(\boldsymbol{I}-\alpha\boldsymbol{W})^{-1}\boldsymbol{A}$ (RWR) and $(1-\alpha)(\boldsymbol{I}-\alpha\boldsymbol{W})^{-1}\boldsymbol{A}^{*}$ (TRWR). We tested the prediction results using the ten-fold cross-validation. It has been observed that across nine real-world networks, TRWR shows enhanced prediction performance compared to RWR. Specifically, the area under the curve (AUC) improves by 6.8%, and the area under the precision-recall curve (AUPR) experiences a significant average increase of 102%. This demonstrates that NR enhances networks, leading to significant performance improvements for RWR in link prediction tasks.

To visualize the enhancement brought by NR, we used t-SNE to compare the spatial distribution of nodes before and after network enhancement[31]. As depicted in Fig.3B, the upper and lower left images display the original networks derived from the PB[32] and Yeast[33] datasets respectively, while the upper and lower right images showcase their respective enhanced counterparts. Notably, the enhanced networks exhibit a more clustered spatial distribution, forming compact clusters. This indicates that NR enhances the network by clustering points that are close in latent hyperbolic space.

To compare NR's performance with existing methods, we evaluated it against linear network enhancement methods: Diffusion State Distances (DSD)[34], network deconvolution (ND)[35], and network enhancement (NE)[36]. DSD identifies data structure through diffusion distance, ND augments both global and local network quality via a deconvolution process, and NE applies random walks combined with regularized information flow for network enhancing. It is noteworthy that the computational formula used in the NR-related experiments is consistent with that of TRWR. In the experiment, we initially enhanced the network, and then applied random walk with restart (RWR) on the reweighted edges for link prediction. We adopted the ten-fold cross-validation to evaluate the prediction outcomes. As shown in Fig.3C, except for DSD, the average prediction performance of the other three algorithms improved. DSD performs poorly because it calculates diffusion state distances using random walk, and performing additional RWR on its distance matrix leads to information loss. Furthermore, we find that NR outperforms the other methods, surpasses the second-best algorithm (NE) by a margin of 22.4%.

The above experiments demonstrate that NR exhibits superior performance in link prediction. Additionally, to explore the applicability of energy distance measures in network enhancement, we compared $|\mathrm{CN}|$ and $d_{topo}$ (Supplementary text 5). The results show that the common neighbor matrix did not significantly enhance prediction performance. This indicates that fine-grained energy distance measure is more suitable for network enhancement.

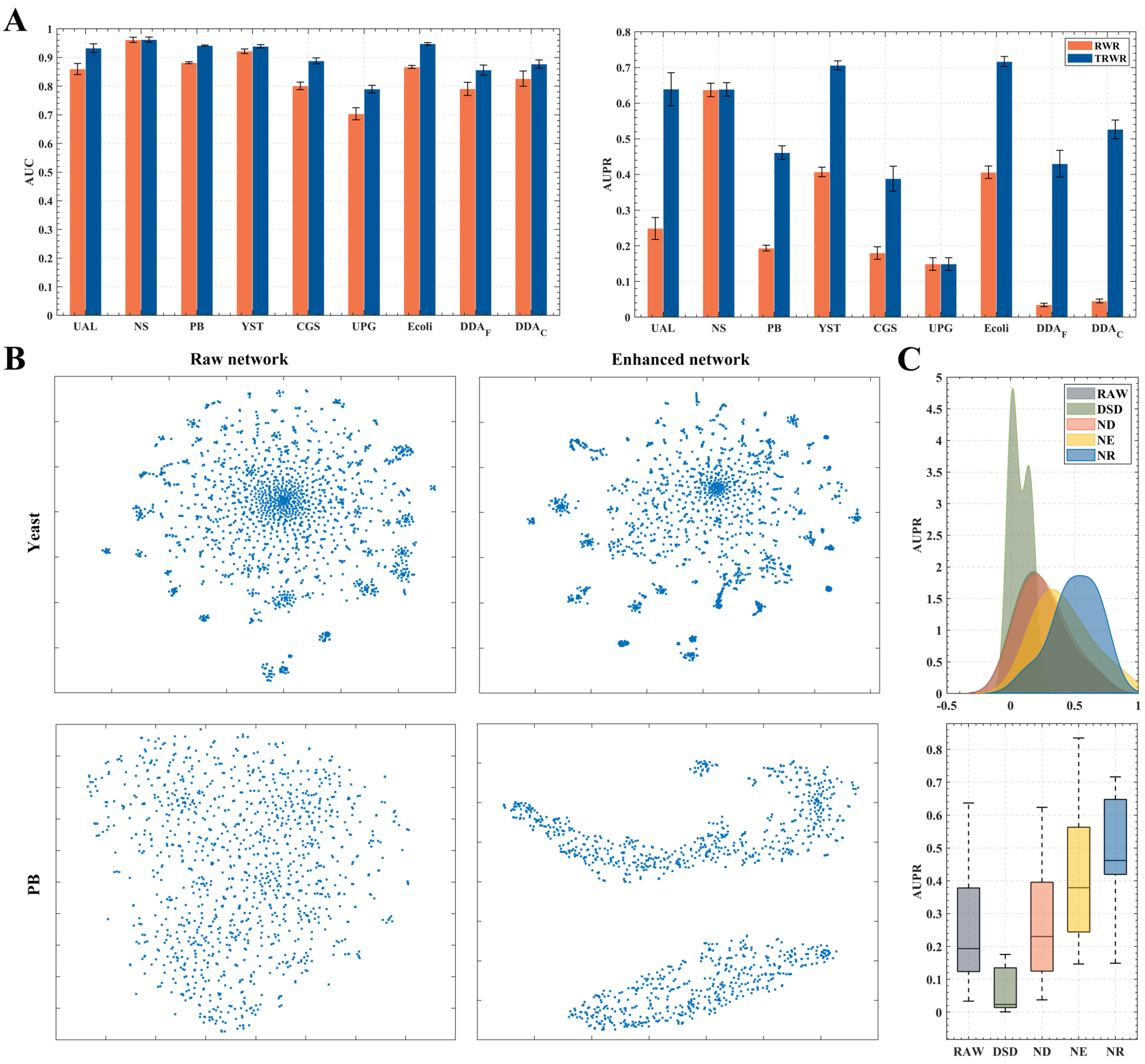

**Fig.3 Link prediction in real-world complex networks.** (**A**) The performance of TRWR and RWR in link prediction across nine distinct complex networks is assessed using two commonly utilized metrics: the area under the curve (AUC) and the precision-recall curve (AUPR). Variability in link prediction outcomes is represented by error bars. (**B**) Visualization of networks before and after enhancement by NR shows the spatial distribution of nodes within two real-world complex networks, PB and Yeast. Initially, the networks are depicted in the upper left and lower left panels. After undergoing an enhancement process, the post-enhancement networks exhibit a more pronounced clustering of nodes in their spatial arrangements. (**C**) Compare the network enhancement performance of DSD, ND, NE, and NR in link prediction tasks. The kernel density plot reflects the distribution of the precision-recall curve (AUPR), while the box plot reflects the median and outlier information of AUPR.

**Single-cell clustering via single-cell RNA-seq data.** The application of single-cell technologies allows us to gain a deeper understanding of the heterogeneity and functional diversity among individual cells. It facilitates studies on processes such as rapid differentiation, evolution into diverse subpopulations, or responses to external perturbations[37].

To confirm the applicability of NR in single-cell RNA-seq analysis, we applied it to single-cell data and assessed the impact of network enhancement on the spatial distribution of the data. We obtained 30 single-cell RNA-seq datasets from various species and tissues, utilizing different sequencing platforms (Supplementary text 1.1). We employed the SIMLR algorithm[38] as a baseline method to evaluate the enhancing performance of the existing methods (DSD, ND, NE) as well as NR.

The results show that on average, DSD, ND, and NE do not effectively improve the clustering quality of single-cell data. In contrast, NR exhibits superior performance, outperforming baseline method, DSD, ND, and NE. Notably, NR improved the Adjusted Rand Index (ARI) by an average of 12.6% compared to the second most effective enhancement method, ND, and by 10.3% compared to the baseline method. Additionally, in terms of Normalized Mutual Information (NMI) (Supplementary text 2.2), NR outperformed the baseline method by 6.2% and the second most effective method, ND, by 7.1%. The results indicate that NR is more suitable for single-cell data analysis compared to existing methods. The comparative performance of the four methods and SIMLR (Raw) across four datasets is illustrated in Fig.4A. Results for an additional 26 datasets are provided in Supplementary text 6.

To further visualize the enhancement effect of NR, we analyzed the Baron_human dataset, as illustrated in Fig. 4B. Here, we employed t-SNE as a visualization tool to depict the similarity network[31]. We observed that various cell types, particularly the 'quiescent_stellate' type, become more spatially clustered. Notably, the 'quiescent_stellate' initially exhibits blurred boundaries, which become significantly distinct from other categories after enhancement through NR. The results further demonstrate the suitability of NR for single-cell data analysis.

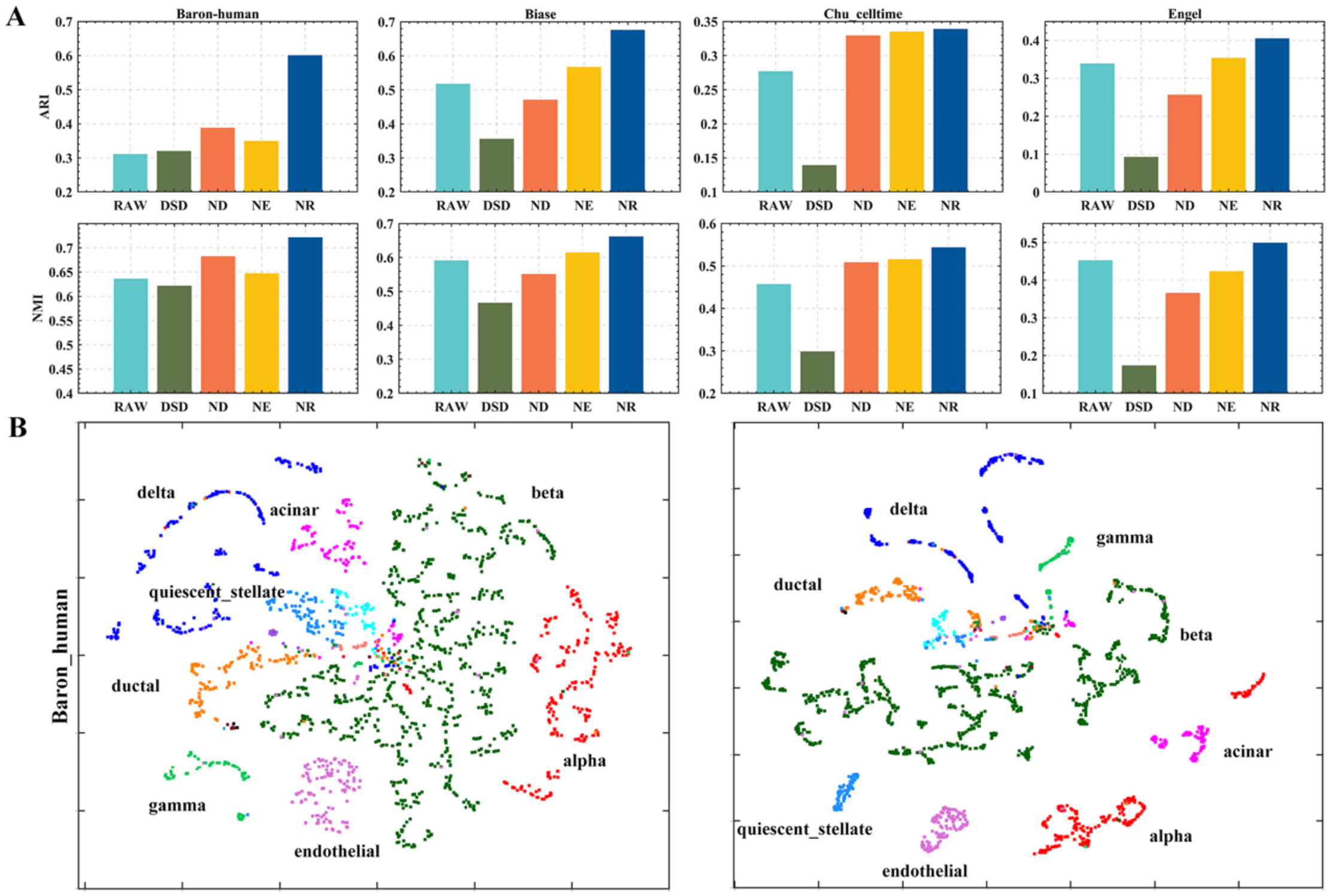

**Fig.4 Identification of cell types using single-cell RNA-seq data**. (**A**) The performance of the original and enhanced cell similarity networks in identifying cell types. The results are shown for four single-cell datasets. Results for an additional 26 datasets are provided in Supplementary text 6. (**B**) Visualization of single-cell RNA-seq data before and after enhancement. The left panel illustrates the spatial distribution of cell types in the Baron_human dataset, using distinct colors for each type; the right panel shows their distribution after NR enhancement.

## Using NR to complement existing methods for network enhancement

Various enhancing methods are available, each tailored from different perspectives. This section assesses the synergistic potential of NR with these varying approaches. By integrating NR, we seek to heighten the efficacy of existing methods. This synergistic effect has been confirmed through applications in two real-world contexts. Furthermore, we have undertaken experiments in link prediction and single-cell analysis (Supplementary text 7).

**Domain identification via Hi-C networks.** In the processes of DNA replication and transcription, the chromatin's 3D structure plays an essential role[39]. The Hi-C technique can be used to determine the three-dimensional structure of the genome. It reveals that large amounts of DNA are folded into discrete 3D domains, which are crucial for understanding the genome's role in gene regulation[40]. Advanced algorithms aggregate sequence reads at a specific resolution and convert them into Hi-C matrices at a specific resolution, called contact maps, where genomic regions are nodes and normalized counts of reads mapping between two regions are weighted edges[39]. Contact maps facilitate the discovery of structural units within chromatin called topologically associated domains (TAD)[41]. Alterations in TAD boundaries have been demonstrated to be linked to cancer or developmental disorders[42]. Accurately detecting the boundaries of TADs is essential for the analysis and interpretation of Hi-C data. However, technical challenges, including a limited number of Hi-C reads, the hierarchical structure of TADs, and the presence of noisy Hi-C networks hinder the accuracy of TAD detection[36].

In this study, we applied NR to complement three existing network enhancement methods and assessed its impact on the detection of TADs. For our experiments, we utilized 1kb resolution Hi-C data of all autosomes from GM12878 cell line[43] compiled by Wang *et al.*[36], which is a synthetic dataset created by concatenating non-overlapping clusters detected in the previous work (Supplementary text 1.1). To compare the impact of different methods on the quality of detected TADs, we used the Louvain community detection method[44]. We first visualized the raw network (RAW, based on Hi-C data from a portion of chromosome 11) and three enhanced networks (DSD, ND, and NE). Additionally, we displayed networks that were first enhanced using the NR method, followed by further enhancement with existing methods (DSD+NR, ND+NR, and NE+NR), as shown in Fig.5A. We find that the network enhanced by NR has clearer boundaries, especially NE+NR.

To quantify the clarity of boundaries, we utilized normalized mutual information (NMI) to compare the communities derived from true clustering with those obtained from the networks after enhancement (Supplementary text 2.2, and Fig.5B). Overall, NE+NR demonstrated the best performance, achieving an average NMI of 0.975 in autosomal data. This represents an 8% improvement over the baseline method and a 1.6% improvement over NE. It is worth noting that ND+NR exhibits a 13.3% improvement over ND. In contrast, although DSD+NR eliminates some noise visually, it has little effect on NMI values.

Our results suggest that using NR-enhanced data can effectively improve TAD detection and the enhancing performance of existing methods. This indicates the potential of the proposed method to enhance network data and identify noise that is not easily distinguishable.

**Fine-grained species identification.** Fine-grained species identification is an essential task that aims to distinguish between individuals of the same species based on subtle differences found in images[45]. Traditional image retrieval methods are only useful for distinguishing between high-level categories, whereas fine-grained image retrieval requires more precise distinctions to differentiate between categories with subtle differences[46]. However, the high similarity between subordinate categories makes this task more challenging. Similar shapes, subtle color differences, or similar colors and textures can lead to misidentification[47]. Furthermore, changes in viewpoint, scale, and occlusion exacerbate the difficulty of this task[48]. To effectively retrieve samples from the correct species in fine-grained species identification, it is necessary to address issues such as noise and inefficiency in similarity networks.

To assess the performance of NR in addressing the aforementioned issues, we performed experiments using a hand-curated dataset of butterfly fine-grained species images comprising 832 pictures that encompass 10 distinct subspecies[48]. We employed encoding methods previously used in related studies to create the initial similarity network[36]. More specifically, we utilized Fisher Vector (FV) and Vector of Linearly Aggregated Descriptors (VLAD) with dense SIFT to encode the images and extract similarity information[49,50] (Supplementary text 1.1). The extracted similarity information is then used to construct the similarity network through an inner product.

To comprehensively evaluate the detection performance of each method, we used mean retrieval accuracy (Supplementary text 2.3) to assess the original network, the networks enhanced by three existing methods (DSD, ND, and NE), and the networks enhanced by NR and then further enhanced by other techniques (DSD+NR, ND+NR, and NE+NR) in Fig.5C. Experiments show that NE+NR achieves the best performance, improving by 50.1% over the baseline method (RAW). In addition, the introduction of NR leads to improved performance for three existing methods. Specifically, NR improves the retrieval accuracy of the ND method by 18.4%, the NE method by 2.2%, and the DSD method by 0.02%. These experiments show that NR, as a network enhancement algorithm, improves the enhancing performance of existing algorithms.

Furthermore, we evaluated the accuracy of the top $k$ identified butterflies (Fig.5D). Specifically, when identifying the top 40 butterflies, NE+NR shows a 20.27% improvement over the original network and a 1.35% improvement over NE, the second-best method. As the number of identified butterflies increases, the improvement provided by the NR algorithm also increases. For the top 80 butterflies, the ND+NR method improves accuracy by 17.9% over ND, while NE+NR improves by 2% over NE. Notably, NE+NR exhibited the highest overall performance improvement, achieving a 45% increase compared to the original network.

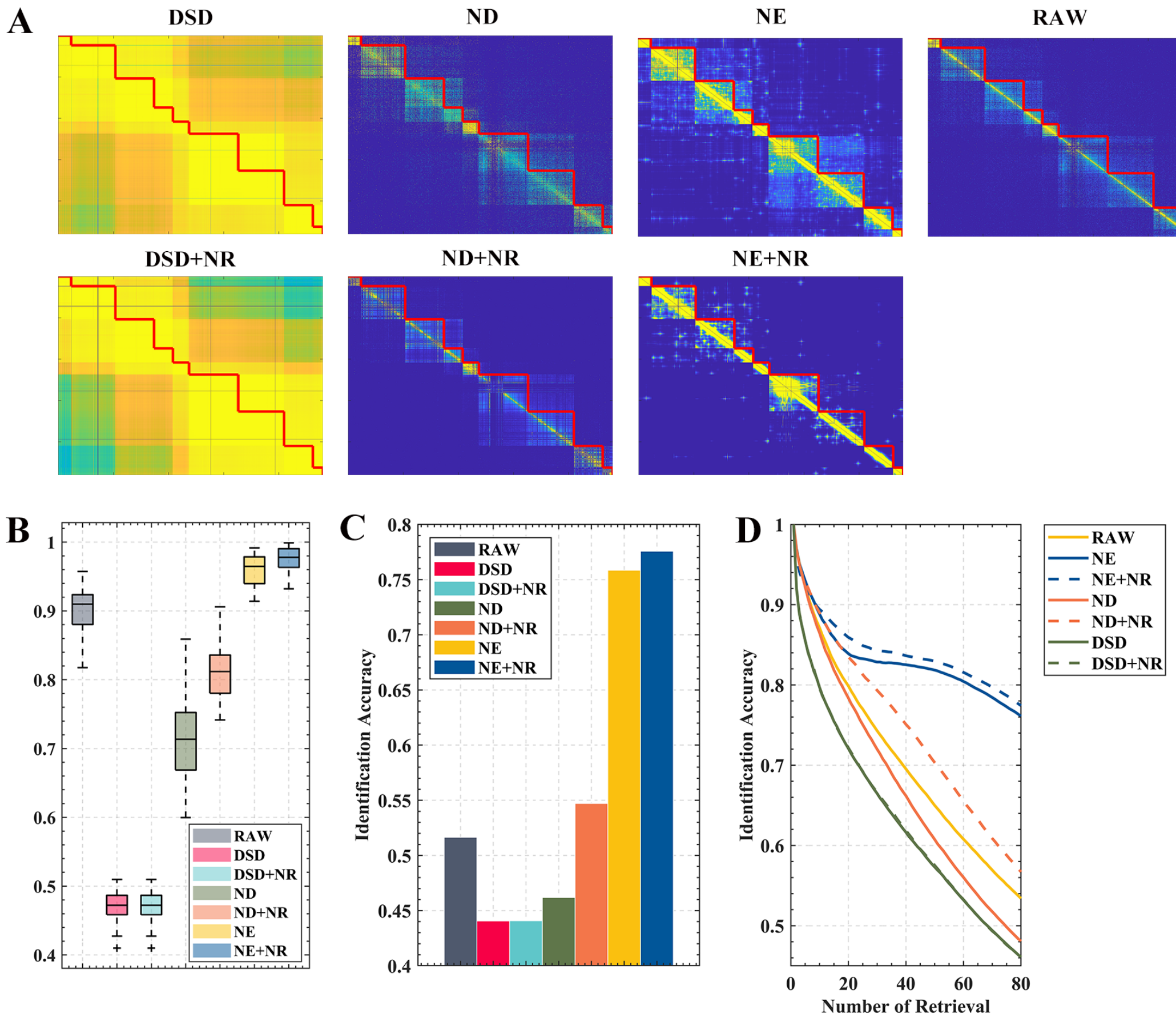

**Fig.5** Using NR to complement existing methods. (**A**) Visualization of Hi-C networks before and after enhancement. Heatmaps depict the enhanced Hi-C contact matrices for a chromosome 11 segment, showcasing the original network, the networks enhanced by three existing methods (DSD, ND, and NE), and the networks enhanced by NR and then further enhanced by other techniques (DSD+NR, ND+NR, and NE+NR). Red lines mark domain boundaries, with nodes and edges representing 1 kb genomic windows and their normalized read counts, respectively. (**B**) Performance evaluation of boundary detection. (**C**) Average species identification accuracy. Identification accuracy was assessed by the average probability that the top N similar butterflies to each target were of the same species (Supplementary text 2.3), where N equals the number of butterflies in the target species category. (**D**) Species identification accuracy for different numbers of retrievals. The original network, the networks enhanced by three existing methods (DSD, ND, and NE), and the networks enhanced by NR and then further enhanced by other techniques (DSD+NR, ND+NR, and NE+NR), were compared. Each curve in the analysis represents the species recognition accuracy for retrieving varying numbers of images.

## Using NR for enhancing low-rank matrix

In this section, we extended NR to low-rank matrix completion, which aims to recover missing entries from partially observed matrices. We used the feature and nuclear norm minimization model (FNNM) as a baseline method [51] and refine the resulting matrices using NR (named as FNNM+NR). Since DSD, ND, and NE cannot enhance asymmetric matrices, they were not included. We validated our approach on two tasks: movie recommendation and multi-label learning.

For movie recommendation task, we utilized the MovieLens dataset (100K)[52], which includes 100,000 ratings from 943 users on 1,682 films. This dataset contains 23 user attributes, such as age and occupation, and 20 film characteristics like genre and release date (Supplementary text 1.2). We designated $sr$ as the sampling rate for known ratings, varying it from 0.1 to 0.9 in 0.2 increments. From the rating matrix, known entries form a training set of $10^5 \times sr$ ratings and a separate test set with $10^5 \times (1 - sr)$ ratings. The test set's performance is quantified using root mean squared error (RMSE, Supplementary text 2.4). As shown in Fig.6A, FNNM can capture the mapping between side information of users and movies a small number of data samples, and increasing the number of samples significantly improves accuracy. FNNM+NR further enhances the relationship between users and movies while retaining FNNM's inherent advantage of integrating inductive and transductive completions[51]. Across varying sampling rates, FNNM+NR consistently yields the lower RMSE values compared to FNNM.

For multi-label learning task, we sourced ten datasets from "yahoo.com" for web page classification[53], covering categories from 'Arts' to 'Social'. Taking the 'Arts' category as an example, it contains 5,000 instances, each with 462 dimensions and 26 labels. From these, we reserved 10% for testing, with the balance allocated for training. During training, we selectively labeled instances; for each label, $\omega$% of both positive and negative instances were chosen, leaving the rest unlabeled. This fraction, $\omega$%, varies from 10% to 90% in 20% increments. Performance is assessed using Average Precision (AP) across all datasets (Supplementary text 2.5). Our observations show that NR consistently improves the performance of FNNM across varying ω%. Remarkably, across all datasets, NR achieves an average performance improvement of 10.5% at a 10% sample size (Fig. 6B). This demonstrates that NR significantly enhances existing methods even with limited data availability.

Our results indicate that NR not only improves the performance of the compared method but also preserves its characteristics. This suggests that the proposed approach has great potential for enhancing low-rank matrices.

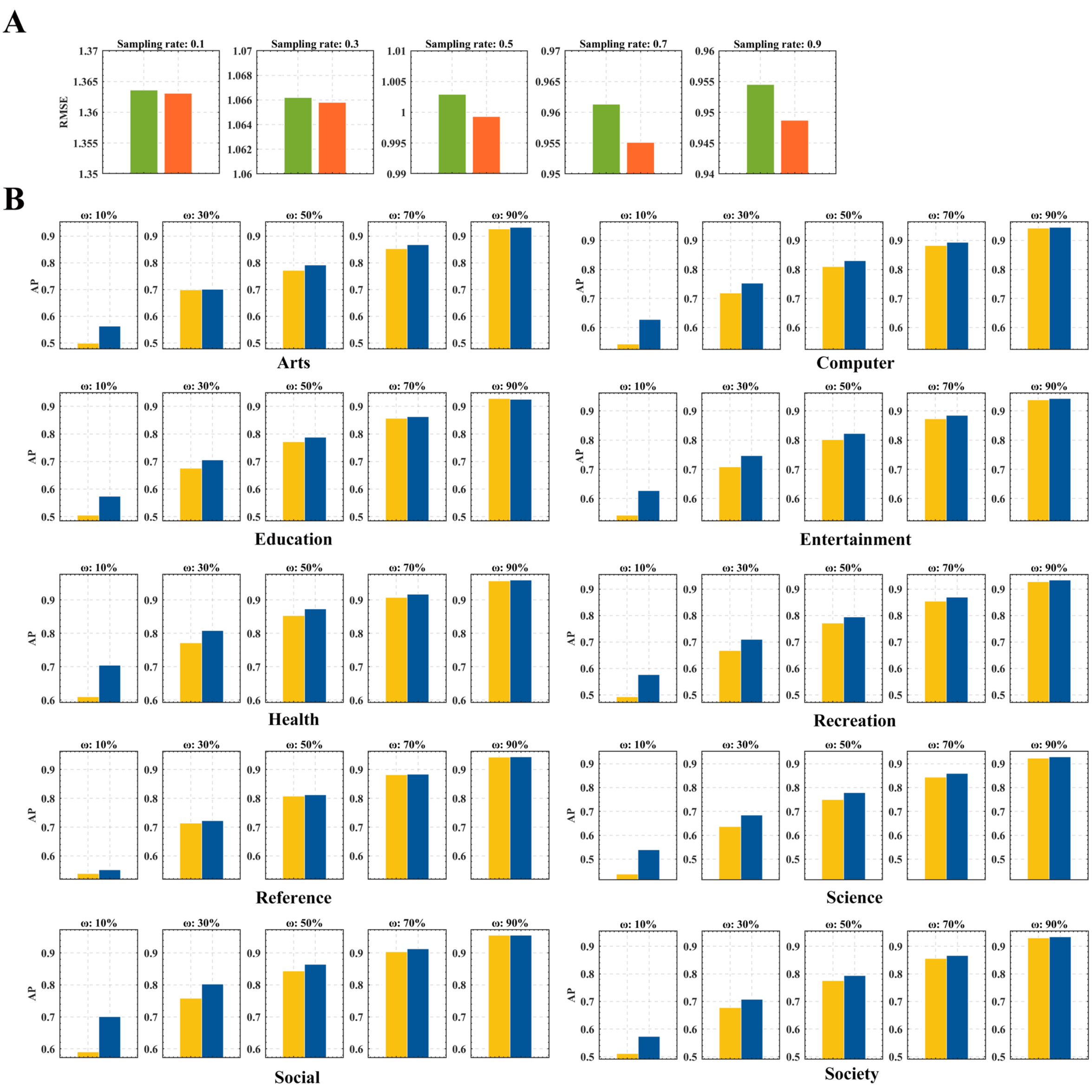

**Fig.6 Utilizing low-rank matrix completion for movie recommendations and multi-label learning.** (**A**) The performance of FNNM and FNNM+NR in movie recommendations. Each bar in the graphical representation denotes the performance of individual models, assessed using RMSE (supplementary text 2.4). Green bars illustrate FNNM's performance and orange bars denote FNNM+NR's. (**B**) The performance of FNNM and FNNM+NR in multi-label learning. Model efficacies are encapsulated in bar charts, measured using the AP (supplementary text 2.5). Yellow bars illustrate FNNM's performance and blue bars denote FNNM+NR's.

## Using the TopoLa distance to improve deep learning models

In this section, we extend the TopoLa distance as a sample relationship measure to the domain of deep learning.

**Knowledge distillation**. Knowledge distillation involves transferring insights from a larger

"teacher" model to a more compact "student" model[54]. Advances in artificial intelligence are significantly anchored by deep neural networks[55]. As such, leading models often impose notable computational and memory costs during inference. Distilling knowledge from these teacher models into streamlined student versions presents a feasible solution to these computational challenges.

In this section, we adopted the Relational Knowledge Distillation (RKD) model as the benchmark, which includes two distinct losses: distance-wise and angle-wise distillation losses. We utilized the TopoLa distance as the relational measure (named RKD-Topola), replacing the Euclidean distance in the distance-wise distillation loss. We then assessed the performance of the adapted model through image retrieval tasks. Specifically, we evaluated our method using the image retrieval datasets CUB-200[56] and Cars-196[57], adhering to the train/test splits proposed in prior research and employing recall@K for assessment[58] (Supplementary text 1.3 and 2.6).

Across the CUB-200 and Cars-196 datasets, we assessed baseline models: ResNet50 (the teacher model) and ResNet18 (the student model), alongside two knowledge distillation techniques: Relationship Knowledge Distillation (RKD)[58] and RKD-Topola (Fig.7A). We illustrate the measurement of relationships between samples using TopoLa distance in Fig. 7B.

In the CUB-200 dataset, the teacher model achieves an accuracy of 61.24%, whereas the student model posts 51.55%. With RKD, the accuracy rises to 55.97%. Significantly, RKD-Topola method enhances the performance to 59.25%, marking a 14.9% enhancement over the baseline performance of the student model and surpassing RKD's performance by 5.8%. In the Cars-196 dataset, the teacher model and student model report accuracies of 77.17% and 64.53%, respectively. RKD advances accuracy to 80.10%. Notably, the RKD-Topola further elevates the accuracy to 81.46%, achieving a 26.2% improvement over the student model.

To visualize the changes in embeddings brought by relational measures, we used t-SNE to perform dimensionality reduction on the embeddings from RKD and RKD-Topola across the two datasets. Observations indicate that the TopoHyper distance effectively delineates relationships between samples (Fig.7C). For instance, in the Cars-196 dataset, RKD-Topola shows superior recognition of "Ford" models compared to RKD. Similarly, in the CUB-200 dataset, RKD-Topola provides more clustered and distinct embeddings for the Louisiana Waterthrush and Bohemian Waxwing compared to other categories. This strategy enhances the compactness of embeddings from identical categories and strengthens the distinction among disparate classes.

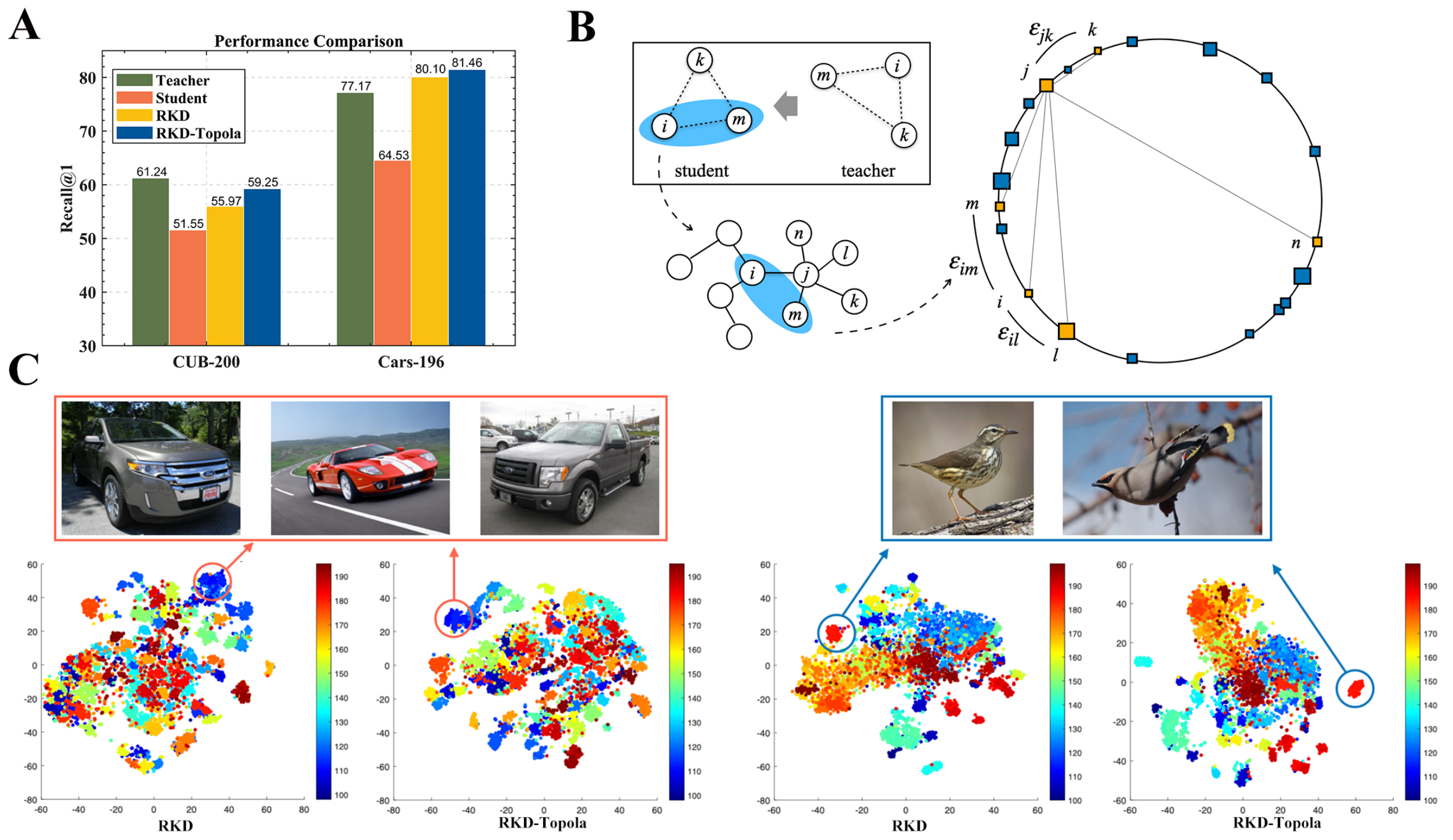

**Fig.7 Utilizing knowledge distillation for image retrieval.** (**A**) The performance of Relational Knowledge Distillation (RKD) with its modified variant, RKD-Topola, as well as teacher (baseline) and student (baseline) models, on the CUB-200 and Cars-196 datasets. (**B**) Using TopoLa distance to measure the relationships between samples. The embeddings of the samples is abstracted into a complex network composed of samples and features. This network is subsequently leveraged to calculate the distance between two samples in the latent space. (**C**) Visual representations of embeddings from RKD and RKD-Topola are presented. The left panel showcases results from the Cars-196 dataset, while the right corresponds to the CUB-200 dataset.

**Contrastive learning**. Contrastive Learning capitalizes on discerning instances by emphasizing similarity within positive pairs and disparity among negative pairs in the embedding space[59,60]. Drawing inspiration from the instance discrimination paradigm, wherein each image is considered its own class, the performance of contrastive learning hinges on the judicious assembly of positive and negative pairs[61]. Techniques such as MoCo, by adopting momentum-based strategies, refine the array of negative samples, thereby elevating performance standards in this field[62].

In this section, we adopte the MoCo v3 as the benchmark[63], introducing the TopoLa distance to replace the vetorial angle in the Euclidean space for delineating relationships between positive and negative samples (named as MoCo-Topola). Figure 8B illustrates the measurement of relationships between positive and negative samples using the TopoLa distance. We applied our methodology to image classification tasks using the CIFAR-10 and CIFAR-100 datasets, adhering to established training/testing splits (Supplementary text 1.3). We evaluated the model using linear evaluation accuracy (Supplementary text 2.7). The experiment result shows that MoCo v3 achieves accuracies of 92.48% and 70.04% on the CIFAR-10 and CIFAR-100 datasets, respectively, the MoCo-Topola model, augmented by the TopoLa distance, posts results of 92.83% and 71.05% (Fig. 8A). The MoCo-Topola model improves accuracy by 0.4% on the CIFAR-10 dataset and by 1.4% on the CIFAR-100 dataset compared to MoCo v3.

To visualize the changes in embeddings brought about by relational measures, we used t-SNE for dimensionality reduction on the embeddings from MoCo and MoCo-Topola across the two datasets. Observations indicate that the TopoLa distance effectively minimizes intra-class discrepancies and accentuates inter-class distinctions (Fig.8C). For instance, in the CIFAR-10 dataset, MoCo-Topola produces notably compact embeddings for bird images, enhancing their semantic alignment. Additionally, in the CIFAR-100 dataset, the embeddings for "cockroach" images by MoCo-Topola stand out distinctly from those of other categories. These findings demonstrate the potential of the TopoLa distance to enhance intra-class compactness of embeddings and magnify inter-class distinctions.

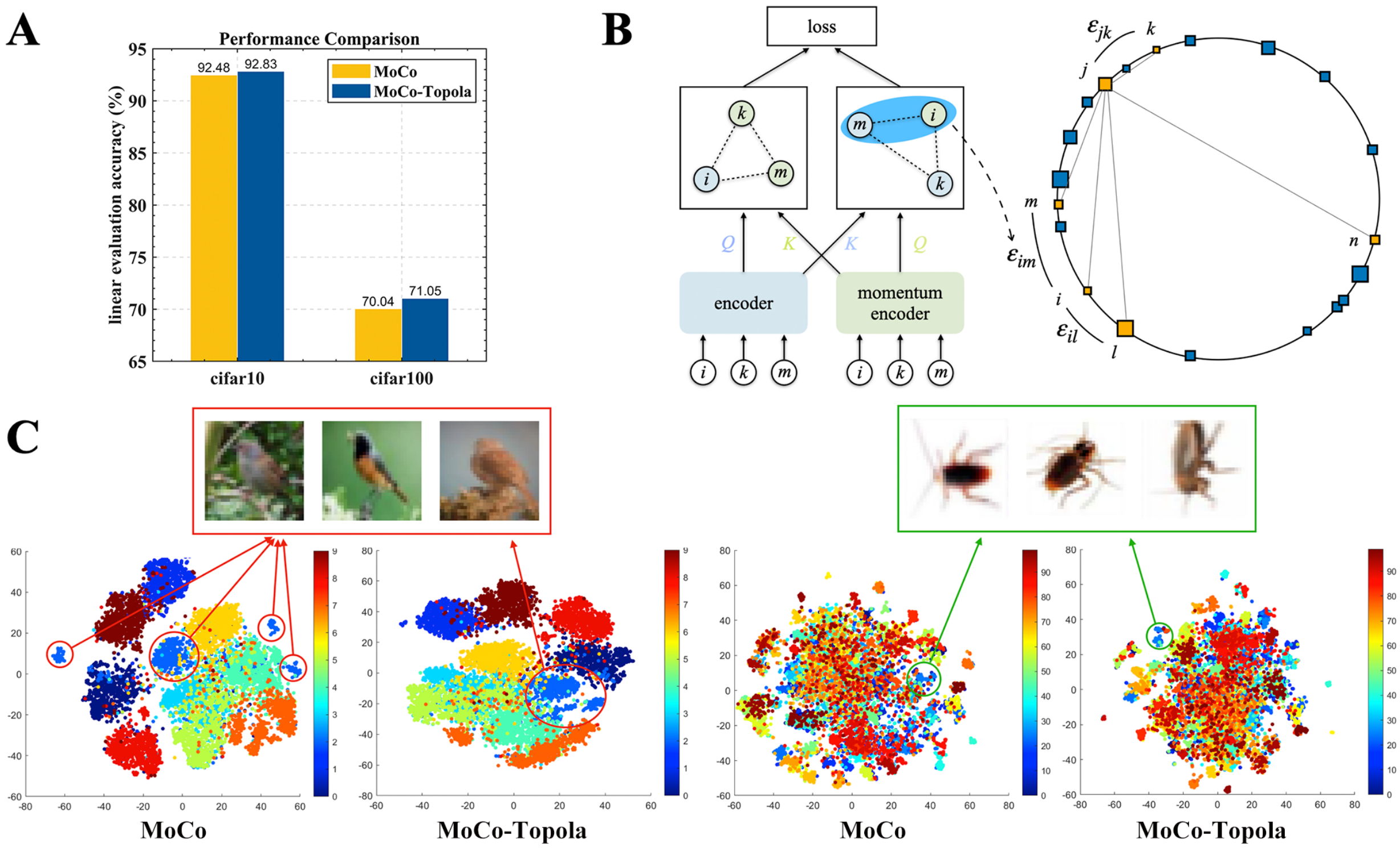

**Fig.8 Utilizing contrastive learning for image classification.** (**A**) We evaluated MoCo v3 against MoCo-Topola, which employed the TopoLa distance, on the cifar10 and cifar100 datasets. Each bar in the graphical representation denotes the performance of individual models, assessed using linear evaluation accuracy (Supplementary text 2.7). (**B**) Utilizing the TopoLa distance to evaluate relationships between positive and negative samples. Through embedding matrices, we forged a complex network of samples and their features. This structure then facilitates the computation of inter-sample distances within the latent space. (**C**) Visual representations of embeddings from MoCo and MoCo-Topola are presented. The left panel showcases results from the cifar10 dataset, while the right corresponds to the cifar100 dataset.

## Discussion

In this study, we demonstrated a direct correlation between a node's global topological structure and its spatial position within the latent space. Specifically, by employing the maximum entropy principle [11,64], we demonstrated that by applying weighted statistics to 2-hop paths, the logarithm of thermodynamic activity increases, resulting in the common-neighbor integral more closely approximating the complement of the energy distance. This suggests inadequacies of the existing 2-hop path statistics for precise depiction of the latent space energy distances.

Subsequently, we integrated even-hop paths into energy distance calculation, leading to the development of a new embedding framework TopoLa. The TopoLa distance not only reflects traditional 2-hop path statistics but also captures the global connectivity information and the degree of topological similarity between two nodes (Supplementary text 3.1, Theorem 1 and Theorem 3). In relational knowledge distillation, we introduced TopoLa distance as an alternative relational measure between samples, replacing the commonly used Euclidean distance. Within the MoCo v3, we specifically leveraged the TopoLa distance as an alternative to vector angles in Euclidean space. Our results confirm that the TopoLa distance enhances image retrieval and classification accuracy, attesting to its effectiveness.

Furthermore, we identified a propensity for nodes to establish connections with neighbors of those nodes possessing similar topological structures. Based on this observation, we propose the NR algorithm. Mathematically, the essence of NR is that the operation process does not change the left and right singular vectors, but only increases the singular value gap. Specifically, the singular value of the enhanced network is $f_\sigma(i) = \sigma_i \frac{\sigma_i^2}{\sigma_i^2+\lambda}$, where $\sigma_i$ is the $i$th singular value and $\lambda > 0$ is the parameter of NR. The singular value is reduced overall, while we proved that the singular value gap increases. It indicates that NR reduces the weight of small singular values more actively, which is beneficial for enhancing. In addition, by matrix perturbation analysis, we demonstrated that the error upper bound of the enhanced network after adding perturbations is less than or equal to that of the original network, where $\boldsymbol{H}$ and $\sigma$ are the perturbation matrix and the singular values, respectively. This result indicates that the enhanced network is more resilient to perturbations, and thereby more robust against noises and data missing. To alleviate the computational burden of matrix inversion, we introduce a novel approach called fastNR that utilizes a random algorithm for low-rank matrix approximation[23]. We demonstrated that this method considerably speeds up the computation.

Topology-encoded Latent Hyperbolic Geometry is a foundational tool in network analysis. It could assist in clarifying inter-node relationships in matrix-focused research, presenting a novel perspective to the methodologies in pertinent fields. Additionally, we similarly expect that this emergent measure (i.e., the TopoLa distance) offers fresh insights into network science, machine learning and other related disciplines in the future.

# Methods

## Data involved in the experiments

In this study, we examined seven tasks from different disciplines, each employing unique datasets. What follows is a description of these datasets.

**Complex networks in link prediction.** This part of the experiment involved nine complex networks collected from various fields, including transportation, biology, social science, and other scientific domains. The datasets used in this study include: UAL - a network of the US air transportation system that consists of attributed nodes (airports) and links between two airports[65]; CGS - a network formed by the neurons present in the C.elegans worm[1]; UPG - an electrical grid network of western US representing the network describing high voltage transmission among generators, transformers, and substations[66]; PB (Political Blog) - a graph representing the network among political blog pages in the US where the blog pages are identified as nodes and hyperlinks between the blog pages are identified as links of the graph[34]; NS (Net Science) - a collaboration network of researchers who publish papers on network science[1]; YST - a protein-protein interaction network present in yeast[67]; Ecoli - a network of metabolite reactions in Escherichia coli bacteria[68]; $\mathrm{DDA_F}$ - a network of associations between drugs and diseases collected by Gottlieb *et al.*[69]; $\mathrm{DDA_C}$ - a network of associations between drugs and diseases collected by Luo *et al.*[70].

**Single-cell RNA-seq data.** Thirty single-cell RNA sequencing (scRNA-seq) datasets from multiple public websites were collected, including NCBI Gene Expression Omnibus (GEO)[71], ArrayExpress[72], and Sequence Read Archive (SRA)[73] (Table S1). We also acquired the 10X PBMC dataset, which was obtained from the 10X genomics website[74].

**Hi-C networks for domain identification.** For each autosome, we utilized the contact matrix (counts per bin) provided by Wang *et al.*[38]. Firstly, they generated clusters that could potentially overlap using the Arrowhead algorithm. To obtain labels with high group confidence, they subsampled the top 15 non-overlapping clusters for each chromosome that did not contain subclusters. Non-overlapping regions were selected because the community detection algorithm Louvain used for post-processing could only identify such regions[58].

**Butterfly image dataset for fine-grained species identification.** In this section, we utilized the relevant data provided by Wang *et al.*[38]. Specifically, they processed a dataset of 10 different categories of butterflies, each containing 55 to 100 images, totaling 832 butterflies [75].

**Datasets for low-rank matrix completion.** The MovieLens dataset comprises 100,000 ratings from 943 users across 1,682 films[52]. Each user rated at least 20 films, with accompanying demographic details including age, gender, occupation, and zip code. In multi-label learning, we assess the performance of both FNNM and rFNNM using eleven datasets sourced from "yahoo.com" designed for web page categorization, spanning categories such as 'Arts', 'Business', and 'Science'[53].

**Image retrieval dataset for knowledge distillation.** We employed the CUB-200-2011[56], Cars196[57], and Stanford Online Products[76] datasets as benchmark datasets for image retrieval. CUB-200-2011 dataset is the most widely used dataset for fine-grained visual categorization task. This dataset encompasses 11,788 bird images spanning 200 subcategories, with a division of 5,994 for training and 5,794 for testing. Each image is meticulously annotated, including a subcategory label, 15 part locations, 312 binary attributes, and a bounding box. The Cars196 dataset consists of 16,185 images across 196 unique car categories, defined by attributes like Make, Model, and Year. With an even distribution of 8,144 training and 8,041 testing images, it is a foundational resource for supervised learning systems targeting the recognition of vehicle types from images. The Stanford Online Products (SOP) dataset, sourced from e-commerce platforms, comprises 120,053 product images spanning 22,634 unique classes. It designates 59,551 images from 11,318 classes for training and allocates the other 60,502 images from 11,316 classes for testing.

**Image classification dataset for contrastive learning.** We utilized the CIFAR-10 and CIFAR-100 datasets as benchmark datasets for image classification[77]. The CIFAR-10 dataset, stemming from the Canadian Institute for Advanced Research, is a subset of the Tiny Images dataset and encapsulates 60,000 32x32 color images distributed across 10 distinct classes. The dataset provides an equal distribution with 6,000 images for each class, further subdivided into 5,000 training and 1,000 testing images. The unique categories delineated in CIFAR-10 include airplane, automobile (excluding truck or pickup truck variants), bird, cat, deer, dog, frog, horse, ship, and truck (again, excluding the pickup truck version). On the other hand, the CIFAR-100 dataset encompasses 60,000 32x32 color images, which are organized into 100 classes. These classes are further aggregated into 20 overarching superclasses. Each image is methodically annotated with both a "fine" label, indicative of its specific class, and a "coarse" label that designates its superclass category. For each of the 100 classes, there are 600 images, with a division of 500 for training and 100 for testing.

## TopoLa framework

Previous studies have established that with a large logarithm of thermodynamic activity, the expected number of common neighbors closely approximates the complement of the energy distance[11]. We demonstrate that, under the same network, applying weighted statistics to 2-hop paths yields more accurate measurements of energy distance compared to unweighted statistics (Supplementary text 3.1,

Theorem 1). Consequently, we incorporate node degree as a weighting factor in the computation of energy distance, and propose the TopoLa distance $\boldsymbol{D}_{topo}$ satisfying the following formula:

$$\boldsymbol{D}_{topo} + \boldsymbol{D}_{topo}^2 + \boldsymbol{D}_{topo}^3 + \boldsymbol{D}_{topo}^4 + \cdots = \frac{\boldsymbol{A}\boldsymbol{A}^T}{\lambda} \tag{5}$$

where $\boldsymbol{A}$ is the adjacency matrix of the original network, $\lambda$ is a coefficient modulating the influence of degree. $\boldsymbol{D}_{topo}^n$ is the matrix power of $\boldsymbol{D}_{topo}$, which represents the $n$-hop path matrix of $\boldsymbol{D}_{topo}$ and $\boldsymbol{D}_{topo}^n(i,j)$ is the weighted sums of all $n$-hop paths between $i$ and $j$ in $\boldsymbol{D}_{topo}$. $\boldsymbol{A}\boldsymbol{A}^T(i,j)$ is the weighted sums of all 2-hop paths between $i$ and $j$ in $\boldsymbol{A}$. Aggregating all hop paths not only introduces both global connectivity and degree information (see section-"Physical significance of the TopoLa distance") but also facilitates the derivation of a closed-form solution. This improves the importance of 2-hop paths for low-degree nodes. Through the Infinite Neumann series, $\boldsymbol{D}_{topo}$ can be solved as:

$$\boldsymbol{D}_{topo}\left(\boldsymbol{I} - \boldsymbol{D}_{topo}\right)^{-1} = \frac{\boldsymbol{A}\boldsymbol{A}^T}{\lambda} \tag{6}$$

$$\boldsymbol{D}_{topo} = \boldsymbol{A}\boldsymbol{A}^T(\lambda\boldsymbol{I} + \boldsymbol{A}\boldsymbol{A}^T)^{-1} \tag{7}$$

where $\boldsymbol{I}$ is the identity matrix. Intriguingly, this representation can be deduced within the convergence conditions of a random walk with restart (RWR) (Supplementary text 6). Therefore, we propose the TopoLa framework, which consists of latent space embedding, energy distance measure, and spatial specificity (such as the triangle inequality). Formula (7) enables the derivation of a closed-form solution for $\boldsymbol{D}_{topo}$, which allows us to calculate its exact value (The process for eigenvalue calculation is depicted in Fig.1B). Therefore, the topologically encoded latent hyperbolic geometry can offer a novel perspective for theoretical analysis in these domains. The TopoLa distance between node $i$ and node $j$ in the network is expressed as:

$$d_{topo}(i,j) = \frac{1}{\lambda}|2-\mathrm{hop}| - \frac{1}{\lambda^2}|4-\mathrm{hop}| + \frac{1}{\lambda^3}|6-\mathrm{hop}| - \cdots \quad (8)$$

Our findings reveal that $d_{topo}$ captures global connectivity information, which can be used to measure topological similarity between two nodes, a concept that we define as 'global connectivity-based topological similarity' (see section-"Physical significance of the TopoLa distance", Fig.1C). This establishes a clear correlation between the topological structure of nodes and their spatial positioning in latent space, notably clustering nodes with similar topology (Supplementary text 3.1, Theorem 3).

In summary, $\boldsymbol{D}_{topo}$ has the following two unique properties which are useful in a variety of applications:

1. $\boldsymbol{D}_{topo}$ captures the information of nodes' global connectivity, topological structure, and degree (see section-"Physical significance of the TopoLa distance").

2. $\boldsymbol{D}_{topo}$ enables the utilization of the chemical potential across various networks for energy distance measure, instead of employing a uniform metric scale for all (see section-"Physical significance of the TopoLa distance").

Through Theorem 3, we demonstrate that $\frac{||\boldsymbol{D}_{topo}(i)-\boldsymbol{D}_{topo}(j)||_F^2}{||\boldsymbol{y}||_F^2} \leq ||\boldsymbol{A}(i)-\boldsymbol{A}(j)||_F^2$, which further proves that $\boldsymbol{D}_{topo}$ reflects the nodes' topological structure (Supplementary text 3.1), where $\boldsymbol{D}_{topo}(i)$ is the $i$-th row of matrix $\boldsymbol{D}_{topo}$, denoting the set of distances between node $i$ and other nodes, $\boldsymbol{y}$ represents any row within matrix $\boldsymbol{A}$ and $\boldsymbol{A}(i)$ is the $i$-th row of matrix $\boldsymbol{A}$, denoting the topological structure of node $i$. Utilizing this framework, we have developed two network enhancement

algorithms called Network Reconstruction via Global Connectivity-Based Topological Similarity (NR) and the Fast Network Reconstruction via Global Connectivity-Based Topological Similarity (fastNR), respectively, for network analysis as follows.

**The network reconstruction algorithm via global connectivity-based topological similarity**

Different from conventional wisdom that node connection likelihood decreases with increasing latent space distance, evidence shows nodes preferentially connect with topologically similar neighbors (Supplementary text 6). To harness this, we introduce a technique, Network Reconstruction via Global Connectivity-Based Topological Similarity (NR), which utilizes TopoLa distance weights for single-step diffusion on networks. This process aims to reconfigure the nodes' topological structure based on latent space distances. NR takes a noisy network as input to generate an enhanced network $\boldsymbol{A}^*$ without supervision or prior knowledge.

In summary, the reconstructed network $\boldsymbol{A}^*$ has the following three properties:

1. The singular value gap of $\boldsymbol{A}^*$ is larger than that of $\boldsymbol{A}$ (Supplementary text 3.2, Theorem 4).

2. The left and right singular vectors of $\boldsymbol{A}^*$ remain the same as those of $\boldsymbol{A}$ (Supplementary text 3.2, Theorem 4).

3. The error upper bound on the distance of $\boldsymbol{A}^*$ is less than or equal to the error upper bound of $\boldsymbol{A}$ (Supplementary text 3.2, Theorem 5), which suggests that $\boldsymbol{A}^*$ is more robust against noises, errors, and data missing.

Specifically, we proved that when $|\sigma_k| \in [\sqrt{\lambda}, +\infty)$, the corresponding singular value gap $(\sigma_{k-1} \frac{\sigma_{k-1}^2}{\sigma_{k-1}^2+\lambda} - \sigma_k \frac{\sigma_k^2}{\sigma_k^2+\lambda})$ of $\boldsymbol{A}^*$ is larger than the singular value gap $(\sigma_{k-1} - \sigma_k)$ of $\boldsymbol{A}$, where $\sigma_k$ is the k-th largest singular value of $\boldsymbol{A}$, and $\sigma_k \frac{\sigma_k^2}{\sigma_k^2+\lambda}$ is the k-th largest singular value of $\boldsymbol{A}^*$. By matrix perturbation theory, when $\sigma_{k-1} - \sigma_k > 2\|\boldsymbol{H}\|$, the error upper bound of the enhanced network is $\min\left\{\frac{2\|\boldsymbol{H}\|}{\frac{\sigma_{k-1}^2}{\sigma_{k-1}^2+\lambda}\sigma_{k-1} - \frac{\sigma_k^2}{\sigma_k^2+\lambda}\sigma_k}, 1\right\}$, which is less than or equal to the error upper bound of the original network $\min\left\{\frac{2\|\boldsymbol{H}\|}{\sigma_{k-1}-\sigma_k}, 1\right\}$ (Fig.S1), where $\boldsymbol{H}$ is the noise matrix, satisfying $\widetilde{\boldsymbol{A}} = \boldsymbol{A} + \boldsymbol{H} \in \mathbb{R}^{n \times m}$, with $\widetilde{\boldsymbol{A}}$ being the perturbed version of $\boldsymbol{A}$. This proves that the same noise has a smaller impact on $\boldsymbol{A}^*$ compared to $\boldsymbol{A}$, which can be used to improve the performance of a number of methods on complex networks (see subsection-"link prediction", "single-cell clustering", "domain identification", and "species identification"). The algorithm for NR is illustrated in Fig.S2A.

**The Fast Network Reconstruction algorithm via Global Connectivity-Based Topological Similarity**

To reduce the computational load of matrix inversion in NR, we introduce the fastNR, demonstrating that the low-rank approximation of any matrix can gain from the disturbance resilience of NR (Supplementary text 3.3, Theorem 6). By using randQB_fp[23] with Theorem 4, we derived a low-rank approximation of matrix $\boldsymbol{A}$ that preserves the top $k$ singular values, denoted as $\boldsymbol{A}_k = \boldsymbol{U}_k \boldsymbol{\Sigma}_k \boldsymbol{V}_k^T$, to reduce network noise.

In addition, we proved that if the matrix $\boldsymbol{A}$ is low rank, the result of NR can be computed through truncated singular value decomposition (SVD) (Supplementary text 3.3, Theorem 7). The algorithm for fastNR is illustrated in Fig.S2B. In summary, fastNR has the following two advantages: fastNR offers a considerable reduction in the time complexity for matrix decomposition from $O(mn^2)$ to $O(mk^2)$ compared to NR, thereby enabling faster computation, where $n$ and $m$ are the numbers of rows and columns of $\boldsymbol{A}$, and $m \geq n$ (Supplementary text 8); fastNR provides comparable accuracy as NR in many applications. In experiments on link prediction, single-cell data analysis, TAD detection, and fine-grained species identification, the performance of $\boldsymbol{A}_k^*$ matches that of $\boldsymbol{A}^*$ (Supplementary text 8).

**Physical significance of the TopoLa distance**

Our way to exploring the physical significance of TopoLa distance is to differentiate various types of paths connecting nodes $i$ and $j$. By counting the number of each type of paths, it allows us to take into consideration not only local and global connectivity, but also node degrees in an implicit manner. Particularly, we focus on the classification of $n$-hop paths connecting nodes $i$ and $j$. Due to the possible existence of loop or loops on these paths, we establish a classification system for all such $n$-hop paths, based on the length of a path actually traversed (i.e., the number of hops after removing all the loops on the path). Clearly, there are a total of ($n$-1) types. To facilitate a clear differentiation, we represent these path types with polygons. For example, a path that actually traverses 2-hop is defined as a $\mathrm{P}_2$ path, which can be viewed as a triangle if adding a direct edge back from node $j$ to node $i$. Similarly, a path that actually traverses 3 hops can be defined as $\mathrm{P}_3$ (represented as a quadrangle), and a path that actually traverses 4 hops can be defined as $\mathrm{P}_4$ (represented as a pentagon; Fig.S3A). The quantity of these types can be expressed as $|\mathrm{P}_l| = loop(a_l)$, where $a_l$ is the set of loop-free $l$-hop paths connecting nodes $i$ and $j$ and $loop(\cdot)$ represents the number of different ways of adding loops to paths in $a_l$ to form $n$-hop paths between nodes $i$ and $j$. While there is a correlation between $|\mathrm{P}_l|$ and $|a_l|$, accurately measuring this relationship remains challenging. Nevertheless, our observations suggest that a proportional relationship exists between the quantity of $\mathrm{P}_l$, characterized by solely loops between node $i$ and its neighbors, or between node $j$ and its neighbors, and $|a_l|$. Paths exhibiting such characteristics are defined as $b_n(l)$ (Fig.S3B). The interplay between $|a_l|$ and $|b_n(l)|$ can be calculated as follows:

$$|b_n(l)| = \sum_{h=0}^{\frac{n-l}{2}} \kappa_i^h \kappa_j^{\frac{n-l}{2}-h} |a_l| \tag{9}$$

where $|b_n(l)|$ is the quantity of $b_n(l)$ in $n$-hop $\mathrm{P}_l$ paths. $\kappa_i$ and $\kappa_j$ represent the degrees of node $i$ and node $j$. Thus, $n$-hop paths between node $i$ and node $j$ in the network are represented as:

$$\begin{aligned} |n-\mathrm{hop}| &= |a_n| + |b_n| + |c_n| \\ &= |a_n| + \sum_{\mathrm{P}_l \in n-\mathrm{hop}} |b_n(\mathrm{P}_l)| + |c_n| \\ &= \begin{cases} \sum_{t=1}^{\frac{n}{2}} \sum_{h=0}^{\frac{n}{2}-t} \kappa_i^h \kappa_j^{\frac{n}{2}-t-h} |a_{2t}| + |c_n| & \text{if } n \text{ is even} \\ \sum_{t=1}^{\frac{n-1}{2}} \sum_{h=0}^{\frac{n-1}{2}-t} \kappa_i^h \kappa_j^{\frac{n}{2}-t-h} |a_{2t+1}| + |c_n| & \text{if } n \text{ is odd} \end{cases} \end{aligned} \tag{10}$$

where $c_n$ denotes the paths distinct from $a_n$ and $b_n$. $|n-\mathrm{hop}|$ refers to the counts of $n$-hop paths. Our analysis for 2-hop, 4-hop, and 6-hop paths reveals that $|c_n|$ does not exhibit a direct correlation with $\kappa_i$ and $\kappa_j$ (Supplementary text 9). Formula (10) reveals that $a_n$ and $b_n$ for even-hop paths consist of polygons with an odd number of sides, while for odd-hop paths, they comprise polygons with an even number of sides.

Subsequently, we delve further into the physical significance of $\boldsymbol{D}_{topo}$, which can also be expressed in the form of a set of paths as shown in Formula (1). The emphasis on even-hop paths arises from their ability to capture the degree information and the global connectivity information of nodes $i$ and $j$. Furthermore, the TopoLa distance $d_{topo}$ is demonstrated to correlate with the topological structure of nodes. Specifically, $a_l$ can be classified into two types: those overlapping with 2-hop paths ($g_l$) and the remaining ones ($s_l$), with $|a_l| = |g_l| + |s_l|$ (Fig.S3C). The $|s_l|$ relates to the topological structure of nodes $i$ and $j$. For instance, identical topological structure between $i$ and $j$ renders $|s_l|$ to zero (which indicates that it measures the topological similarity of nodes $i$ and $j$). Thus, we can express $d_{topo}$ as:

$$d_{topo}(i,j) = \lim_{n\to\infty}\left[\left(\sum_{t=1}^{\frac{n}{2}}\sum_{h=0}^{\frac{n}{2}-t}\frac{\kappa_i^h\kappa_j^{\left(\frac{n}{2}-t-h\right)}}{-(-\lambda)^{\left(\frac{n}{2}-t+1\right)}}\right)\left(\sum_{t=1}^{\frac{n}{2}}\frac{|a_{2t}|}{(-\lambda)^{t-1}}\right)+\sum_{t=1}^{\frac{n}{2}}\frac{c_{(2t)}}{\lambda^t}\right]$$

$$= \lim_{n\to\infty}\left[\left(\sum_{t=1}^{\frac{n}{2}}\sum_{h=0}^{\frac{n}{2}-t}\frac{\kappa_i^h\kappa_j^{\left(\frac{n}{2}-t-h\right)}}{-(-\lambda)^{\left(\frac{n}{2}-t+1\right)}}\right)\left[\sum_{t=1}^{\frac{n}{2}}\frac{|g_{2t}|}{(-\lambda)^{t-1}}-\sum_{t=2}^{\frac{n}{2}}\frac{|s_{2t}|}{(-\lambda)^{t-1}}\right]+\sum_{t=1}^{\frac{n}{2}}\frac{c_{(2t)}}{\lambda^t}\right] \quad (11)$$

where $\sum_{t=1}^{\frac{n}{2}}\frac{|g_{2t}|}{(-\lambda)^{t-1}}$ and $\sum_{t=1}^{\frac{n}{2}}\frac{c_{(2t)}}{\lambda^t}$ are unaffected by node's topological structure, while $\sum_{t=2}^{\frac{n}{2}}\frac{|s_{2t}|}{(-\lambda)^{t-1}}$ acts as a topology-dependent penalty term.

## Data availability

The data generated and analyzed in the study have been deposited on Zenodo at https://zenodo.org/records/11160439. A detailed source data file is provided with the manuscript. The publicly available datasets used in this study can be found on their associated websites: Complex networks in link prediction (https://noesis.ikor.org/datasets/link-prediction), Single-cell RNA-seq data (https://www.ncbi.nlm.nih.gov/geo/), Hi-C networks and Butterfly image dataset (http://snap.stanford.edu/ne/), Datasets for low-rank matrix completion (https://github.com/BioinformaticsCSU/FNNM), Image retrieval dataset (https://github.com/lenscloth/RKD) and Image classification dataset (https://www.cs.toronto.edu/~kriz/cifar.html).

## Code availability

All code produced in this study are freely available in supplementary materials or open repositories, with the code for methods development and validation accessible on GitHub (https://github.com/kaizheng-academic/TopoLa)

## Acknowledgments

We express our gratitude to Q. Lei for insightful guidance on the proofs about TopoLa framework, X. Liang for valuable direction on NR and fastNR proofs, and J. Chen for assistance with contrastive learning experiments. We also thank Y. Li for valuable suggestions on fastNR and M. Yang for proofreading the manuscript.

## Funding

This work has been supported by the National Natural Science Foundation of China (No. 62072473, No.61772552, No.61972423 and No.61832019). This work was carried out in part using computing resources at the High Performance Computing Center of Central South University.

## Author contributions

K.Z., J.W., and J.X. conceived of the presented idea. K.Z., Q.Z. wrote computer code. K.Z., Q.F. performed experiments, and analyzed data. K.Z. and J.X. wrote the manuscript with input from all authors. All authors discussed the results and contributed to the final manuscript. J.W. provided supervision and resources for this study.

## Competing interests

The authors declare no competing interests.

**Supplementary Information** for

# Deciphering complex networks via topology-encoded latent hyperbolic geometry

Zheng *et al.*

**Supplementary Text**

**Overview**

In this document, we offer an exhaustive discussion of the datasets utilized and their analytical methodologies. The structure of the document is as follows: Section 1 elaborates on the experimental details comprehensively. Section 2 describes the Definition of evaluation metrics involved in the experiments. Section 3 explicates the foundational principles of Topology-encoded Latent Hyperbolic Geometry (TopoLa) distance and undertakes theoretical analyses of two derived methods: NR and fastNR. Section 4 encompasses a comparative study of link prediction methods predicated on latent space. Section 5 conducts a comparative analysis of two energy distance measures concerning their efficacy on network enhancement. Section 6 provides an extensive comparison of NR's efficacy on single-cell RNA-seq datasets beyond the primary text. Section 7 details additional experiments utilizing NR to augment network enhancement methods.. In Section 8, a detailed comparison between NR and fastNR is presented, focusing on their performance and computational efficiency. Finally, Section 9 analyses the n-hop path statistic and introduces an innovative path classification methodology.

# 1 Further Information on Datasets

## *1.1 Conventional networks*

### *1.1.1 Complex networks in link prediction*

To enable the evaluation and comparison of different network analysis techniques, we carefully selected a diverse set of datasets that represent various fields. Fig.3C illustrates the effect of different denoising or enhancement methods on random walk with restart (RWR). To do this, we first denoised or enhanced the network and then applied RWR. We used a grid search to determine the optimal probability $\alpha$, with a search range of [0.1, 0.2, 0.3, 0.4, 0.5, 0.6, 0.7, 0.8, 0.9]. We computed the closed-form solution at convergence via $\boldsymbol{P}^{\infty} = (1-\alpha)(\boldsymbol{I}-\alpha\boldsymbol{W})^{-1}\boldsymbol{P}^{0}$ and predicted the access probabilities for all nodes in the test set.

To assess the predictive performance of the model, we employed ten-fold cross-validation. This technique involves dividing the original dataset into ten equally sized subsets, with each subset used in turn as a validation set, while the remaining nine subsets are used as training sets. This process is repeated ten times to ensure that each subset is used once as a validation set. The final performance metric of the model is the average of the ten validation results. Ten-fold cross-validation is a widely used technique in machine learning and statistics for model selection and parameter tuning. It effectively evaluates the generalization ability of a model and provides reliable guidance, making it an important tool for researchers.

### *1.1.2 Single-cell RNA-seq data*

We utilized SIMLR, a similarity-based learning method, to identify cell types since DSD, ND, and NE can only denoise symmetric matrices[1]. Specifically, in SIMLR, the initial similarity network is obtained from the distance matrix by network diffusion. Each similarity network is used as an input to network denoising (or reinforcing) method to change the noise weights. Then, the resulting network is used as the initial similarity in the SIMLR method to identify cell types.

In Figure 3B, the left panel shows the spatial distribution of the initial similarity network of Baron_human, which is projected in 2D using t-SNE[2]. The right panel shows the similarity network that is enhanced using the NR method and then projected to visualize all cells.

### *1.1.3 Hi-C networks for domain identification*

For each autosome, we utilized the contact matrix (counts per bin) provided by Wang *et al.*[3] Firstly, they generated clusters that could potentially overlap using the Arrowhead algorithm. To obtain labels with high group confidence, they subsampled the top 15 non-overlapping clusters for each chromosome that did not contain subclusters. Non-overlapping regions were selected because the community detection algorithm Louvain used for post-processing could only identify such regions[4]. For visualization, we displayed only the top 9 communities in the heatmap shown in Figure 4A.

### *1.1.4 Butterfly image dataset for fine-grained species identification*

In this section, we utilized the relevant data provided by Wang *et al.*[3] Specifically, they processed a dataset of 10 different categories of butterflies, each containing 55 to 100 images, totaling 832 butterflies[5]. They used two encoding methods, Fisher Vector (FV)[6, 7] and Vector of Linearly Aggregated Descriptors (VLAD)[8] with dense SIFT[9], to generate descriptors for these images, and calculated the similarity networks for the two representations separately using the Gaussian kernel method:

$$\mathcal{G}(i,j) = \exp\left(-\frac{\|x_i - x_j\|^2}{2\sigma^2}\right). \tag{1}$$

where the feature set of images denotes as $X = \{x_1, x_2, x_3, \ldots, x_n\}$. Then, to overcome the sensitivity to σ, they used the local scale of distance to estimate the variance [10], which can be expressed as:

$$\epsilon_i = \frac{\sum_{j=KNN(i)} \|x_i - x_j\|}{k}, \tag{2}$$

where $k$ represents the number of neighbors, and $KNN(i)$ represents the top-$k$ neighbors of the $i$-th butterfly. Thus, the new similarity network can be calculated as:

$$\mathcal{G}_k^{\sigma}(i,j) = \exp\left(-\frac{\|x_i - x_j\|^2}{\sigma^2(\epsilon_i + \epsilon_j)^2}\right). \tag{3}$$

where $k$=20 and $\sigma$=0.5. The final similarity matrix is computed from the inner product of the two computed similarities.

## ***1.2 The low-rank approximation networks***

### *1.2.1 Movie recommendation and multi-label learning dataset for low-rank matrix completion*

The MovieLens dataset comprises 100,000 ratings from 943 users across 1,682 films. Each user rated at least 20 films, with accompanying demographic details including age, gender, occupation, and zip code. Sourced from the MovieLens website between September 1997 and April 1998, the data was subsequently curated to exclude users with fewer than 20 ratings or incomplete information. In multi-label learning, we assess the performance of both FNNM and rFNNM using eleven datasets sourced from "yahoon.com" designed for web page categorization, spanning categories such as ‘Arts,’ ‘Business,’ and ‘Science.’ Notably, the 'Arts' dataset features 5,000 instances, with each instance having 462 dimensions and 26 labels, where a randomized 10% is earmarked as the test set, and the remaining 90% is allocated for training. During the training phase, we deploy a strategy of partial label assignment. For every label, we randomly selected a ω% mix of positive and negative instances, leaving the residual instances without labels. This selection percentage, ω%, ranges from 10% to 90%, incremented by 20%.

## *1.3 Deep learning models*

### *1.3.1 Image retrieval dataset for knowledge distillation*

We employed the CUB-200-2011[11], Cars196[12], and Stanford Online Products[13] datasets as benchmark datasets for image retrieval. CUB-200-2011 dataset is the most widely used dataset for fine-grained visual categorization task. This dataset encompasses 11,788 bird images spanning 200 subcategories, with a division of 5,994 for training and 5,794 for testing. Each image is meticulously annotated, including a subcategory label, 15 part locations, 312 binary attributes, and a bounding box. The Cars196 dataset consists of 16,185 images across 196 unique car categories, defined by attributes like Make, Model, and Year. With an even distribution of 8,144 training and 8,041 testing images, it is a foundational resource for supervised learning systems targeting the recognition of vehicle types from images. The Stanford Online Products (SOP) dataset, sourced from e-commerce platforms, comprises 120,053 product images spanning 22,634 unique classes. It designates 59,551 images from 11,318 classes for training and allocates the other 60,502 images from 11,316 classes for testing. The data splits align with previous research[14]. We used a machine with one NVIDIA A100 GPU for Image retrieval. The student model is ResNet18 with an embedding size of 64, while the teacher model is ResNet50 with an embedding size of 512. $\lambda_{\mathrm{RKD-D}}$ and $\lambda_{\mathrm{RKD-A}}$ are set to 1 and 2, respectively.

### *1.3.2 Image classification dataset for contrastive learning*

We utilized the CIFAR-10 and CIFAR-100 datasets as benchmark datasets for image classification[15]. The CIFAR-10 dataset, stemming from the Canadian Institute for Advanced Research, is a subset of the Tiny Images dataset and encapsulates 60,000 32x32 color images distributed across 10 distinct classes. The dataset provides an equal distribution with 6,000 images for each class, further subdivided into 5,000 training and 1,000 testing images. The unique categories delineated in CIFAR-10 include airplane, automobile (excluding truck or pickup truck variants), bird, cat, deer, dog, frog, horse, ship, and truck (again, excluding the pickup truck version). On the other hand, the CIFAR-100 dataset, also a derivative of the Canadian Institute for Advanced Research and a subset of the Tiny Images dataset, encompasses 60,000 32x32 color images, which are organized into 100 classes. These classes are further aggregated into 20 overarching superclasses. Each image is methodically annotated with both a "fine" label, indicative of its specific class, and a "coarse" label that designates its superclass category. For each of the 100 classes, there are 600 images, with a division of 500 for training and 100 for testing. We use a machine with one NVIDIA A100 GPU for Image retrieval. The MoCo model consists of a base encoder, a momentum encoder, and two MLPs, which jointly share parameters for data representations. These representations are crucial for computing both contrastive and transform matrix losses. The model utilizes specific hyperparameters, including a feature dimension of 256, MLP hidden dimension of 4096, and a softmax temperature of 1.0. In the forward pass, input images undergo encoding, and losses are determined for contrastive and transform matrix objectives.

## 2 Definition of Evaluation Metrics

### *2.1 The Adjusted Rand Index*

In this article, we used the Adjusted Rand Index (ARI)[16] to assess the degree of correspondence between the clustering results and the true class labels. The ARI is calculated using the following formula:

$$ARI = \frac{\sum_i \sum_j \binom{n_{ij}}{2} - \left[\sum_i \binom{a_i}{2} \sum_j \binom{b_j}{2}\right] / \binom{n}{2}}{\frac{1}{2}\left[\sum_i \binom{a_i}{2} + \sum_j \binom{b_j}{2}\right] - \left[\sum_i \binom{a_i}{2} \sum_j \binom{b_j}{2}\right] / \binom{n}{2}}. \quad (4)$$

where $X = \{X_1, X_2, X_3, \dots, X_r\}$ represents the set of $r$ clusters obtained by clustering, and $Y = \{Y_1, Y_2, Y_3, \dots, Y_s\}$ represents the set of the correct labels, and $n_{ij}$ refers the number of samples at the intersection of $X_i$ and $Y_j$ that is, $n_{ij} = |X_i \cap Y_j|$. $a_i$ represents $\sum_k n_{ik}$ and $b_j$ represents $\sum_l n_{lj}$. The value range of ARI is $[-1, 1]$, the larger the value, the higher the consistency between the two sets.

### *2.2 Normalized Mutual Information*

We introduced Normalized Mutual Information (NMI)[17] to evaluate the quality of clustering from another perspective. The NMI is defined as $I(U,V)/\max\{H(U), H(V)\}$ given two clustering results $U$ and $V$ on a set of data points. Here, $I(U,V)$ represents the mutual information between $U$ and $V$, while $H(U)$ represents the entropy of the clustering $U$. To compute the mutual information, we assume that $U$ has $p$ clusters and $V$ has $q$ clusters, and the calculation is performed as follows:

$$I(U,V) = \sum_{i=1}^{p} \sum_{j=1}^{q} \frac{|U_i \cap V_j|}{N} log \frac{|U_i \cap V_j|}{|U_i|/N \times |V_j|/N}, \quad (5)$$

The cardinality of the $p$-th cluster in $U$ is denoted by $|U_i|$, and $N$ represents the number of points. The entropy of each cluster assignment is calculated using the following formula:

$$H(U) = -\sum_{i=1}^{p} \frac{|U_i|}{N} log \frac{|U_j|}{N}, \quad (6)$$

$$H(V) = -\sum_{j=1}^{q} \frac{|V_j|}{N} log \frac{|V_j|}{N}, \quad (7)$$

The value range of NMI is $[0, 1]$, the larger the value, the higher the consistency between the two sets.

### 2.3 Retrieval Accuracy

We adopted the retrieval accuracy[3] proposed by Wang *et al.* as the evaluation metric for fine-grained species recognition. For a given butterfly $q$, the retrieval accuracy can be calculated as the proportion of correct matches among the top k retrieved butterflies:

$$acc_q(k) = \frac{H_q}{min(k, N_q)}, \quad (8)$$

where $H_q$ is number of correct retrievals which refers to the number of butterflies that have the same type as q. Therefore, we can define the average identification accuracy as:

$$Acc = \frac{1}{n}\sum_{i=1}^{n} acc_i(N_q). \quad (9)$$

where the $n$ is the number of butterflies in the dataset. In addition, the accuracy for different numbers of retrievals can be defined as:

$$Acc_k = \frac{1}{n}\sum_{i=1}^{n} acc_i(k). \quad (10)$$

### 2.4 Root Mean Square Error

For the movie recommendation task, we employed the Root Mean Square Error (RMSE)[18]. RMSE quantifies the differences between observed and predicted values, representing them as the square root of the average of squared discrepancies. These discrepancies are termed 'residuals' when calculated on the estimation dataset and 'errors' or 'prediction errors' when determined out-of-sample. RMSE consolidates the magnitudes of prediction deviations for various data points into a singular measure of predictive efficacy. To assess the performance of various methods, we employed RMSE on the test set, defined as:

$$RMSE = \frac{\mathcal{P}_{\widetilde{\Omega}}(\boldsymbol{A} - \boldsymbol{B})}{\sqrt{|\widetilde{\Omega}|}}, \quad (11)$$

where we utilize the original rating matrix, denoted as $\boldsymbol{A}$, and the completed matrix, represented as $\boldsymbol{B}$. The set of indices in the test set is signified by $\widetilde{\Omega}$. $\mathcal{P}_{\widetilde{\Omega}}(\boldsymbol{A} - \boldsymbol{B})$ is the projection operator projecting matrix $(\boldsymbol{A} - \boldsymbol{B})$ onto $\widetilde{\Omega}$ such that

$$\left(\mathcal{P}_{\widetilde{\Omega}}(\boldsymbol{A} - \boldsymbol{B})\right)_{ij} = \begin{cases} (\boldsymbol{A} - \boldsymbol{B})_{ij}, & (i,j) \in \widetilde{\Omega} \\ 0, & (i,j) \notin \widetilde{\Omega}. \end{cases} \quad (12)$$

For every method employed, we randomly extracted 10% of the known elements from the training data to form a validation set, which assists in determining the optimal parameters. After executing each trial five times, we declare the average RMSE on the test set as the result.

### 2.5 *Average Precision*

We use Average Precision (AP) on the multi-label learning data as the evaluation metric. AP, akin to the Area Under the Precision-Recall Curve (AUPR), offers a concise summary of the PR curve as a single value[19]. Specifically, AP represents the weighted mean of precision scores at each threshold of the PR curve, where the increment in recall from the preceding threshold serves as the weighting factor. The formula is as follows:

$$AP(f) = \frac{1}{p}\sum_{i=1}^{p}\frac{1}{|\Upsilon_i|}\sum_{y\in\Upsilon_i}\frac{\left|\left\{rank_f(\boldsymbol{x}_i, y') \leq rank_f(\boldsymbol{x}_i, y), y' \in \Upsilon_i\right\}\right|}{rank_f(\boldsymbol{x}_i, y)}. \tag{13}$$

where $\Upsilon_i$ is the set of labels associated with $\boldsymbol{x}_i$, $\boldsymbol{x}_i$ is a feature vector. $rank_f(\boldsymbol{x}_i, y)$ returns the rank of $y$ in $\Upsilon_i$ based on the descending order induced from $f(\boldsymbol{x}, y)$. $f(\boldsymbol{x}, y)$ returns the confidence of $y$ being proper label of $\boldsymbol{x}$.

### 2.6 *Recall@K*

For the image retrieval task, we employed the recall@K[20]. After embedding all test images using the model, each image serves as a query to retrieve its top K nearest neighbors from the test set, excluding the query image itself. A recall value of 1 is assigned if the retrieved images share the same category as the query. The overall recall@K is then determined by averaging recall values across the entire test set.

### 2.7 *Linear Evaluation Accuracy*

We substantiated the efficacy of contrastive learning model by evaluating the model's linear evaluation accuracy[21]. Specifically, we maintained the encoder parameters in a fixed state, conducted training for a linear classifier, and scrutinized its performance in classification.

## 3 Theoretical Analysis

### 3.1 *The Topology-encoded Latent Hyperbolic Geometry*

This section mainly presents the proofs of the relevant theorems for TopoLa.

**Theorem 1.** Given a network $\boldsymbol{X} \in \mathbb{R}^{d\times n}$, let $\langle t(x)\rangle$ be the expected numbers of triangles, and $\langle t'(x)\rangle$ be the expected numbers of weighted triangles. We have $\alpha' > \alpha$, where $\alpha'$ is the logarithm of thermodynamic activity corresponding to $\langle t'(x)\rangle$ and $\alpha$ is the logarithm of thermodynamic corresponding to $\langle t(x)\rangle$.

***Proof.*** In network geometry, the expected numbers of triangles can be obtained through the integration of the graphon[22]:

$$\langle t(x)\rangle = \frac{1}{2}\iint_{\mathbb{R}^2} p(x, y)\, p(y, z) p(z, x)\, dydz, \tag{14}$$

Under the assumption of redundancy in the current statistics, the non-redundant expectation $\langle t'(x)\rangle$, is anticipated to be less than its redundant counterpart, $\langle t(x)\rangle$. In this context, $t(x)$ denotes the quantity of triangles associated with node $x$. Utilizing the maximum-entropy principle[22], we derive the following formula:

$$\langle t'(x)\rangle = \frac{1}{2}\iint_{\mathbb{R}^2} p'(x,y)\,p'(y,z)p'(z,x)\,dydz = \bar{t}' < \bar{t}. \tag{15}$$

Therefore,

$$\frac{1}{2}\iint_{\mathbb{R}^2} p'(x,y)\,p'(y,z)p'(z,x)\,dydz < \frac{1}{2}\iint_{\mathbb{R}^2} p(x,y)\,p(y,z)p(z,x)\,dydz, \tag{16}$$

When the network size is sufficiently large, the approximate solution for the graphon that maximizes entropy is the Fermi-Dirac graphon:

$$p^*(x,y) = \begin{cases} \dfrac{1}{1+e^{2\alpha\left(r-\frac{1}{2}\right)}} & \text{if } 0 \le r \le 1, \\ \dfrac{1}{1+e^{\alpha}} & \text{if } r > 1, \end{cases} \tag{17}$$

where $\alpha$ and $r$ are the rescaled inverse temperature and energy distance, respectively. Inserting the terms from Formula (17) into Formula (16), we obtain:

$$\iint_{\mathbb{R}^2} \left[\frac{1}{1+e^{2\alpha'\left(r-\frac{1}{2}\right)}}\right]^3 dydz < \iint_{\mathbb{R}^2} \left[\frac{1}{1+e^{2\alpha\left(r-\frac{1}{2}\right)}}\right]^3 dydz \tag{18}$$

Consequently, we deduce that $\alpha' > \alpha$. Based on prior research, the common neighbor integral corresponding to $\alpha'$ provides a more accurate representation of the latent space distance than that associated with $\alpha$ [22]. From this, we demonstrate that the current triangle statistics exhibit redundancy. Given that the common-neighbor integral is defined as $\int_{\mathbb{R}} \left[\frac{1}{1+e^{2\alpha\left(r-\frac{1}{2}\right)}}\right]^2 dz$, the 2-hop paths (common neighbors) are also found to be imprecise when representing distances within the latent space and **Theorem 1** holds.

**Theorem 2.** Given a network $\boldsymbol{P}_0 \in \mathbb{R}^{n\times m}$, transition matrix $\boldsymbol{W} \in \mathbb{R}^{n\times n}$. Suppose $\boldsymbol{P}_t$ is a random walk with restart after $t$-step diffusion. We have

$$\boldsymbol{P}_t = (1-\alpha)(\boldsymbol{I} - \alpha\boldsymbol{W})^{-1}\boldsymbol{C}\boldsymbol{P}_0 \tag{19}$$

where its convergence condition is expressible as $||\boldsymbol{P}_{t+1} - \boldsymbol{P}_t||_F^2 = ||\boldsymbol{P}_0 - \boldsymbol{C}\boldsymbol{P}_0||_F^2 < \varepsilon$.

***Proof.*** The formula for RWR is $\boldsymbol{P}_t = (1-\alpha)\boldsymbol{P}_0 + \alpha\boldsymbol{W}\boldsymbol{P}_{t-1}$. We expand the formula:

$$\begin{aligned}\boldsymbol{P}_t &= (1-\alpha)\boldsymbol{P}_0 + \alpha\boldsymbol{W}[(1-\alpha)\boldsymbol{P}_0 + \alpha\boldsymbol{W}\boldsymbol{P}_{t-1}] \\ &= (1-\alpha)\boldsymbol{P}_0 + \alpha\boldsymbol{W}(1-\alpha)\boldsymbol{P}_0 + \alpha^2\boldsymbol{W}^2\boldsymbol{P}_{t-2} \\ &= (1-\alpha)\boldsymbol{P}_0 + \alpha\boldsymbol{W}(1-\alpha)\boldsymbol{P}_0 + \alpha^2\boldsymbol{W}^2[(1-\alpha)\boldsymbol{P}_0 + \alpha\boldsymbol{W}\boldsymbol{P}_{t-3}] \\ &= (1-\alpha)\boldsymbol{P}_0 + \alpha\boldsymbol{W}(1-\alpha)\boldsymbol{P}_0 + \alpha^2\boldsymbol{W}^2(1-\alpha)\boldsymbol{P}_0 + \alpha^3\boldsymbol{W}^3\boldsymbol{P}_{t-3}\end{aligned}$$

$$= \left[(1-\alpha)\sum_{k=0}^{t-1}\alpha^k \boldsymbol{W}^k + \alpha^t \boldsymbol{W}^t\right]\boldsymbol{P}_0 \tag{20}$$

Let $[(1-\alpha)\sum_{k=0}^{t-1}\alpha^k \boldsymbol{W}^k + \alpha^t \boldsymbol{W}^t]$ be the matrix $\boldsymbol{B}$, then $\boldsymbol{P}_t = \boldsymbol{B}\boldsymbol{P}_0$. From this, we can get that the result of random walk with restart for each iteration can be equivalently viewed as a single-step diffusion. Then, we expand the termination condition as:

$$||\boldsymbol{P}_{t+1} - \boldsymbol{P}_t||_F^2 < \varepsilon \tag{21}$$

$$||\,(1-\alpha)\boldsymbol{P}_0 + \alpha\boldsymbol{W}\boldsymbol{P}_t - \boldsymbol{P}_t||_F^2 < \varepsilon \tag{22}$$

Substituting $\boldsymbol{P}_t = \boldsymbol{B}\boldsymbol{P}_0$ into the above formula, we get:

$$||\ (1-\alpha)\boldsymbol{P}_0 + \alpha\boldsymbol{W}\boldsymbol{B}\boldsymbol{P}_0 - \boldsymbol{B}\boldsymbol{P}_0||_F^2 < \varepsilon \tag{23}$$

According to the associative law of matrix product, we get:

$$||\ (1-\alpha)\boldsymbol{P}_0 + (\alpha\boldsymbol{W}\boldsymbol{B} - \boldsymbol{B})\boldsymbol{P}_0||_F^2 < \varepsilon \tag{24}$$

Then, we have

$$||\ \boldsymbol{P}_0 - \frac{(\boldsymbol{B} - \alpha\boldsymbol{W}\boldsymbol{B})}{(1-\alpha)}\boldsymbol{P}_0||_F^2 < \varepsilon \tag{25}$$

Let $\frac{(\boldsymbol{B}-\alpha\boldsymbol{W}\boldsymbol{B})}{(1-\alpha)}$ be $\boldsymbol{C}$, then the above formula can be transformed into:

$$||\boldsymbol{P}_0 - \boldsymbol{C}\boldsymbol{P}_0||_F^2 < \varepsilon \tag{26}$$

According to the derivation, the matrix $\boldsymbol{B}$ can be obtained as

$$\frac{(\boldsymbol{B} - \alpha\boldsymbol{W}\boldsymbol{B})}{(1-\alpha)} = \boldsymbol{C} \tag{27}$$

$$\boldsymbol{B} = (1-\alpha)(\boldsymbol{I} - \alpha\boldsymbol{W})^{-1}\boldsymbol{C} \tag{28}$$

Substituting $\boldsymbol{B}$ into the random walk with restart formulation, we get:

$$\boldsymbol{P}_t = (1-\alpha)(\boldsymbol{I} - \alpha\boldsymbol{W})^{-1}\boldsymbol{C}\boldsymbol{P}_0 \tag{29}$$

Thus, **Theorem 2** holds.

**Theorem 3.** Given a topological structure of node $y \in \mathbb{R}^d$, network $\boldsymbol{X} \in \mathbb{R}^{n\times d}$ and a parameter $\lambda$. Let $\boldsymbol{c}^*$ be the optimal solution to the least squares optimization with regularization (in vector form):

$$\min \frac{1}{\lambda}\|\boldsymbol{y} - \boldsymbol{c}\boldsymbol{X}\|_F^2 + \|\boldsymbol{c}\|_F^2 \tag{30}$$

Then, we have

$$\frac{\left\|\boldsymbol{c}_i^* - \boldsymbol{c}_j^*\right\|_F^2}{\|\boldsymbol{y}\|_F^2} \le \frac{1}{\lambda}\left\|\boldsymbol{x}_i^T - \boldsymbol{x}_j^T\right\|_F^2 \tag{31}$$

***Proof.*** Inspired by prior work, we conducted the following inference[23]. Let $L(c) = \frac{1}{\lambda}\|\boldsymbol{y} - \boldsymbol{c}\boldsymbol{X}\|_F^2 + \|\boldsymbol{c}\|_F^2$. Since $\boldsymbol{c}^*$ is the optimal solution to problem [30], it satisfies

$$\frac{\partial L(c)}{\partial c_k}\Big|_{c\,=\,c^*} = 0 \tag{32}$$

Thus, we have

$$-\frac{2}{\lambda}\boldsymbol{x}_i^T(\boldsymbol{y} - \boldsymbol{c}^* X) + 2\boldsymbol{c}_i^* = 0, \tag{33}$$

$$-\frac{2}{\lambda}\boldsymbol{x}_j^T(\boldsymbol{y} - \boldsymbol{c}^* X) + 2\boldsymbol{c}_j^* = 0. \tag{34}$$

Formulas (33) and (34) give us

$$\boldsymbol{c}_i^* - \boldsymbol{c}_j^* = \frac{1}{\lambda}\left(\boldsymbol{x}_i^T - \boldsymbol{x}_j^T\right)(\boldsymbol{y} - \boldsymbol{c}^*\boldsymbol{X}). \tag{35}$$

Since $\boldsymbol{c}^*$ is optimal to Formula (30), we get

$$\frac{1}{\lambda}\|\boldsymbol{y} - \boldsymbol{c}^*\boldsymbol{X}\|_F^2 + \|\boldsymbol{c}^*\|_F^2 = L(\boldsymbol{c}^*) \le L(0) = \|\boldsymbol{y}\|_F^2 \tag{36}$$

Thus, we have $\frac{1}{\lambda}\|\boldsymbol{y} - \boldsymbol{c}^*\boldsymbol{X}\|_F^2 < \|\boldsymbol{y}\|_F^2$. Then Formula (36) implies

$$\frac{\left\|\boldsymbol{c}_i^* - \boldsymbol{c}_j^*\right\|_F^2}{\|\boldsymbol{y}\|_F^2} \le \frac{1}{\lambda}\left\|\boldsymbol{x}_i^T - \boldsymbol{x}_j^T\right\|_F^2 \tag{37}$$

From Formula (37), we discern that the difference in distances between two nodes in the latent space relative to other nodes correlates with the divergence in their topological structures. That is, nodes with highly similar topological structures occupy proximate positions within the latent space and **Theorem 3** holds.

### *3.2 The Network Reconstruction via Global Connectivity-Based Topological Similarity*

This section mainly presents the proofs of the relevant theorems for NR.

**Theorem 4.** If $|\sigma_j| \in [\sqrt{\lambda}, +\infty)$, the singular value gap $(\sigma_{j-1} - \sigma_j)$ of $\boldsymbol{A}$ is smaller than the singular value gap $(\sigma_j^* - \sigma_{j+1}^*)$ of $\boldsymbol{A}^*$.

***Proof.*** Let $\boldsymbol{A}$ represent the original network and $\boldsymbol{A}^*$ represent the enhanced network. Since the matrix $\boldsymbol{A} \in \mathbb{R}^{n\times m}$, where $n$ and $m$ are not necessarily equal, we can express it as $\boldsymbol{A} = \boldsymbol{U\Sigma V}^T$, where $\boldsymbol{U}$ is the left singular vector and the eigenvector of $\boldsymbol{AA}^{\mathrm{T}}$, $\boldsymbol{V}$ is the right singular vector and the eigenvector of $\boldsymbol{A}^T\boldsymbol{A}$, and $\boldsymbol{\Sigma}$ is the singular value matrix with $\boldsymbol{\Sigma}_{\mathrm{i,i}} = \sigma_{\mathrm{i}}$. Consequently, we can obtain the change in singular values as follows:

$$\begin{aligned}\boldsymbol{A}^* &= \boldsymbol{AA}^T(\boldsymbol{AA}^T + \boldsymbol{\lambda I})^{-1}\boldsymbol{A}\\ &= [\boldsymbol{U\Sigma}^T\boldsymbol{\Sigma U}^T][(\boldsymbol{U}^T)^{-1}(\boldsymbol{\Sigma}^T\boldsymbol{\Sigma} + \lambda\boldsymbol{I})^{-1}\boldsymbol{U}^{-1}]\boldsymbol{U\Sigma V}^T\\ &= \boldsymbol{U\Sigma}^T\boldsymbol{\Sigma}(\boldsymbol{\Sigma}^T\boldsymbol{\Sigma} + \lambda\boldsymbol{I})^{-1}\boldsymbol{\Sigma V}^T\end{aligned} \tag{38}$$

As shown in the formula above, the left and right singular vectors remain unchanged, while the singular values become $\boldsymbol{\Sigma}^*{}_{i,i} = \sigma_i \frac{\sigma_i^2}{\sigma_i^2+\lambda}$. We need to prove that $|\sigma_{j-1} - \sigma_j| \le |\sigma_{j-1}^* - \sigma_j^*|$, where $j \ge 2$ and $\sigma_j^*$ is a singular value of $\boldsymbol{A}^*$. Since $\sigma_j^* = \sigma_j \frac{\sigma_j^2}{\sigma_j^2+\lambda}$, the inequality can be turned into:

$$\sigma_{j-1} - \sigma_{j-1}\frac{\sigma_{j-1}^2}{\sigma_{j-1}^2+\lambda} \le \sigma_j - \sigma_j \frac{\sigma_j^2}{\sigma_j^2+\lambda}. \tag{39}$$

Since $\sigma_{j-1} \ge \sigma_j$, the remaining task is to prove that $g(x) = x - \frac{x^3}{x^2+\lambda}$ is a decreasing function. We obtain the derivative of $g(x)$ as follows:

$$\begin{aligned}\frac{\partial g(x)}{\partial x} &= 1 - \frac{3x^2(x^2+\lambda) - 2x^4}{(x^2+\lambda)^2} \le 0 \\ &= 1 - \frac{x^4 + 3\lambda x^2}{(x^2+\lambda)^2} \le 0\,.\end{aligned} \tag{40}$$

When $\lambda \ge 0$, we can get $|x| \ge \sqrt{\lambda}$.

$$1 - \frac{x^4 + 3\lambda x^2}{(x^2+\lambda)^2} \le 0, \tag{41}$$

$$1 - \frac{3x^2(x^2+\lambda) - 2x^4}{(x^2+\lambda)^2} \le 0. \tag{42}$$

We set $g(x) = x - \frac{x^3}{x^2+\lambda}$, and its derivative is $g'(x) = 1 - \frac{3x^2(x^2+\lambda)-2x^4}{(x^2+\lambda)^2}$. Since $1 - \frac{3x^2(x^2+\lambda)-2x^4}{(x^2+\lambda)^2} \le 0$, $g(x)$ is a decreasing function. We can get $\sigma_{j-1} - \sigma_j \le \sigma_{j-1}\frac{\sigma_{j-1}^2}{\sigma_{j-1}^2+\lambda} - \sigma_j \frac{\sigma_j^2}{\sigma_j^2+\lambda}$. Therefore, we derive $|\sigma_{j-1} - \sigma_j| \le |\sigma_{j-1}^* - \sigma_j^*|$ from $|x| \ge \sqrt{\lambda}$ and **Theorem 4** holds.

**Theorem 5.** Suppose that $\widetilde{\boldsymbol{A}} = \boldsymbol{A} + \boldsymbol{H} \in \mathbb{R}^{n\times m}$, $\widetilde{\boldsymbol{A}}^* = \boldsymbol{A}^* + \boldsymbol{H} \in \mathbb{R}^{n\times m}$, $\boldsymbol{H}$ has *i.i.d.* $N(0,\sigma^2)$ entries, $\boldsymbol{A}^* = \boldsymbol{A}\boldsymbol{A}^T(\boldsymbol{A}\boldsymbol{A}^T + \lambda\boldsymbol{I})^{-1}\boldsymbol{A}$. Then

$$sup\{dist(\boldsymbol{A}, \widetilde{\boldsymbol{A}})\} \ge sup\{dist(\boldsymbol{A}^*, \widetilde{\boldsymbol{A}}^*)\} \tag{43}$$

where $\sigma_k \ge \sqrt{\lambda}$, $\sigma_k$ is the singular value of $\boldsymbol{A}$, and $dist(\boldsymbol{A}, \widetilde{\boldsymbol{A}})$ refers to the sin Θ distance between singular spaces of $\boldsymbol{A}$ and $\widetilde{\boldsymbol{A}}$ [24]. $\widetilde{\boldsymbol{A}} := \boldsymbol{A} + \boldsymbol{H}$, where $\boldsymbol{H}$ is the noise. Similarly, $\widetilde{\boldsymbol{A}}^* := \boldsymbol{A}^* + \boldsymbol{H}$.

***Proof.*** We prove **Theorem 5** in terms of **Lemma 1**:

For rectangular matrices $\boldsymbol{A}, \boldsymbol{H} \in \mathbb{R}^{n\times m}$, we can write the conformal SVD of the original matrix $\boldsymbol{A}$ and its perturbed version $\widetilde{\boldsymbol{A}} = \boldsymbol{A} + \boldsymbol{H}$ as

$$\boldsymbol{A} = \boldsymbol{U}\boldsymbol{\Sigma}\boldsymbol{V}^T = (\boldsymbol{U}_k \;\; \boldsymbol{U}_\perp)\begin{pmatrix}\boldsymbol{\Sigma}_k & \\ & \boldsymbol{\Sigma}_\perp\end{pmatrix}\begin{pmatrix}\boldsymbol{V}_k^T \\ \boldsymbol{V}_\perp^T\end{pmatrix}, \tag{44}$$

$$\widetilde{\boldsymbol{A}} = \widetilde{\boldsymbol{U}}\widetilde{\boldsymbol{\Sigma}}\widetilde{\boldsymbol{V}}^T = \begin{pmatrix}\widetilde{\boldsymbol{U}}_k & \widetilde{\boldsymbol{U}}_\perp\end{pmatrix}\begin{pmatrix}\widetilde{\boldsymbol{\Sigma}}_k & \\ & \widetilde{\boldsymbol{\Sigma}}_\perp\end{pmatrix}\begin{pmatrix}\widetilde{\boldsymbol{V}}_k^T \\ \widetilde{\boldsymbol{V}}_\perp^T\end{pmatrix}. \tag{45}$$

where $\boldsymbol{U}_k \in \mathbb{R}^{n\times(k-1)}$, $\boldsymbol{U}_\perp \in \mathbb{R}^{n\times(n-k+1)}$, $\boldsymbol{V}_k \in \mathbb{R}^{m\times(k-1)}$, $\boldsymbol{V}_\perp \in \mathbb{R}^{m\times(m-k+1)}$, $[\boldsymbol{U}_k, \boldsymbol{U}_\perp] \in \mathbb{R}^{n\times n}$, $[\boldsymbol{V}_k, \boldsymbol{V}_\perp] \in \mathbb{R}^{m\times m}$ are orthogonal matrices. $\boldsymbol{\Sigma}_k = \mathrm{diag}\{\sigma_1, \sigma_2, \dots, \sigma_{k-1}\} \in \mathbb{R}^{(k-1)\times(k-1)}$, $\boldsymbol{\Sigma}_\perp = \mathrm{diag}\{\sigma_k, \sigma_{k+1}, \dots, \sigma_{\min(m,n)}\} \in \mathbb{R}^{(n-k+1)\times(m-k+1)}$. When $n \neq m$, $\boldsymbol{\Sigma}_\perp$ is rectangular, and the extra columns/rows are padded with $0s$. The decomposition of $\widetilde{\boldsymbol{A}}$ has a similar structure $\widetilde{\boldsymbol{\Sigma}}_k = \mathrm{diag}\{\tilde{\sigma}_1, \tilde{\sigma}_2, \dots, \tilde{\sigma}_{k-1}\} \in \mathbb{R}^{(k-1)\times(k-1)}$, $\widetilde{\boldsymbol{\Sigma}}_\perp = \mathrm{diag}\{\tilde{\sigma}_k, \tilde{\sigma}_{k+1}, \dots, \tilde{\sigma}_{\min(m,n)}\} \in \mathbb{R}^{(n-k+1)\times(m-k+1)}$. Then,

$$sup\{dist(\boldsymbol{A}, \widetilde{\boldsymbol{A}})\} = \max\{\|\sin\Theta(\boldsymbol{U}_k, \widetilde{\boldsymbol{U}}_k)\|, \|\sin\Theta(\boldsymbol{V}_k, \widetilde{\boldsymbol{V}}_k)\|\} \leq min\left\{\frac{2\|\widetilde{\boldsymbol{A}} - \boldsymbol{A}\|}{\sigma_{k-1} - \sigma_k}, 1\right\}. \tag{46}$$

where $\max\{\|\sin\Theta(\boldsymbol{U}_k, \widetilde{\boldsymbol{U}}_k)\|, \|\sin\Theta(\boldsymbol{V}_k, \widetilde{\boldsymbol{V}}_k)\|\}$ is the maximum value of the distance between the $\widetilde{\boldsymbol{V}}_k$ and $\boldsymbol{V}_k$, $\widetilde{\boldsymbol{U}}_k$ and $\boldsymbol{U}_k$, $\sin\Theta(\boldsymbol{U}_k, \widetilde{\boldsymbol{U}}_k) = \sin(\Delta\Theta_2)$, $\sin\Theta(\boldsymbol{V}_k, \widetilde{\boldsymbol{V}}_k) = \sin(\Delta\Theta_1)$. Suppose $\widetilde{\boldsymbol{A}}^* = \boldsymbol{A}^* + \boldsymbol{H}$. Then the error upper bound of $\boldsymbol{A}^*$ can be expressed as:

$$\max\{\|\sin\Theta(\boldsymbol{U}_k^*, \widetilde{\boldsymbol{U}}_k^*)\|, \|\sin\Theta(\boldsymbol{V}_k^*, \widetilde{\boldsymbol{V}}_k^*)\|\} \leq \min\left\{\frac{\|\boldsymbol{H}\|}{\frac{\sigma_{k-1}^2}{\sigma_{k-1}^2 + \lambda}\sigma_{k-1} - \frac{\sigma_k^2}{\sigma_k^2 + \lambda}\sigma_k}, 1\right\}. \tag{47}$$

Therefore, when $\sigma_{k-1} - \sigma_k > 2\|\boldsymbol{H}\|$, we can infer that the error upper bound of the enhanced network is smaller than the error upper bound of the original network. That is, the error upper bound on the distance between $\widetilde{\boldsymbol{A}}^*$ and $\boldsymbol{A}^*$ is less than or equal to the error upper bound on the distance between $\widetilde{\boldsymbol{A}}$ and $\boldsymbol{A}$. **Theorem 5** holds.

**Lemma 1**. We have the following uniform error bound on $\sin\Theta$ distance

$$\max\{\|\sin\Theta(\boldsymbol{U}_k, \widetilde{\boldsymbol{U}}_k)\|, \|\sin\Theta(\boldsymbol{V}_k, \widetilde{\boldsymbol{V}}_k)\|\} \leq \min\left\{\frac{2\|\boldsymbol{H}\|}{\sigma_{k-1} - \sigma_k}, 1\right\} \tag{48}$$

***Proof.*** To calculate error bound, we use a modified version of the $sin\Theta$ distance, as presented in Lemma 4.7 in [24]. If $\sigma_{k-1} = \sigma_k$, then $max\{\|\sin\Theta(\boldsymbol{U}_k, \widetilde{\boldsymbol{U}}_k)\|, \|\sin\Theta(\boldsymbol{V}_k, \widetilde{\boldsymbol{V}}_k)\|\} \leq 1$, and the formula holds. When $\sigma_{k-1} > \sigma_k$, there are two cases: 1. $\sigma_{k-1} - \sigma_k > 2\|\boldsymbol{H}\|$; 2. $\sigma_{k-1} - \sigma_k \leq 2\|\boldsymbol{H}\|$. When $\sigma_{k-1} - \sigma_k > 2\|\boldsymbol{H}\|$, combining with the Weyl inequality ($|\tilde{\sigma}_k - \sigma_k| \leq \|\boldsymbol{H}\|$), we can obtain:

$$\tilde{\sigma}_{k-1} - \sigma_k > \sigma_{k-1} - \sigma_k - \|\boldsymbol{H}\| > \frac{1}{2}(\sigma_{k-1} - \sigma_k) > 0. \tag{49}$$

This satisfies the assumptions of **Lemma 2** and thus we can conclude that:

$$\|\sin\Theta(\boldsymbol{U}_k, \widetilde{\boldsymbol{U}}_k)\| = \|\boldsymbol{U}_\perp^T \widetilde{\boldsymbol{U}}_k\| \leq \frac{\|\boldsymbol{H}\|}{\sigma_{k-1} - \sigma_k - \|\boldsymbol{H}\|} \leq \frac{2\|\boldsymbol{H}\|}{\sigma_{k-1} - \sigma_k}. \tag{50}$$

When $\sigma_{k-1} - \sigma_k \leq 2\|\boldsymbol{H}\|$, we can obtain:

$$\left\|\boldsymbol{U}_{\perp}^{T}\widetilde{\boldsymbol{U}}_{k}\right\| \leq 1 \leq \frac{2\|\boldsymbol{H}\|}{\sigma_{k-1}-\sigma_{k}}, \tag{51}$$

Combining the two cases, we can get the following formula:

$$\left\|\sin\Theta\left(\boldsymbol{U}_{k},\widetilde{\boldsymbol{U}}_{k}\right)\right\|\} \leq \min\left\{\frac{2\|\boldsymbol{H}\|}{\sigma_{k-1}-\sigma_{k}},1\right\}. \tag{52}$$

Similarly, we can show that

$$\left\|\sin\Theta\left(\boldsymbol{V}_{k},\widetilde{\boldsymbol{V}}_{k}\right)\right\|\} \leq \min\left\{\frac{2\|\boldsymbol{H}\|}{\sigma_{k-1}-\sigma_{k}},1\right\}. \tag{53}$$

Therefore, **Lemma 1** holds.

**Lemma 2**. If $\tilde{\sigma}_{k-1}-\sigma_{k}>0$ and na, then:

$$max\left\{\left\|\sin\Theta\left(\boldsymbol{U}_{k},\widetilde{\boldsymbol{U}}_{k}\right)\right\|,\left\|\sin\Theta\left(\boldsymbol{V}_{k},\widetilde{\boldsymbol{V}}_{k}\right)\right\|\right\} \leq \min\left\{\frac{1}{\sigma_{k-1}-\tilde{\sigma}_{k}},\frac{1}{\tilde{\sigma}_{k-1}-\sigma_{k}}\right\}\|\boldsymbol{H}\| \tag{54}$$

***Proof.*** According to **Lemma 3**, we can obtain the following formula:

$$\boldsymbol{U}_{\perp}^{T}\widetilde{\boldsymbol{U}}_{k}=\boldsymbol{F}_{\boldsymbol{U}}^{21}\circ\left(\boldsymbol{U}_{\perp}^{T}\boldsymbol{H}\widetilde{\boldsymbol{V}}_{k}\widetilde{\boldsymbol{\Sigma}}_{k}^{T}+\boldsymbol{\Sigma}_{\perp}\boldsymbol{V}_{\perp}^{T}\boldsymbol{H}^{T}\widetilde{\boldsymbol{U}}_{k}\right), \tag{55}$$

$$\boldsymbol{U}_{k}^{T}\widetilde{\boldsymbol{U}}_{\perp}=\boldsymbol{F}_{\boldsymbol{U}}^{12}\circ\left(\boldsymbol{U}_{k}^{T}\boldsymbol{H}\widetilde{\boldsymbol{V}}_{\perp}\boldsymbol{\Sigma}_{\perp}^{T}+\boldsymbol{\Sigma}_{k}\boldsymbol{V}_{k}^{T}\boldsymbol{H}^{T}\widetilde{\boldsymbol{U}}_{\perp}\right). \tag{56}$$

According to **Lemma 4**, we have

$$\left\|\boldsymbol{U}_{\perp}^{T}\widetilde{\boldsymbol{U}}_{k}\right\| \leq \frac{\tilde{\sigma}_{k-1}}{\tilde{\sigma}_{k-1}^{2}-\sigma_{k}^{2}}\left\|\boldsymbol{U}_{\perp}^{T}\boldsymbol{H}\widetilde{\boldsymbol{V}}_{k}\right\|+\frac{\sigma_{k}}{\tilde{\sigma}_{k-1}^{2}-\sigma_{k}^{2}}\left\|\boldsymbol{V}_{\perp}^{T}\boldsymbol{H}^{T}\widetilde{\boldsymbol{U}}_{k}\right\|, \tag{57}$$

$$\left\|\boldsymbol{U}_{\perp}^{T}\widetilde{\boldsymbol{U}}_{k}\right\| \leq \frac{\tilde{\sigma}_{k-1}}{\tilde{\sigma}_{k-1}^{2}-\tilde{\sigma}_{k}^{2}}\left\|\boldsymbol{H}\widetilde{\boldsymbol{V}}_{k}\right\|+\frac{\sigma_{k}}{\tilde{\sigma}_{k-1}^{2}-\sigma_{k}^{2}}\left\|\boldsymbol{H}^{T}\widetilde{\boldsymbol{U}}_{k}\right\|, \tag{58}$$

$$\left\|\boldsymbol{U}_{\perp}^{T}\widetilde{\boldsymbol{U}}_{k}\right\| \leq \frac{\|\boldsymbol{H}\|}{\tilde{\sigma}_{k-1}-\sigma_{k}}. \tag{59}$$

Similarly, we can get

$$\left\|\boldsymbol{U}_{k}^{T}\widetilde{\boldsymbol{U}}_{\perp}\right\| \leq \frac{\tilde{\sigma}_{k}}{\sigma_{k-1}^{2}-\tilde{\sigma}_{k}^{2}}\left\|\boldsymbol{U}_{k}^{T}\boldsymbol{H}\right\|+\frac{\sigma_{k-1}}{\tilde{\sigma}_{k-1}^{2}-\tilde{\sigma}_{k}^{2}}\left\|\boldsymbol{H}\boldsymbol{V}_{k}\right\|, \tag{60}$$

$$\left\|\boldsymbol{U}_{k}^{T}\widetilde{\boldsymbol{U}}_{\perp}\right\| \leq \frac{\|\boldsymbol{H}\|}{\sigma_{k-1}-\tilde{\sigma}_{k}}. \tag{61}$$

Substituting the above formula into $\left\|\sin\Theta(\boldsymbol{U}_{k},\widetilde{\boldsymbol{U}}_{k})\right\|=\min\left\{\left\|\boldsymbol{U}_{\perp}^{T}\widetilde{\boldsymbol{U}}_{k}\right\|,\left\|\boldsymbol{U}_{k}^{T}\widetilde{\boldsymbol{U}}_{\perp}\right\|\right\}$[25], and **Lemma 2** holds.

**Lemma 3**. Let $\widetilde{\boldsymbol{A}} = \boldsymbol{A} + \boldsymbol{H}$, where $\boldsymbol{A}, \boldsymbol{H} \in \mathbb{R}^{n\times m}$. The conformal SVD of matrices $\boldsymbol{A}$ and $\widetilde{\boldsymbol{A}}$ are defined as:

$$\boldsymbol{A} = \boldsymbol{U\Sigma V}^T = (\boldsymbol{U}_k\ \boldsymbol{U}_\perp)\begin{pmatrix}\boldsymbol{\Sigma}_k & \\ & \boldsymbol{\Sigma}_\perp\end{pmatrix}\begin{pmatrix}\boldsymbol{V}_k^T \\ \boldsymbol{V}_\perp^T\end{pmatrix}. \tag{62}$$

$$\widetilde{\boldsymbol{A}} = \widetilde{\boldsymbol{U}}\widetilde{\boldsymbol{\Sigma}}\widetilde{\boldsymbol{V}}^T = (\widetilde{\boldsymbol{U}}_k\widetilde{\boldsymbol{U}}_\perp)\begin{pmatrix}\widetilde{\boldsymbol{\Sigma}}_k & \\ & \widetilde{\boldsymbol{\Sigma}}_\perp\end{pmatrix}\begin{pmatrix}\widetilde{\boldsymbol{V}}_k^T \\ \widetilde{\boldsymbol{V}}_\perp^T\end{pmatrix}. \tag{63}$$

If the singular values of $\boldsymbol{A}$ and $\widetilde{\boldsymbol{A}}$ satisfy $\sigma_{k-1} - \tilde{\sigma}_k > 0$ and $\tilde{\sigma}_{k-1} - \sigma_k > 0$, then the following expression holds:

$$\boldsymbol{U}_\perp^T\widetilde{\boldsymbol{U}}_k = \boldsymbol{F}_{\boldsymbol{U}}^{21} \circ \left(\boldsymbol{U}_\perp^T\boldsymbol{H}\widetilde{\boldsymbol{V}}_k\widetilde{\boldsymbol{\Sigma}}_k^T + \boldsymbol{\Sigma}_\perp\boldsymbol{V}_\perp^T\boldsymbol{H}^T\widetilde{\boldsymbol{U}}_k\right), \tag{64}$$

$$\boldsymbol{U}_k^T\widetilde{\boldsymbol{U}}_\perp = \boldsymbol{F}_{\boldsymbol{U}}^{12} \circ \left(\boldsymbol{U}_k^T\boldsymbol{H}\widetilde{\boldsymbol{V}}_\perp\widetilde{\boldsymbol{\Sigma}}_\perp^T + \boldsymbol{\Sigma}_k\boldsymbol{V}_k^T\boldsymbol{H}^T\widetilde{\boldsymbol{U}}_\perp\right), \tag{65}$$

$$\boldsymbol{V}_\perp^T\widetilde{\boldsymbol{V}}_k = \boldsymbol{F}_{\boldsymbol{V}}^{21} \circ \left(\widetilde{\boldsymbol{\Sigma}}_\perp^T\widetilde{\boldsymbol{U}}_\perp\boldsymbol{H}\widetilde{\boldsymbol{V}}_k + \boldsymbol{V}_\perp^T\boldsymbol{H}^T\widetilde{\boldsymbol{U}}_k\boldsymbol{\Sigma}_k\right), \tag{66}$$

$$\boldsymbol{V}_k^T\widetilde{\boldsymbol{V}}_\perp = \boldsymbol{F}_{\boldsymbol{V}}^{12} \circ \left(\boldsymbol{\Sigma}_k^T\boldsymbol{U}_k^T\boldsymbol{H}\widetilde{\boldsymbol{V}}_\perp + \boldsymbol{V}_k^T\boldsymbol{H}^T\widetilde{\boldsymbol{U}}_\perp\widetilde{\boldsymbol{\Sigma}}_\perp\right). \tag{67}$$

where $\circ$ is Hadamard product. $\boldsymbol{F}_{\boldsymbol{U}}^{21} \in \mathbb{R}^{n-k+1,k-1}$ has entries $(\boldsymbol{F}_{\boldsymbol{U}}^{21})_{i,j} = \frac{1}{\tilde{\sigma}_j^2 - \sigma_{i+k-1}^2}, 1 \le i \le n-k+1, 1 \le j \le k-1$ . $\boldsymbol{F}_{\boldsymbol{U}}^{12} \in \mathbb{R}^{k-1,n-k+1}$ has entries $(\boldsymbol{F}_{\boldsymbol{V}}^{12})_{i,j} = \frac{1}{\tilde{\sigma}_j^2 - \sigma_{i+k-1}^2}, 1 \le i \le m-k+1, 1 \le j \le k-1$. $\mathbf{F}_{\boldsymbol{V}}^{12} \in \mathbb{R}^{k-1,m-k+1}$ has entries $(\boldsymbol{F}_{\boldsymbol{V}}^{12})_{i,j} = \frac{1}{\tilde{\sigma}_{j+k-1}^2 - \sigma_i^2}$, $1 \le i \le k-1, 1 \le j \le m-k+1$.Since $i > \min\{n, m\}$, we set $\sigma_i$ and $\tilde{\sigma}_i$ to be 0.

***Proof.*** Firstly, we will decompose the perturbation $\boldsymbol{H}$ in the following two ways:

$$\begin{aligned}\boldsymbol{H} &= \widetilde{\boldsymbol{A}} - \boldsymbol{A} = \widetilde{\boldsymbol{U}}\widetilde{\boldsymbol{\Sigma}}\widetilde{\boldsymbol{V}}^T - \boldsymbol{U\Sigma V}^T \\ &= (\boldsymbol{U} + \Delta\boldsymbol{U})\widetilde{\boldsymbol{\Sigma}}\widetilde{\boldsymbol{V}}^T - \boldsymbol{U\Sigma}\left(\widetilde{\boldsymbol{V}} - \Delta\boldsymbol{V}\right)^T \\ &= \boldsymbol{U}\widetilde{\boldsymbol{\Sigma}}\widetilde{\boldsymbol{V}}^T + (\Delta\boldsymbol{U})\widetilde{\boldsymbol{\Sigma}}\widetilde{\boldsymbol{V}}^T - \boldsymbol{U\Sigma}\widetilde{\boldsymbol{V}} + \boldsymbol{U\Sigma}(\Delta\boldsymbol{V})^T \\ &= \boldsymbol{U}(\Delta\boldsymbol{\Sigma})\widetilde{\boldsymbol{V}}^T + (\Delta\boldsymbol{U})\widetilde{\boldsymbol{\Sigma}}\widetilde{\boldsymbol{V}}^T + \boldsymbol{U\Sigma}(\Delta\boldsymbol{V})^T.\end{aligned} \tag{68}$$

Multiplying (68) with $\boldsymbol{U}^T$ on the left and $\widetilde{\boldsymbol{V}}$ on the right leads to:

$$\boldsymbol{U}^T\boldsymbol{H}\widetilde{\boldsymbol{V}} = \Delta\boldsymbol{\Sigma} + \boldsymbol{U}^T(\Delta\boldsymbol{U})\widetilde{\boldsymbol{\Sigma}} + \boldsymbol{\Sigma}(\Delta\boldsymbol{V})^T\widetilde{\boldsymbol{V}}. \tag{69}$$

Thus,

$$\begin{aligned}\boldsymbol{H} &= \widetilde{\boldsymbol{A}} - \boldsymbol{A} = \widetilde{\boldsymbol{U}}\widetilde{\boldsymbol{\Sigma}}\widetilde{\boldsymbol{V}}^T - \boldsymbol{U\Sigma V}^T \\ &= \widetilde{\boldsymbol{U}}\widetilde{\boldsymbol{\Sigma}}(\boldsymbol{V} + \Delta\boldsymbol{V})^T - \left(\widetilde{\boldsymbol{U}} - \Delta\boldsymbol{U}\right)\boldsymbol{\Sigma V}^T \\ &= \widetilde{\boldsymbol{U}}\widetilde{\boldsymbol{\Sigma}}\boldsymbol{V}^T + \widetilde{\boldsymbol{U}}\widetilde{\boldsymbol{\Sigma}}(\Delta\boldsymbol{V})^T - \widetilde{\boldsymbol{U}}\boldsymbol{\Sigma V}^T + (\Delta\boldsymbol{U})\boldsymbol{\Sigma V}^T \\ &= \widetilde{\boldsymbol{U}}(\Delta\boldsymbol{\Sigma})\boldsymbol{V}^T + \widetilde{\boldsymbol{U}}\widetilde{\boldsymbol{\Sigma}}(\Delta\boldsymbol{V})^T + (\Delta\boldsymbol{U})\boldsymbol{\Sigma V}^T.\end{aligned} \tag{70}$$

Similarly, Multiplying (70) with $\widetilde{\boldsymbol{U}}^T$ on the left and $\boldsymbol{V}$ on the right leads to:

$$\widetilde{\boldsymbol{U}}^T\boldsymbol{H}\boldsymbol{V} = \Delta\boldsymbol{\Sigma} + \widetilde{\boldsymbol{\Sigma}}(\Delta\boldsymbol{V})^T\boldsymbol{V} + \widetilde{\boldsymbol{U}}^T(\Delta\boldsymbol{U})\boldsymbol{\Sigma}. \tag{71}$$

Denote $dP = \boldsymbol{U}^T\boldsymbol{H}\widetilde{\boldsymbol{V}}$, $d\bar{P} = \widetilde{\boldsymbol{U}}^T\boldsymbol{H}\boldsymbol{V}, \Delta\boldsymbol{\Omega}_U = \boldsymbol{U}^T(\Delta\boldsymbol{U})$, $\Delta\boldsymbol{\Omega}_V = \boldsymbol{V}^T(\Delta\boldsymbol{V})$. Notice that $\boldsymbol{I} = \widetilde{\boldsymbol{U}}^T\widetilde{\boldsymbol{U}} = \boldsymbol{U}^T\boldsymbol{U}$ gives $(\boldsymbol{U} + \Delta\boldsymbol{U})^T\widetilde{\boldsymbol{U}} = \boldsymbol{U}^T(\widetilde{\boldsymbol{U}} - \Delta\boldsymbol{U})$. Hence $\boldsymbol{U}^T(\Delta\boldsymbol{U}) = -(\Delta\boldsymbol{U})^T\widetilde{\boldsymbol{U}}$. Similarly, we also have $\boldsymbol{V}^T(\Delta\boldsymbol{V}) = -(\Delta\boldsymbol{V})^T\widetilde{\boldsymbol{V}}$. Plugging these into (68) and (70), we have

$$\begin{cases} dP = \boldsymbol{U}^T\boldsymbol{H}\widetilde{\boldsymbol{V}} = \Delta\boldsymbol{\Sigma} + \Delta\boldsymbol{\Omega}_U\widetilde{\boldsymbol{\Sigma}} - \boldsymbol{\Sigma}\Delta\boldsymbol{\Omega}_V, \\ d\bar{P} = \widetilde{\boldsymbol{U}}^T\boldsymbol{H}\boldsymbol{V} = \Delta\boldsymbol{\Sigma} + \widetilde{\boldsymbol{\Sigma}}\Delta\boldsymbol{\Omega}_V - \Delta\boldsymbol{\Omega}_U^T\boldsymbol{\Sigma}, \end{cases} \tag{72}$$

Next, from (72) we can cancel $\Delta\boldsymbol{\Omega}_V$ by

$$\begin{aligned} \mathbf{G}_\mathrm{U} &:= dP\widetilde{\boldsymbol{\Sigma}}^\mathrm{T} + \boldsymbol{\Sigma}d\bar{P}^\mathrm{T} \\ &= \widetilde{\boldsymbol{\Sigma}}\widetilde{\boldsymbol{\Sigma}}^T - \boldsymbol{\Sigma}\boldsymbol{\Sigma}^T + \Delta\boldsymbol{\Omega}_U\widetilde{\boldsymbol{\Sigma}}\widetilde{\boldsymbol{\Sigma}}^T - \boldsymbol{\Sigma}\boldsymbol{\Sigma}^T\Delta\boldsymbol{\Omega}_U. \end{aligned} \tag{73}$$

Let $\Delta\boldsymbol{\Omega}_\mathrm{U} = \{w_{ij}\}_{i,j=1}^n$. Then for all $1 \le i, j \le n$, the following formulas hold

$$(\mathbf{G}_U)_{ij} = \begin{cases} (\tilde{\sigma}_j{}^2 - \sigma_i^2)w_{ij}, & i \ne j, \\ (\tilde{\sigma}_j{}^2 - \sigma_i^2)(w_{ij} + 1), & i = j\,. \end{cases} \tag{74}$$

Here if $i > \min\{n, m\}$, we define $\sigma_i$ or $\tilde{\sigma}_i$ to be 0. Also, define $\boldsymbol{F}_{\boldsymbol{U}}^{21}$, $\boldsymbol{F}_{\boldsymbol{U}}^{12}$, $\boldsymbol{F}_{\boldsymbol{V}}^{21}$, $\boldsymbol{F}_{\boldsymbol{V}}^{12}$ as in the statement of **Lemma 2**. By assumption, $\sigma_{k-1} - \tilde{\sigma}_k > 0$ and $\tilde{\sigma}_{k-1} - \sigma_k > 0$, we can directly check that the denominators in these four matrices only have nonzero entries, thus are well defined. Consider the upper right part in $\Delta\boldsymbol{\Omega}_U = \boldsymbol{U}^T\boldsymbol{H}$, that is, $1 \le i \le k-1, k \le j \le n$, from (74) we have

$$w_{ij} = \frac{1}{\tilde{\sigma}_j^2 - \sigma_i^2}(\mathbf{G}_U)_{ij}, 1 \le i \le k-1, k \le j \le n. \tag{75}$$

Therefore,

$$\boldsymbol{U}_k^T\widetilde{\boldsymbol{U}}_\perp = \boldsymbol{U}_k^T(\Delta\boldsymbol{U}_\perp) = \boldsymbol{F}_{\boldsymbol{U}}^{12} \circ \boldsymbol{G}_{\boldsymbol{U}}^{12} = \boldsymbol{F}_{\boldsymbol{U}}^{12} \circ (\boldsymbol{U}_k^T\boldsymbol{H}\widetilde{\boldsymbol{V}}_\perp\widetilde{\boldsymbol{\Sigma}}_\perp^T + \boldsymbol{\Sigma}_k\boldsymbol{V}_k^T\boldsymbol{H}^T\widetilde{\boldsymbol{U}}_\perp) \tag{76}$$

Following the same reasoning, we also obtain

$$\boldsymbol{U}_\perp^T\widetilde{\boldsymbol{U}}_k = \boldsymbol{U}_\perp^T(\Delta\boldsymbol{U}) = \boldsymbol{F}_{\boldsymbol{U}}^{21} \circ (\boldsymbol{U}_\perp^T\boldsymbol{H}\widetilde{\boldsymbol{V}}_k\widetilde{\boldsymbol{\Sigma}}_k^T + \boldsymbol{\Sigma}_\perp\boldsymbol{V}_\perp^T\boldsymbol{H}^T\widetilde{\boldsymbol{U}}_k), \tag{77}$$

$$\boldsymbol{V}_\perp^T\widetilde{\boldsymbol{V}}_k = \boldsymbol{V}_\perp^T(\Delta\boldsymbol{V}_\perp) = \boldsymbol{F}_{\boldsymbol{V}}^{21} \circ (\boldsymbol{\Sigma}_\perp^T\widetilde{\boldsymbol{U}}_\perp\boldsymbol{H}\widetilde{\boldsymbol{V}}_k + \boldsymbol{V}_\perp^T\boldsymbol{H}^T\widetilde{\boldsymbol{U}}_k\widetilde{\boldsymbol{\Sigma}}_k), \tag{78}$$

$$\boldsymbol{V}_k^T\widetilde{\boldsymbol{V}}_\perp = \boldsymbol{V}_k^T(\Delta\boldsymbol{V}_\perp) = \boldsymbol{F}_{\boldsymbol{V}}^{12} \circ (\boldsymbol{\Sigma}_k^T\boldsymbol{U}_k^T\boldsymbol{H}\widetilde{\boldsymbol{V}}_\perp + \boldsymbol{V}_k^T\boldsymbol{H}^T\widetilde{\boldsymbol{U}}_\perp\widetilde{\boldsymbol{\Sigma}}_\perp). \tag{79}$$

**Lemma 4**. Assume $\sigma_{k-1} - \tilde{\sigma}_k > 0$ and $\tilde{\sigma}_{k-1} - \sigma_k > 0$, and let $\mathbf{B}_1 \in \mathbb{R}^{n-k+1,k-1}, \mathbf{B}_2 \in \mathbb{R}^{m-k+1,k-1}, \mathbf{B}_3 \in \mathbb{R}^{k-1,m-k+1}, \mathbf{B}_4 \in \mathbb{R}^{k-1,n-k+1}$ be some arbitrary matrices.

Then, we have

$$\left\|\boldsymbol{F}_{\boldsymbol{U}}^{21} \circ (\boldsymbol{B}_1\widetilde{\boldsymbol{\Sigma}}_k)\right\|_F \le \frac{\tilde{\sigma}_{k-1}}{\tilde{\sigma}_{k-1}^2 - \sigma_k^2}\|\boldsymbol{B}_1\|_F, \|\boldsymbol{F}_{\boldsymbol{U}}^{21} \circ (\boldsymbol{\Sigma}_\perp\boldsymbol{B}_2)\|_F \le \frac{\sigma_k}{\tilde{\sigma}_{k-1}^2 - \sigma_k^2}\|\boldsymbol{B}_2\|_F, \tag{80}$$

$$\left\|\boldsymbol{F}_{\boldsymbol{U}}^{12} \circ (\boldsymbol{B}_3\widetilde{\boldsymbol{\Sigma}}_\perp{}^T)\right\|_F \le \frac{\tilde{\sigma}_k}{\sigma_{k-1}^2 - \tilde{\sigma}_k^2}\|\boldsymbol{B}_3\|_F, \|\boldsymbol{F}_{\boldsymbol{U}}^{12} \circ (\boldsymbol{\Sigma}_k\boldsymbol{B}_4)\|_F \le \frac{\sigma_{k-1}}{\sigma_{k-1}^2 - \tilde{\sigma}_k^2}\|\boldsymbol{B}_4\|_F. \tag{81}$$

where $\|\cdot\|_F$ is the Frobenius norm. The same inequality holds for the spectral norm. Similar results also hold for $\boldsymbol{F}_{\boldsymbol{V}}^{12}$ and $\boldsymbol{F}_{\boldsymbol{V}}^{21}$.

***Proof.*** Here we only prove the first inequality in the above formulas, and the other three inequalities can be proved similarly. Recall that the definition of $\boldsymbol{F}_{\boldsymbol{U}}^{21}$ is $(\boldsymbol{F}_{\boldsymbol{U}}^{21})_{i-k+1,j} = \frac{\mathbf{1}}{\tilde{\sigma}_j^2-\sigma_{i+k-1}^2}$, $1 \le i \le n,\ 1 \le j \le k-1$. We directly have

$$\boldsymbol{F}_{\boldsymbol{U}}^{21} \circ \left(\boldsymbol{B}_1 \widetilde{\boldsymbol{\Sigma}}_k\right) = \overline{\boldsymbol{F}}_{\boldsymbol{U}}^{21} \circ \boldsymbol{B}_1 \tag{82}$$

where

$$(\overline{\boldsymbol{F}}_{\boldsymbol{U}}^{21})_{i-k+1,j} = \frac{\tilde{\sigma}_j}{\tilde{\sigma}_j^2 - \sigma_i^2}, k \le i \le n, 1 \le j \le k-1 \tag{83}$$

Let $\mathbf{D}_1 = \boldsymbol{F}_{\boldsymbol{U}}^{21} \circ \left(\boldsymbol{B}_1 \widetilde{\boldsymbol{\Sigma}}_k\right)$. Then $\boldsymbol{B}_1 = \widetilde{\boldsymbol{F}}_{\boldsymbol{U}}^{21} \circ \boldsymbol{D}_1$, where

$$\left(\widetilde{\boldsymbol{F}}_{\boldsymbol{U}}^{21}\right)_{i-k+1,j} = \frac{\tilde{\sigma}_j^2 - \sigma_i^2}{\tilde{\sigma}_j} = \tilde{\sigma}_j - \frac{\sigma_i^2}{\tilde{\sigma}_j}, k \le i \le n, 1 \le j \le k-1 \tag{84}$$

Inserting the above expression of $\widetilde{\boldsymbol{F}}$ into $\boldsymbol{B}_1 = \widetilde{\boldsymbol{F}}_{\boldsymbol{U}}^{21} \circ \boldsymbol{D}_1$, we have

$$\boldsymbol{B}_1 = \boldsymbol{D}_1 \begin{pmatrix} \tilde{\sigma}_1 & & & \\ & \tilde{\sigma}_2 & & \\ & & \ddots & \\ & & & \tilde{\sigma}_{k-1} \end{pmatrix} - \begin{pmatrix} \tilde{\sigma}_k^2 & & & \\ & \tilde{\sigma}_{k+1}^2 & & \\ & & \ddots & \\ & & & \tilde{\sigma}_n^2 \end{pmatrix} \mathbf{D}_1 \begin{pmatrix} \frac{1}{\tilde{\sigma}_1} & & & \\ & \frac{1}{\tilde{\sigma}_2} & & \\ & & \ddots & \\ & & & \frac{1}{\tilde{\sigma}_{k-1}} \end{pmatrix} \tag{85}$$

Taking norm on both sides, we obtain

$$|||\boldsymbol{B}_1||| \le \tilde{\sigma}_{k-1} |||\boldsymbol{D}_1||| - \frac{\sigma_k^2}{\tilde{\sigma}_{k-1}} |||\boldsymbol{D}_1||| = \frac{\tilde{\sigma}_{k-1}^2 - \sigma_k^2}{\tilde{\sigma}_{k-1}} |||\boldsymbol{D}_1|||, \tag{86}$$

which further gives

$$|||\boldsymbol{D}_1||| \le \frac{\tilde{\sigma}_{k-1}}{\tilde{\sigma}_{k-1}^2 - \sigma_k^2} |||\boldsymbol{B}_1||| \tag{87}$$

Similarly, other formulas can be obtained.

### *3.3 The fast Network Reconstruction via Global Connectivity-Based Topological Similarity*

This section mainly presents the proofs of the relevant theorems for fastNR.

**Theorem 6.** For any $\boldsymbol{A} \in \mathbb{R}^{n \times m}$ and noisy version $\widetilde{\boldsymbol{A}} \coloneqq \boldsymbol{A} + \boldsymbol{H}$, their low-rank approximations $\boldsymbol{A}_k$ and $\widetilde{\boldsymbol{A}}_k$ satisfy $sup\ \{\|\boldsymbol{A}_k - \widetilde{\boldsymbol{A}}_k\|\} \ge \sup\ \{\|\boldsymbol{A}_k^* - \widetilde{\boldsymbol{A}}_k^*\|\}$, where $\boldsymbol{A}_k^*$ and $\widetilde{\boldsymbol{A}}_k^*$ are the enhanced $\boldsymbol{A}_k$ and $\widetilde{\boldsymbol{A}}_k$.

***Proof*.** According to perturbation result on singular value thresholding (**Theorem 2.10** in [24]), we can get:

$$\left\|\boldsymbol{A}_k - \widetilde{\boldsymbol{A}}_k\right\| \leq 2\|\boldsymbol{H}\| + 2\sigma_{k+1}\min\left\{\frac{2\|\boldsymbol{H}\|}{\sigma_k - \sigma_{k+1}}, 1\right\}, \tag{88}$$

$$\left\|\boldsymbol{A}_k - \widetilde{\boldsymbol{A}}_k\right\| \leq 2\|\boldsymbol{H}\| + \min\left\{\frac{4\|\boldsymbol{H}\|}{\frac{\sigma_k}{\sigma_{k+1}} - 1}, 2\sigma_{k+1}\right\}. \tag{89}$$

According to **Theorem 4**,

$$\left\|\boldsymbol{A}_k^* - \widetilde{\boldsymbol{A}}_k^*\right\| \leq 2\|\boldsymbol{H}\| + \min\left\{\frac{4\|\boldsymbol{H}\|}{\frac{\sigma_k}{\sigma_{k+1}}\frac{\sigma_k^2\sigma_{k+1}^2+\lambda\sigma_k^2}{\sigma_k^2\sigma_{k+1}^2+\lambda\sigma_{k+1}^2} - 1}, 2\sigma_{k+1}\frac{\sigma_{k+1}^2}{\sigma_{k+1}^2+\lambda}\right\}. \tag{90}$$

We demonstrate that the perturbation on the augmented network $\boldsymbol{A}_k^* = \mathrm{NR}(\boldsymbol{A}_k)$ has a smaller impact compared to a low-rank approximation $\boldsymbol{A}_k$ of any matrix $\boldsymbol{A}$, as $sup\left\{\left\|\boldsymbol{A}_k - \widetilde{\boldsymbol{A}}_k\right\|\right\} \geq \sup\left\{\left\|\boldsymbol{A}_k^* - \widetilde{\boldsymbol{A}}_k^*\right\|\right\}$.

As the matrix obtained by randQB_fp decomposition is in the form of truncated SVD ($\boldsymbol{A}_k = \boldsymbol{U}_k\boldsymbol{\Sigma}_k\boldsymbol{V}_k^T$), while according to the previous derivation, NR can only be calculated in the form of full SVD. To enable direct computation of NR, we introduce **Theorem 7.**

**Theorem 7.** When $\boldsymbol{A} \in \mathbb{R}^{n\times m}$ is low-rank, the result of NR ($\boldsymbol{A}^* = \boldsymbol{U}\boldsymbol{\Sigma}^T\boldsymbol{\Sigma}(\boldsymbol{\Sigma}^T\boldsymbol{\Sigma} + \lambda\boldsymbol{I})^{-1}\boldsymbol{\Sigma}\boldsymbol{V}^T$) can be calculated by truncated SVD. The full SVD of matrix $\boldsymbol{A}$ can be expressed as $(\boldsymbol{U}' \quad \boldsymbol{U}'_\perp)\begin{pmatrix}\boldsymbol{\Sigma}' & \boldsymbol{0}\\ \boldsymbol{0} & \boldsymbol{0}\end{pmatrix}(\boldsymbol{V}' \quad \boldsymbol{V}'_\perp)^T$ ,and the truncated SVD decomposition of matrix A can be expressed as $\mathbf{U}'\boldsymbol{\Sigma}'\mathbf{V}'$.

***Proof.*** The NR is calculated by full SVD:

$$\begin{aligned}
\mathbf{A}^* &= (\boldsymbol{U}' \quad \boldsymbol{U}'_\perp)\begin{pmatrix}\boldsymbol{\Sigma}' & \boldsymbol{0}\\ \boldsymbol{0} & \boldsymbol{0}\end{pmatrix}^T\begin{pmatrix}\boldsymbol{\Sigma}' & \boldsymbol{0}\\ \boldsymbol{0} & \boldsymbol{0}\end{pmatrix}\left(\begin{pmatrix}\boldsymbol{\Sigma}' & \boldsymbol{0}\\ \boldsymbol{0} & \boldsymbol{0}\end{pmatrix}^T\begin{pmatrix}\boldsymbol{\Sigma}' & \boldsymbol{0}\\ \boldsymbol{0} & \boldsymbol{0}\end{pmatrix} + \lambda\boldsymbol{I}\right)^{-1}\begin{pmatrix}\boldsymbol{\Sigma}' & \boldsymbol{0}\\ \boldsymbol{0} & \boldsymbol{0}\end{pmatrix}(\boldsymbol{V}' \quad \boldsymbol{V}'_\perp)^T \\
&= (\boldsymbol{U}' \quad \boldsymbol{U}'_\perp)\begin{pmatrix}\boldsymbol{\Sigma}'^T\boldsymbol{\Sigma}' & \boldsymbol{0}\\ \boldsymbol{0} & \boldsymbol{0}\end{pmatrix}\left(\begin{pmatrix}\boldsymbol{\Sigma}'^T\boldsymbol{\Sigma}' & \boldsymbol{0}\\ \boldsymbol{0} & \boldsymbol{0}\end{pmatrix} + \lambda\boldsymbol{I}\right)^{-1}\begin{pmatrix}\boldsymbol{\Sigma}' & \boldsymbol{0}\\ \boldsymbol{0} & \boldsymbol{0}\end{pmatrix}(\boldsymbol{V}' \quad \boldsymbol{V}'_\perp)^T \\
&= (\boldsymbol{U}' \quad \boldsymbol{U}'_\perp)\begin{pmatrix}\boldsymbol{\Sigma}'^T\boldsymbol{\Sigma}' & \boldsymbol{0}\\ \boldsymbol{0} & \boldsymbol{0}\end{pmatrix}\begin{pmatrix}\boldsymbol{\Sigma}'^T\boldsymbol{\Sigma}' + \lambda\boldsymbol{I} & \boldsymbol{0}\\ \boldsymbol{0} & \lambda\boldsymbol{I}\end{pmatrix}^{-1}\begin{pmatrix}\boldsymbol{\Sigma}' & \boldsymbol{0}\\ \boldsymbol{0} & \boldsymbol{0}\end{pmatrix}(\boldsymbol{V}' \quad \boldsymbol{V}'_\perp)^T \\
&= (\boldsymbol{U}' \quad \boldsymbol{U}'_\perp)\begin{pmatrix}\boldsymbol{\Sigma}'^T\boldsymbol{\Sigma}' & \boldsymbol{0}\\ \boldsymbol{0} & \boldsymbol{0}\end{pmatrix}\begin{pmatrix}(\boldsymbol{\Sigma}'^T\boldsymbol{\Sigma}' + \lambda\boldsymbol{I})^{-1} & \boldsymbol{0}\\ \boldsymbol{0} & \frac{1}{\lambda}\boldsymbol{I}\end{pmatrix}\begin{pmatrix}\boldsymbol{\Sigma}' & \boldsymbol{0}\\ \boldsymbol{0} & \boldsymbol{0}\end{pmatrix}(\boldsymbol{V}' \quad \boldsymbol{V}'_\perp)^T \\
&= (\boldsymbol{U}' \quad \boldsymbol{U}'_\perp)\begin{pmatrix}\boldsymbol{\Sigma}'^T\boldsymbol{\Sigma}'(\boldsymbol{\Sigma}'^T\boldsymbol{\Sigma}' + \lambda\boldsymbol{I})^{-1} & \boldsymbol{0}\\ \boldsymbol{0} & \boldsymbol{0}\end{pmatrix}\begin{pmatrix}\boldsymbol{\Sigma}' & \boldsymbol{0}\\ \boldsymbol{0} & \boldsymbol{0}\end{pmatrix}(\boldsymbol{V}' \quad \boldsymbol{V}'_\perp)^T \\
&= \boldsymbol{U}'\boldsymbol{\Sigma}'^T\boldsymbol{\Sigma}'(\boldsymbol{\Sigma}'^T\boldsymbol{\Sigma}' + \lambda\boldsymbol{I})^{-1}\boldsymbol{V}'
\end{aligned} \tag{91}$$

Therefore, if the matrix A is low rank, the result of NR can be computed by truncated SVD.

## 4 Comparison of link prediction methods based on latent space

In investigating link prediction methods rooted in latent hyperbolic spaces, our analysis revealed a lack of substantive superiority over traditional techniques. Such methods predominantly rely on the assumption that nodes in closer proximity within the latent space exhibit a higher propensity to form links. In contrast, we present the Tikhonov random walk with restart (TRWR), a strategy single-step diffusion predicated on latent space energy distances. This underpins our novel hypothesis: nodes exhibit a tendency to form connections with neighbors of those nodes with analogous topological structure. Specifically, by applying the convergence conditions proven in **Theorem 2**, formulated as a self-expressive problem, we utilize Tikhonov regularization to determine the value of $\boldsymbol{C}$, thus deriving a new closed-form solution for Random Walk with Restart (RWR).

The convergence of the iterative process occurs when $\boldsymbol{C}$ equals the identity matrix, that is, $\boldsymbol{P}^{\infty} = (1-\alpha)(\boldsymbol{I}-\alpha\boldsymbol{W})^{-1}\boldsymbol{P^0}$, as suggested by the Frobenius norm $||\boldsymbol{P^0}-\boldsymbol{CP^0}||_F^2 < \varepsilon$. This iterative process can be seen as a self-expressive problem[23]. In solving $\boldsymbol{CP}^0 = \boldsymbol{P}^0$, if $\boldsymbol{P}^0$ is an ill-conditioned matrix, the solution $\boldsymbol{C}$ will deviate from the true value. To determine whether complex networks are ill-conditioned, we computed the condition number of 132 complex networks (**Table S2 and S3**).

Among them, 22 link prediction networks were downloaded from https://noesis.ikor.org/datasets/link-prediction; 45 real-world networks were downloaded from https://networkrepository.com/index.php; 44 Hi-C data were obtained from http://snap.stanford.edu/ne/; 18 biological association networks, and 3 link prediction networks were obtained from published literatures. Our observations reveal that the condition numbers of all complex networks are significantly high. A substantial number of these networks lack full rank properties, leading to the computed value of "inf," indicating infinity. As a result, these networks $\boldsymbol{P^0}$render the matrix equation ($\boldsymbol{CP}_0 = \boldsymbol{P}_0$) unsolvable. Therefore, we can conclude that a common characteristic among many complex networks is "ill-conditioned".

Here, we solve for $\boldsymbol{P}_0 - \boldsymbol{CP}_0 = \boldsymbol{0}$ using the Tikhonov regularization[26] to find a $\boldsymbol{C}$ that is not affected by the ill-conditioned matrix.

$$min \;\; \frac{1}{\lambda}||\,\boldsymbol{P}_0 - \boldsymbol{CP}_0||_F^2 + ||\boldsymbol{C}||\,_F^2 \tag{92}$$

where $||\boldsymbol{C}||\,_F^2 = \mathrm{Tr}(\boldsymbol{C^T C})$. The above formula can be expressed as:

$$\begin{aligned} E &= \frac{1}{\lambda}||\boldsymbol{P}_0 - \boldsymbol{CP}_0||_F^2 + ||\boldsymbol{C}||\,_F^2 \\ &= \frac{1}{\lambda}\mathrm{Tr}[(\boldsymbol{P}_0 - \boldsymbol{CP}_0)^T(\,\boldsymbol{P}_0 - \boldsymbol{CP}_0)] + Tr(\boldsymbol{C^T C}) \\ &= \frac{1}{\lambda}\mathrm{Tr}\big[\boldsymbol{P}_0{}^T\boldsymbol{P}_0 - \boldsymbol{P}_0{}^T\boldsymbol{CP}_0 - \boldsymbol{C}^T\boldsymbol{P}_0^T\boldsymbol{P}_0 + \boldsymbol{C}^T\boldsymbol{P}_0^T\boldsymbol{CP}_0\big] + \mathrm{Tr}(\boldsymbol{C}^T\boldsymbol{C}) \end{aligned} \tag{93}$$

Then,

$$\frac{\partial E}{\partial \boldsymbol{C}} = \frac{1}{\lambda}\left[-2\boldsymbol{P}_0{}^T\boldsymbol{P}_0 + \boldsymbol{2}\boldsymbol{P}_0{}^T\boldsymbol{P}_0\boldsymbol{C}\right] + 2\boldsymbol{C} \tag{94}$$

Let $\frac{\partial E}{\partial \boldsymbol{C}} = 0$. Then we can get

$$\boldsymbol{C} = \boldsymbol{P}_0\boldsymbol{P}_0{}^T\left(\boldsymbol{P}_0\boldsymbol{P}_0{}^T + \lambda\boldsymbol{I}\right)^{-1} \tag{95}$$

Substituting $\boldsymbol{C}$ into the analytical solution:

$$\boldsymbol{B} = (1-\alpha)(\boldsymbol{I} - \alpha\boldsymbol{W})^{-\boldsymbol{1}}\,\boldsymbol{P}_0\boldsymbol{P}_0{}^T\left(\,\boldsymbol{P}_0\boldsymbol{P}_0{}^T + \lambda\boldsymbol{I}\right)^{-1} \tag{96}$$

$$\boldsymbol{P}_\infty = (1-\alpha)(\boldsymbol{I} - \alpha\boldsymbol{W})^{-1}\,\boldsymbol{P}_0\boldsymbol{P}_0{}^T\left(\,\boldsymbol{P}_0\boldsymbol{P}_0{}^T + \lambda\boldsymbol{I}\right)^{-1}\boldsymbol{P}^{\boldsymbol{0}} \tag{97}$$

Compared with the basic random walk with restart, Tikhonov random walk with restart (TRWR) converts $\boldsymbol{P}_0$ into $\boldsymbol{P}_0\boldsymbol{P}_0{}^T\left(\boldsymbol{P}_0\boldsymbol{P}_0{}^T + \lambda\boldsymbol{I}\right)^{-1}\boldsymbol{P}_0$ and then performs the basic random walk with restart to reduce the impact of noise on the algorithm. For the network $\boldsymbol{A}$, TRWR can be defined as:

$$\boldsymbol{A}^* = (1-\alpha)(\boldsymbol{I} - \alpha\boldsymbol{W})^{-1}\boldsymbol{A}\boldsymbol{A}^T(\boldsymbol{A}\boldsymbol{A}^T + \lambda\boldsymbol{I})^{-1}\boldsymbol{A} \tag{98}$$

where $\boldsymbol{A}^*$ represents the network after random walk. $\boldsymbol{W}$ is the transition probability matrix. To investigate the performance of methods based on two different assumptions, we compared TRWR with Hyperlink, the state-of-the-art model based on the traditional assumption. Empirical test demonstrated that TRWR persistently outclasses the HyperLink over nine distinct link prediction datasets (**Fig.S4**). TRWR exhibits a remarkable enhancement in average Area Under the Curve (AUC) performance, registering an average AUC of roughly 0.8790 versus HyperLink's 0.6399, this represents an impressive increase of 37.38% for TRWR. Delving into individual datasets, the 'Ecoli' dataset merits particular attention. While HyperLink posts a commendable AUC of 0.9065, it remains eclipsed by TRWR's result of 0.9367, underscoring TRWR's sustained edge across varied data contexts. These results validate our hypothesis that nodes exhibit a tendency to form connections with neighbors of those nodes with analogous topological structure.

## 5 Comparison of energy-distance-based methods for network enhancement

Previously, we theoretically demonstrated that the TopoLa distance, compared to the common neighbor algorithm, exhibits a higher rescaled inverse temperature under the same network, offering a more precise description of the energy distance (Supplementary text 3.1, Theorem 1 and Theorem 3). Here, we verify it by network enhancement tasks, indirectly validating the performance differences between the two methods in measurements. Specifically, we enhance nine complex networks using the common neighbor (CN) algorithm and TopoLa distance, subsequently employing Random Walk with Restart (RWR) for link prediction on these enhanced networks. The prediction accuracy thus reflects the precision of the energy distance measurement (**Fig.S5**).

In a comparative analysis of algorithmic efficacy across diverse datasets, TopoLa distance consistently outperformed the CN algorithm, as evidenced by superior AUC and AUPR metrics. Notably, TopoLa distance's performance on the NS dataset, with an AUC of 0.9621 and an AUPR of 0.6387, underscores its robust predictive capabilities. In contrast, the CN algorithm demonstrated modest predictive ability, as reflected in its AUC of 0.5232 and AUPR of 0.0398 on the USAir dataset. Moreover, the relatively low standard deviations associated with TopoLa distance's performance suggest a high degree of result reproducibility. The experimental outcomes indirectly illuminate the precision discrepancies between the two methods in measuring energy distance.

## 6 Identification of cell types using single-cell RNA-seq data

In the main text, we applied NR on four single-cell RNA-seq datasets. To complement our experimental findings, this section assesses the performance of the four methods on the remaining 26 datasets (**Fig.S6 and S7**). The outcomes reveal that, on average, DSD, ND, and NE fail to significantly enhance the clustering quality of single-cell data. In contrast, NR exhibits superior performance, outperforming DSD, ND, and NE, with an average increase of 10.3% in the adjusted Rand index (ARI). The second-best denoising method, ND, demonstrates an average reduction of 2%, while the normalized mutual information (NMI) shows an average increase of 6.2% (NR) compared to the average 1% decrease with ND.

## 7 Using NR to complement network enhancement methods

To validate the complementary enhancement of NR with existing methods, we conducted additional experiments in link prediction and single-cell analysis (**Table S4** and **S5**). Specifically, we first enhanced the original networks with NR, then applied existing methods. Results show improvements in both experiments: for link prediction, ND's AUPR increased by an average of 107.4%, and NE's by 7.1%. The improvement shown by DSD is insignificant. In single-cell analysis, NR significantly boosted DSD, with an average NMI increase of 11.7% across thirty datasets, ND by 4.4%, and NE by 5.5%. These outcomes across diverse real-world cases demonstrate the synergistic potential of NR with existing methods, suggesting Global Connectivity-Based Topological Similarity as a novel network enhancement strategy.

## 8 Comparison of performance between NR and fastNR

The computational efficiency of large networks is compromised by the need to perform matrix inversions. To address this issue, we propose a novel approach, fastNR, to improve the efficiency of matrix inversions. Our approach leverages randQB_fp[27] to obtain a low-rank approximation $\boldsymbol{A}_k = \boldsymbol{QB} = \boldsymbol{U}_k \boldsymbol{\Sigma}_k \boldsymbol{V}_k^T$ of the matrix $\boldsymbol{A}$. We then employ the corollary of **Theorem 4** to calculate the fastNR, which can enhance networks (**Theorem 6**). Firstly, we evaluate the impact of low-rank approximation on its performance across four experiments: link prediction, single-cell data analysis, TAD detection, and fine-grained species identification (**Fig.S8A**). We observed that the performance of NR and fastNR is consistent in link prediction (Figure left 1) and fine-grained species identification (Figure right 1), and the low-rank approximation did not affect NR's performance on these datasets. In the case of single-cell data analysis, seven datasets saw an improvement in performance, seventeen saw a decrease, and six remained consistent.

Furthermore, we investigated the effect of random algorithms on the model's computational efficiency and requirements (**Fig.S8B**). The time complexity for singular value decomposition stands at $O(mn^2)$, where $m > n$; however, within the randQB_fp algorithm, this complexity reduces to $O(mk^2)$ for decomposing $\boldsymbol{B}$, thereby accelerating computation. To simulate complex networks in real-world scenarios, we generated a randomly sized matrix of $n \times n$ with geometrically distributed singular values. In the left figure, we compared the computational requirements of the two methods and observed that fastNR's advantage in computational requirements became increasingly apparent with an increase in the number of nodes. Notably, the improvement in computational efficiency was more significant than that in computational requirements. At 4000 nodes, fastNR's computational efficiency improved by 680%, and at 6000 nodes, it improved by 1024%. These experiments demonstrate that the proposed fastNR not only outperforms NR in computational efficiency and requirements but also outperforms NR in performance in certain scenarios.

## 9 Analysis of $n$-hop path statistics

In the statistical analysis of n-hop paths, paths with loops are also included. By removing loops from these paths, $n$-hop paths can be categorized into ($n$-1) types. Based on the number of hops after loop removal, these paths can be abstracted as polygons. For instance, 2-hop paths can be represented as triangles ($P_2$), while 4-hop paths correspond to pentagons ($P_4$). We observed a direct correlation between $|b_n(l)|$ and the degrees $\kappa_i$ and $\kappa_j$ . Consequently, all n-hop paths are categorized into three types: $a_n$, $b_n$, and $c_n$. In this section, our focus is on $c_n$, primarily through the analysis of 2-hop, 4-hop, and 6-hop paths between nodes D and E (**Fig.S9**). Paths without loop in the network are:

$a_2$: *D-C-E.*

$a_3$: *D-A-B-E*; *D-A-C-E.*

$a_4$: *D-A-B-G-E*; *D-C-A-B-E*; *D-C-H-G-E.*

$a_6$: *D-A-B-G-H-C-E*; *D-A-C-H-G-B-E.*

According to our proposed classification system, the statistics of n-hop paths can be represented as:

$$|n-\text{hop}| = \left(\sum_{t=2}^{n-2} |\mathrm{P}_t|\right) + |\mathrm{P}_n| \tag{99}$$

Consequently, we can determine the quantity of 2-hop paths:

$$|2-\text{hop}| = |\mathrm{P}_2| = |a_2| + |b_2| + |c_2| = |a_2| + 0 + 0 = 1 \tag{100}$$

where $|\mathrm{P}_2|$ represents the quantity of $\mathrm{P}_2$-type paths. The quantity of 4-hop paths is:

$$\begin{aligned}|4-\text{hop}| &= |\mathrm{P}_1| + |\mathrm{P}_2| + |\mathrm{P}_4| = 0 + \left[(\kappa_D + \kappa_E)|a_2| + \left(\sum_{t\in\mathcal{N}_2, t\notin\{D,E\}} \kappa_t\right) - 2|a_3|\right] + |a_4| \\ &= |\mathrm{P}_4| + (\kappa_D + \kappa_E)|a_2| + \left[\left(\sum_{t\in\mathcal{N}_2, t\notin\{D,E\}} \kappa_t\right) - 2|a_3|\right] \\ &= |a_4| + |b_4| + |c_4| = 3 + (5\times 1) + (4-2) = 10\end{aligned} \tag{101}$$

where $\mathcal{N}_2$ represents the set of nodes involved in all 2-hop paths, while $\kappa_t$ denotes the degree of node $t$. Additionally, $\left[\left(\sum_{t\in\mathcal{N}_2, t\notin\{D,E\}} \kappa_t\right) - 2|\mathrm{P}_3|\right]$ represents the count of $\mathrm{P}_3$-type paths related to the degrees of other nodes. The quantity of 6-hop paths is:

$$
\begin{aligned}
|6-\text{hop}| &= |\mathrm{P}_1| + |\mathrm{P}_2| + |\mathrm{P}_3| + |\mathrm{P}_4| + |\mathrm{P}_6| \\
&= 0 + |\mathrm{P}_2| + |\mathrm{P}_3| + \left[\left(\sum_{t\in\mathcal{N}_4, t\notin\{D,E\}} \kappa_t\right) - 2|a_4| - \eta_1\right] + (\kappa_D + \kappa_E)|a_4| + |a_6| \\
&= 0 + |\mathrm{P}_2| + |\mathrm{P}_3| + (27-6-8) + 5|a_4| + |a_6| \\
&= |a_6| + |\mathrm{P}_2| + |\mathrm{P}_3| + 5|a_4| + 13 \\
&= |a_6| + |\mathrm{P}_2| + 2[l_3(D) + l_3(E)]|a_3| + \left(\sum_{t\in\mathcal{N}_3, t\notin\{D,E\}} l_3(t)\right) + 5|a_4| + 13 \\
&= |a_6| + |\mathrm{P}_2| + 8 + 0 + 5|a_4| + 13 \\
&= |a_6| + |\mathrm{P}_2| + 5|a_4| + 21 \\
&= 2 + 15 + |\mathrm{P}_2| + 21 \\
&= 2 + 15 + p_1 + p_2 + p_3 + p_4 + 21 \\
&= 2 + 15 + (\kappa_D^2 + \kappa_E^2 + \kappa_D\kappa_E)|a_2| + p_2 + p_3 + p_4 + 21 \\
&= 2 + 15 + 19 + p_2 + p_3 + p_4 + 21 \\
&= |a_6| + |b_6| + p_2 + p_3 + p_4 + 21 \\
&= |a_6| + |b_6| + \sum_{t\in\mathcal{N}_2, t\notin\{B,D\}} (\kappa_t^2 - 3) + p_3 + p_4 + 21 \\
&= |a_6| + |b_6| + 13 + p_3 + p_4 + 21 \\
&= |a_6| + |b_6| + p_3 + p_4 + 34 \\
&= |a_6| + |b_6| + \left(\sum_{t\in\mathcal{N}_{D.n}, t\notin\{D,E\}} \kappa_t + \sum_{t\in\mathcal{N}_{E.n}, t\notin\{D,E\}} \kappa_t\right)|a_2| - \eta_3 + p_4 + 34 \\
&= |a_6| + |b_6| + 9 - 3 + p_4 + 34 \\
&= |a_6| + |b_6| + 2[l_4(D) + l_4(E)]|a_2| + \sum_{t\in\mathcal{N}_2, t\notin\{B,D\}} l_4(t) - \eta_4 + 40 \\
&= |a_6| + |b_6| + 2(1+3) + 27 - 22 + 40 \\
&= |a_6| + |b_6| + 53 \\
&= |a_6| + |b_6| + |c_6| = 89
\end{aligned}
\tag{102}
$$

where $\mathcal{N}_4$ denotes the list of nodes, which may be repeated, involved in all 4-hop paths. $l_3(\cdot)$ measures the quantity of 3-hop paths returning to itself. $p_1$ represents $|b_n|$. $p_2$ denotes the number of paths starting from common neighbors, jumping to non-D, non-E nodes, and then returning to themselves. $p_3$ accounts for paths initiating from D or E, jumping to non-common neighbors, and then looping back. $p_4$ counts 4-hop paths looping back to themselves via D, E, or common neighbors. $\mathcal{N}_{D.n}$ denotes the list of neighbor nodes of D. $\mathcal{N}_{E.n}$ denotes the list of neighbor nodes of E. $\eta_1$, $\eta_2$, $\eta_3$, $\eta_4$ represent the statistical redundancies of paths that satisfy multiple classification criteria within their respective terms. The analysis reveals that the calculation of $|c_n|$ is predominantly related to the degrees of nodes other than $\kappa_i$ and $\kappa_j$. Hence, the study does not investigate the influence of $|c_n|$ on the topological structure and degree information captured by $\boldsymbol{D}_{topo}$.

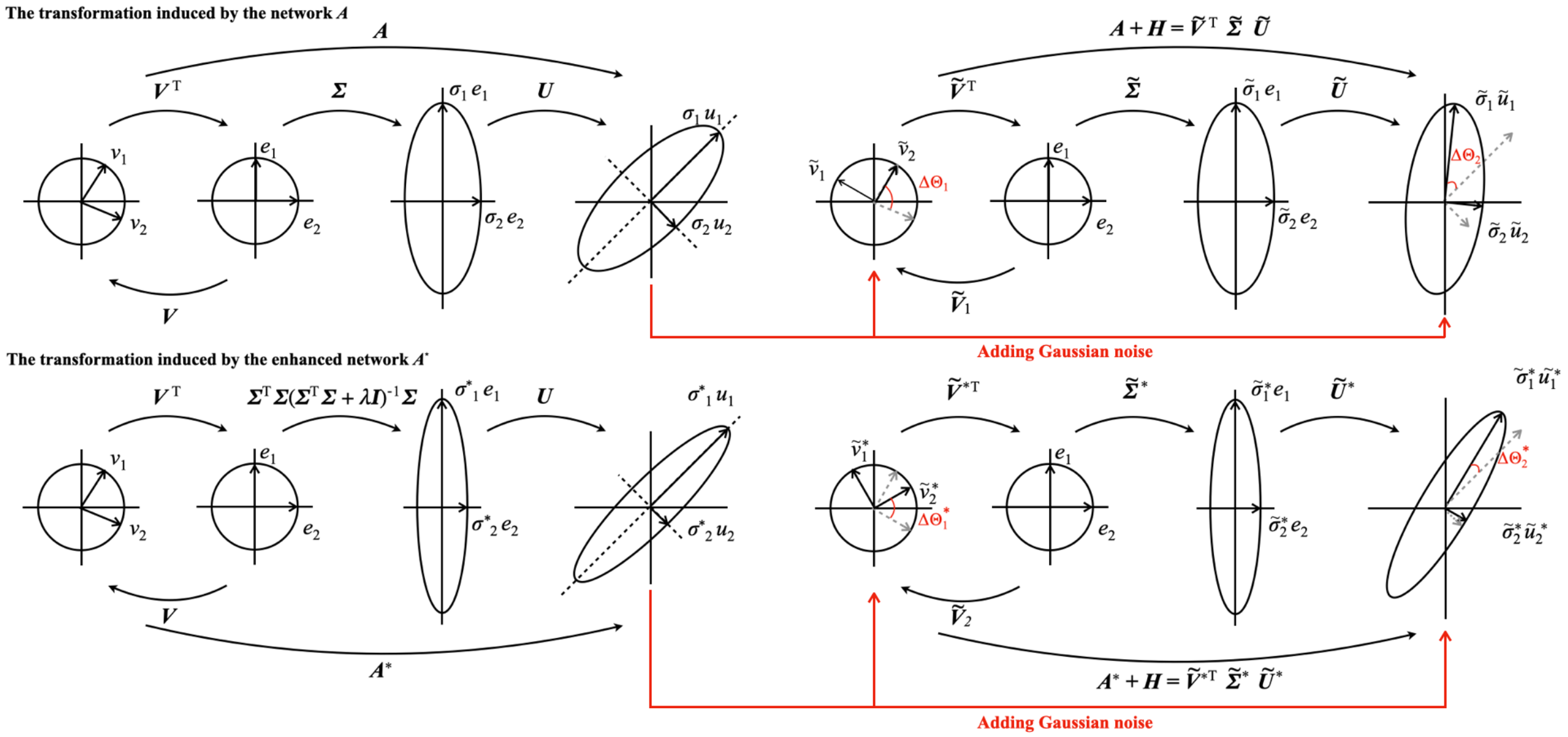

**Fig.S1 NR enhances complex networks by adding singular value gaps.** The principle is illustrated using a graphical representation of linear transformation. In this case, increasing the singular value gap reduces the weight of non-principal component data, as depicted in the figure where $\sigma_2^*$ is smaller than $\sigma_2$, leading to a narrower $\boldsymbol{A}^*$. Importantly, TopoLa preserves the left and right singular vectors of the matrix, ensuring that the underlying structure of the matrix is not altered. Through matrix perturbation analysis, when identical Gaussian noise $\boldsymbol{H}$ is added to both $\boldsymbol{A}$ and $\boldsymbol{A}^*$, the error upper bound of $\boldsymbol{A}^*$ is less than or equal to that of $\boldsymbol{A}$, that is $\max\{\sin(\Delta\Theta_1^*), \sin(\Delta\Theta_2^*)\} \leq \max\{\sin(\Delta\Theta_1), \sin(\Delta\Theta_2)\}$ (supplementary text 3.2, **Theorem 5**).

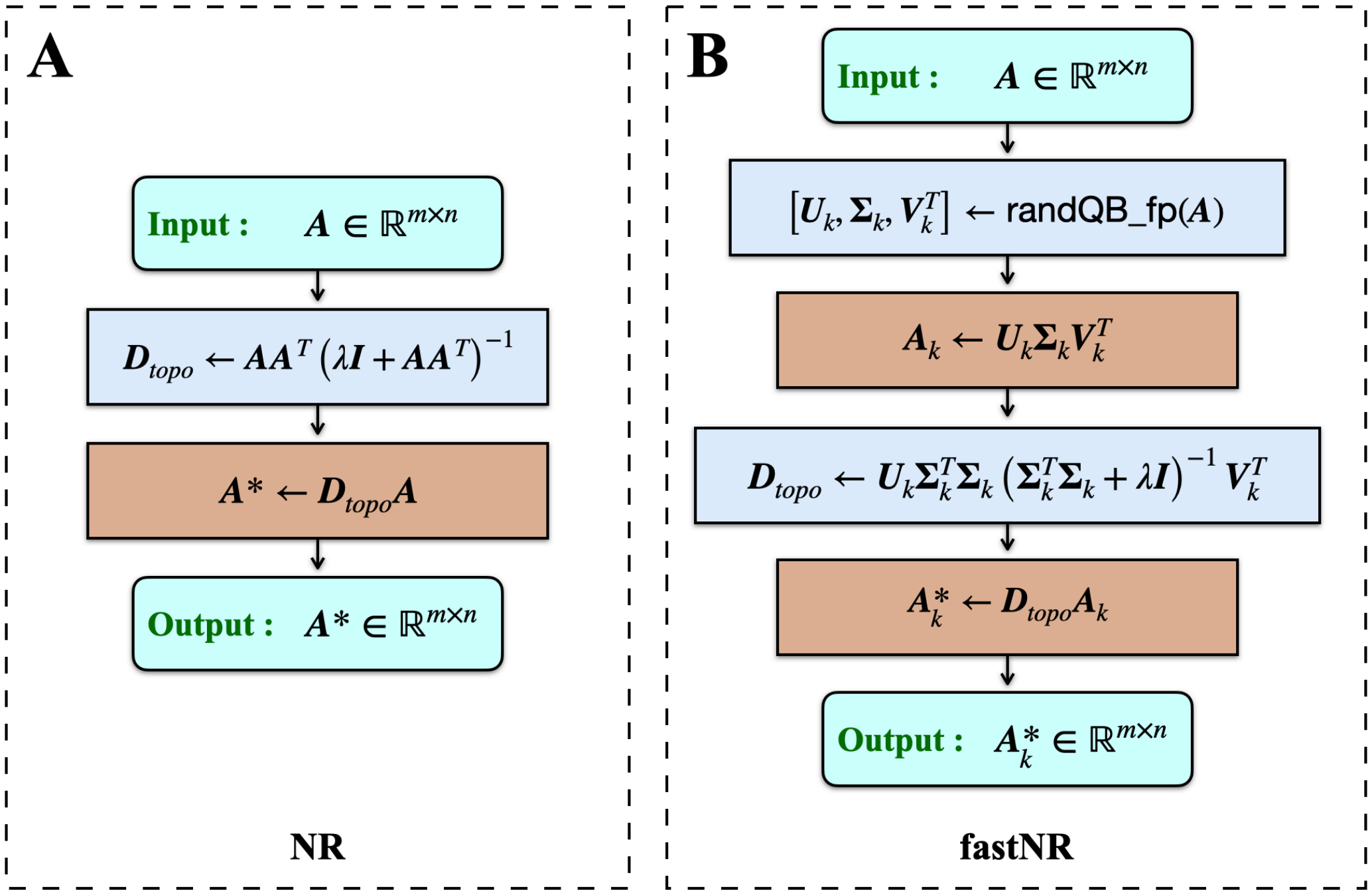

**Fig.S2 Illustration of the algorithms for NR and fastNR.** (**A**) NR inputs any network $\boldsymbol{A}$ and outputs the enhanced network $\boldsymbol{A}^*$. (**B**) fastNR takes any network $\boldsymbol{A}$ as input, applies a low-rank approximation to generate $\boldsymbol{A}_k$, and then enhances $\boldsymbol{A}_k$. Its output is the enhanced low-rank approximated network $\boldsymbol{A}_k^*$.

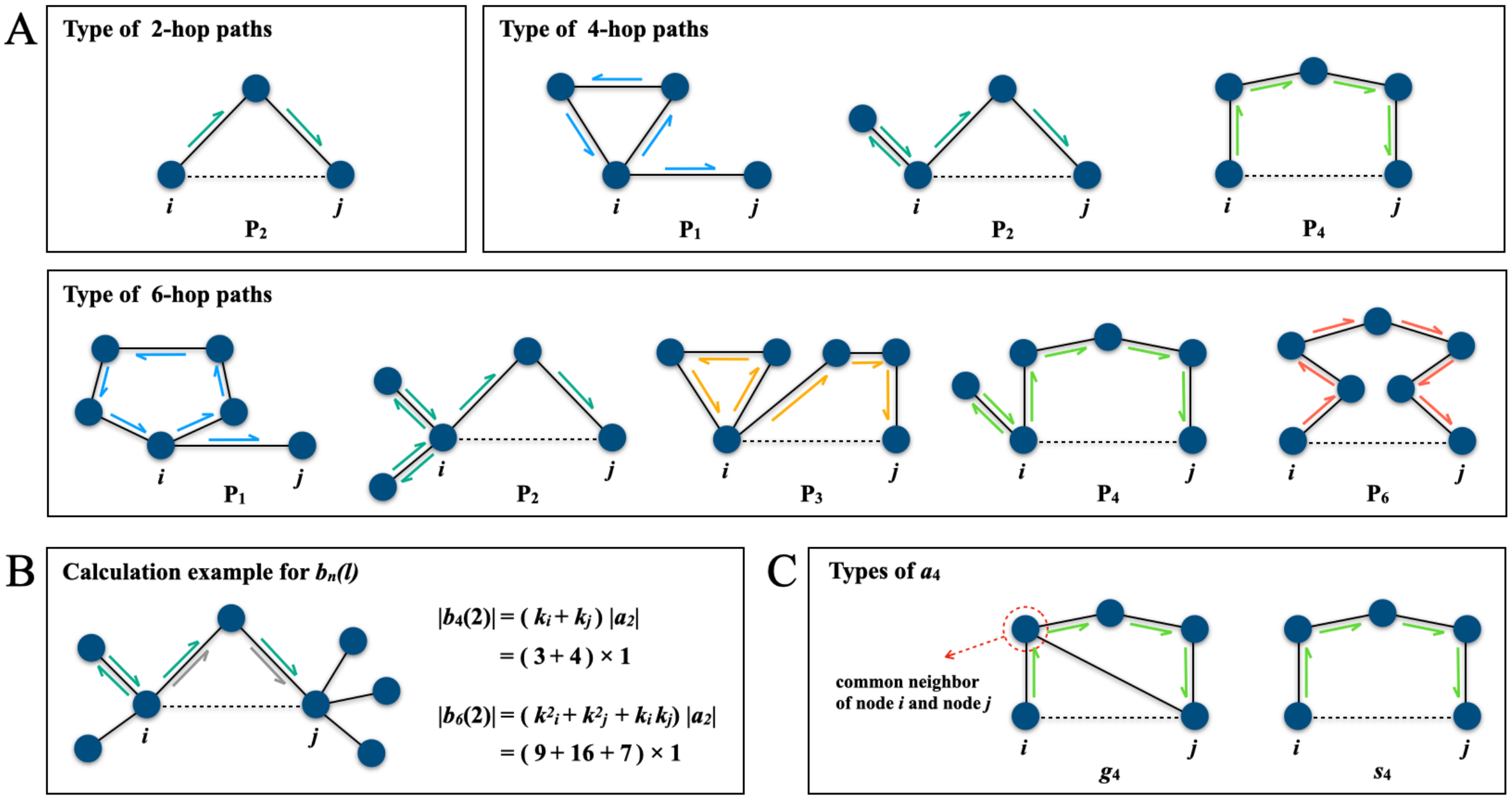

**Fig.S3 Types of even-hop paths.** (**A**) Paths can be abstracted into polygons. For instance, a 2-hop path can be visualized as a triangle $P_2$. When considering loops in paths, multi-hop path can be depicted as a collection of various polygons. 4-hop paths consist of straight lines, triangles, and pentagons. 6-hop paths consist of straight lines, triangles, quadrangles, pentagons, and heptagons (**B**) The calculation example for $|b_n(l)|$. $b_n(l)$ denotes $n$-hop $P_l$-type paths that solely loop between node $i$ and its neighbors, or between node $j$ and its neighbors. $\kappa_i$ and $\kappa_j$ are the degree of node $i$ and $j$. $|a_l|$ is the quantity of $l$-hop paths without loop. When $l$ is fixed, $|b_n(l)|$ can be computed using $\kappa_i$ and $\kappa_j$, along with $|a_l|$. This implies that $|b_n(l)|$ encapsulates information regarding $\kappa_i$ and $\kappa_j$. (**C**) Polygons $a_l$ can be classified into two types: $a_l$ overlapping with $P_2$ $(g_l)$ and the remainder $(s_l)$. The quantity of $s_l$ exhibits a direct correlation with the topological similarity of the two nodes.

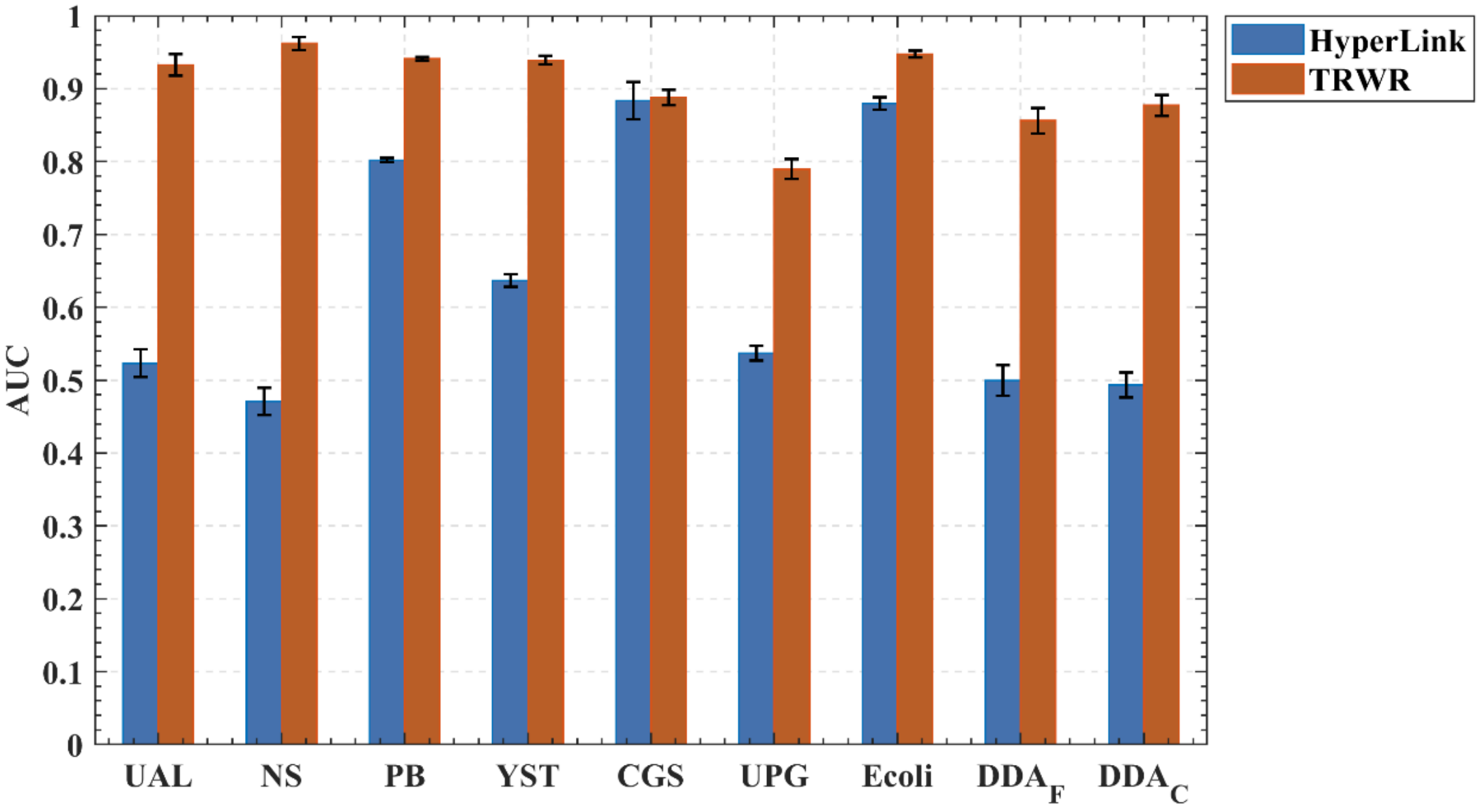

**Fig.S4 Comparison of link prediction methods based on latent space**. Our experimental observations indicate that methods grounded on the premise "nodes closer in latent space are more likely to form links" typically underperform compared to those built on the assumption that "nodes tend to form connections with neighbors of those nodes that have similar topological structures" Here, we selected the state-of-the-art latent space-based link prediction technique, HyperLink (*22*), which operates on the former assumption. In contrast, our proposed TRWR model is predicated on the latter. Empirical results suggest that our TRWR approach outperforms HyperLink.

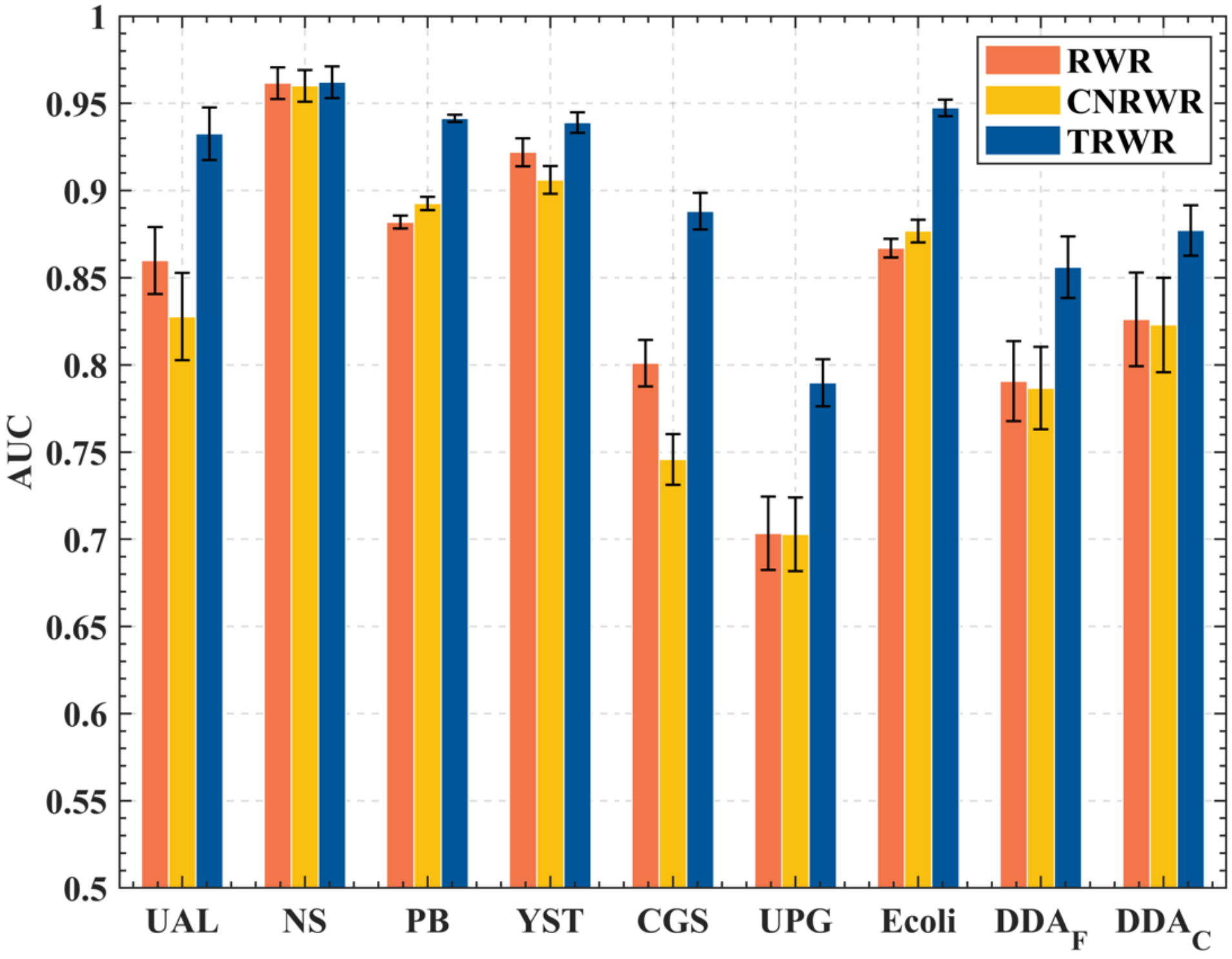

**Fig.S5 Comparison of energy-distance-based methods for network enhancement**. To confirm the performance differential between the common neighbor matrix and $\boldsymbol{D}_{topo}$, we undertake network enhancement tasks. This involves enhancing nine complex networks with these methods and applying Random Walk with Restart (RWR) for link prediction (named as CNRWR and TRWR), indirectly indicates the precision of energy distance measurements.

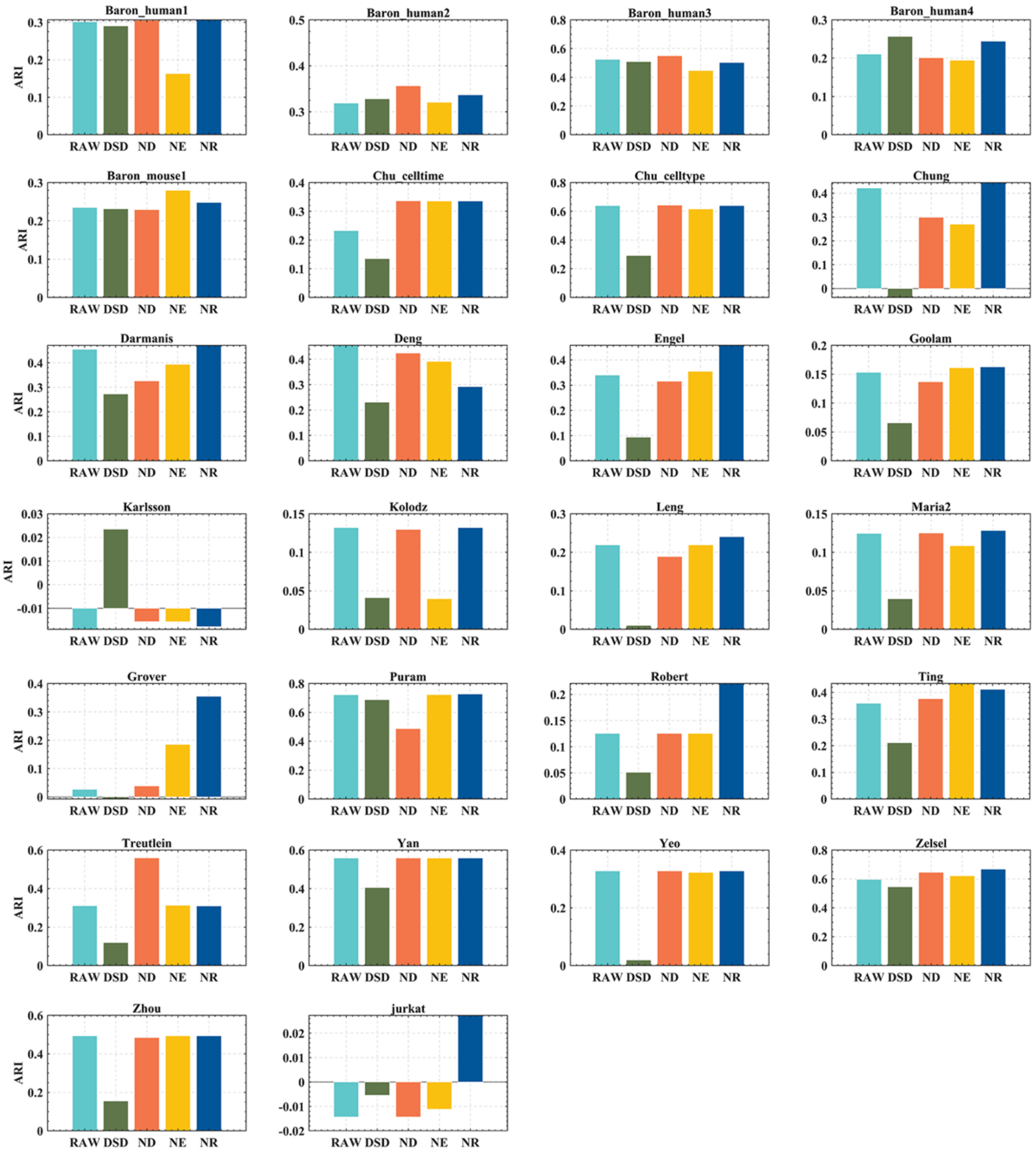

**Fig.S6 The clustering quality of single-cell RNA-seq data was assessed by ARI**. We evaluated the utility of the original and enhanced cell similarity networks in identifying cell types. Each bar graph represents the performance of kernel-based methods applied to the original or enhanced similarity networks obtained from a specific single-cell dataset and used for cell type identification. The prediction performance was calculated using ARI, where higher values of ARI indicate better agreement between the identified cell types and the ground truth. The average ARI values achieved by the methods across 30 datasets are: NR: 0.4381, NE: 0.3844, ND: 0.3890, DSD: 0.2347, and RAW: 0.3971.

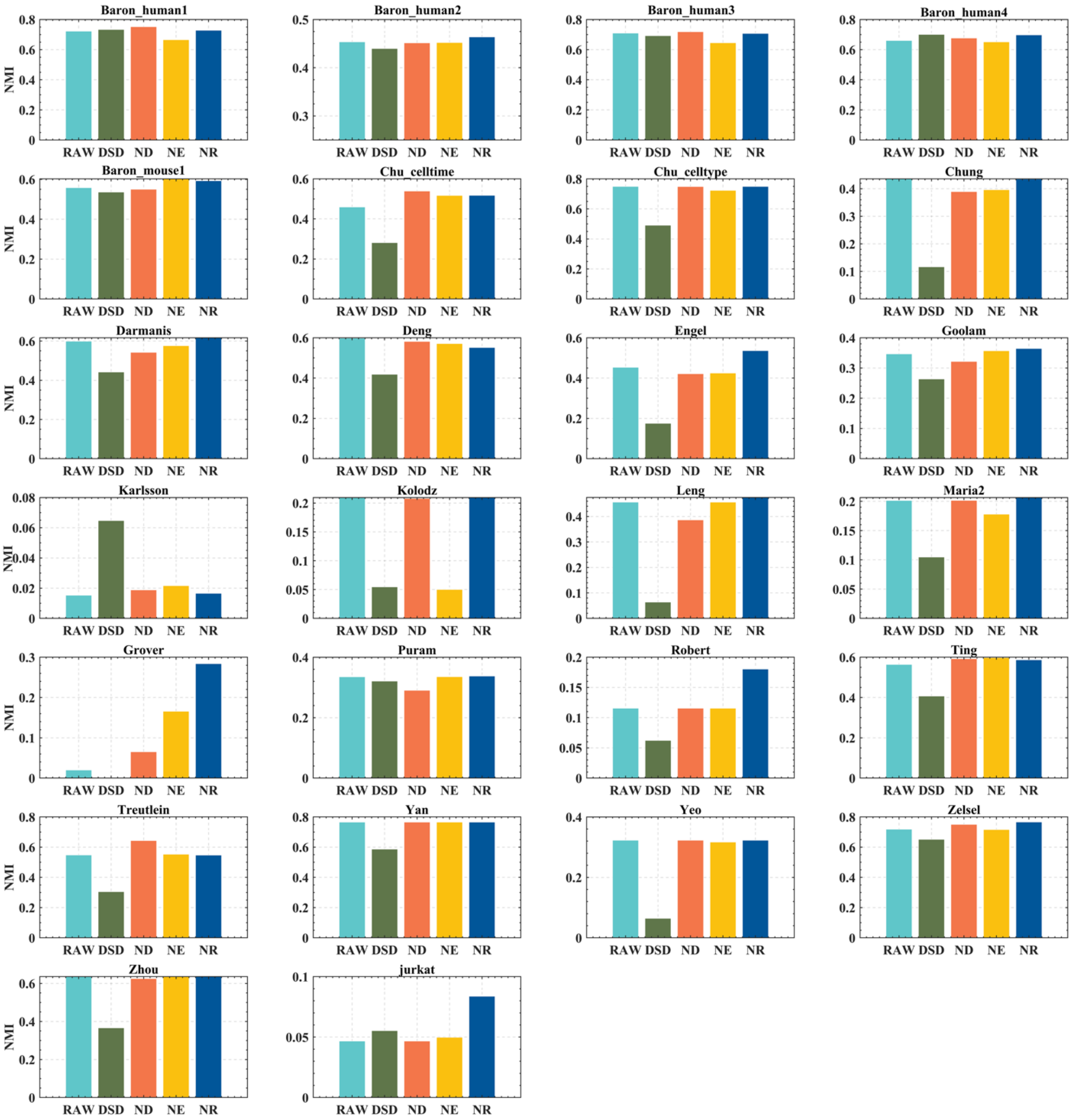

**Fig.S7 The clustering quality of single-cell RNA-seq data was assessed by NMI.** We evaluated the utility of the original and enhanced cell similarity networks in identifying cell types. Each bar graph represents the performance of kernel-based methods applied to the original or enhanced similarity networks obtained from a specific single-cell dataset and used for cell type identification. The prediction performance was calculated using NMI, where higher values of NMI indicate better agreement between the identified cell types and the ground truth. The average NMI values achieved by the methods across 30 datasets are: NR: 0.5450, NE: 0.5076, ND: 0.5087, DSD: 0.3682, and RAW: 0.5131.

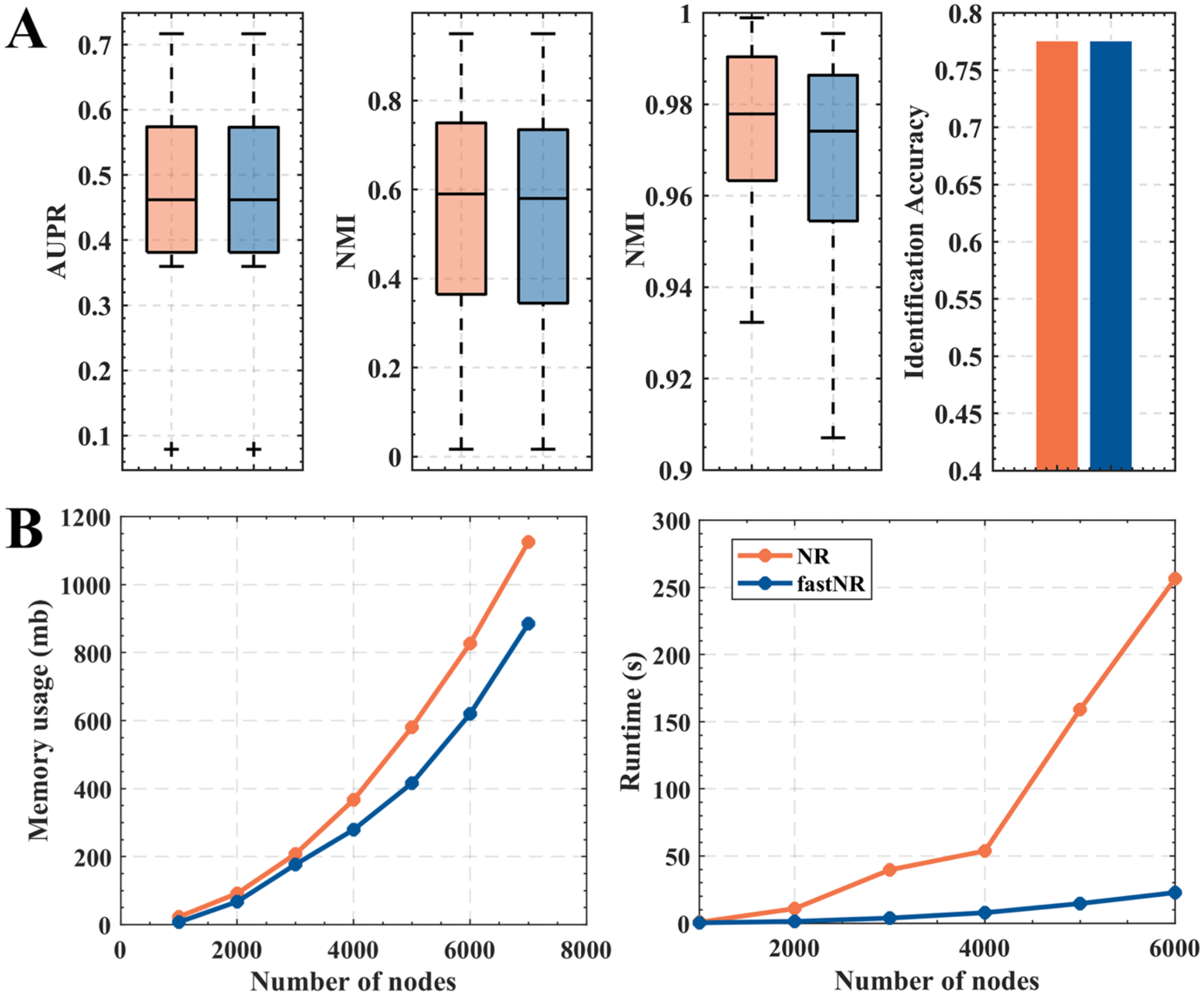

**Fig.S8 Comparison of performance between NR and fastNR.** (**A**) Performance of NR (red) and fastNR (blue) on four tasks: link prediction, single-cell data analysis, TAD detection, and fine-grained species identification. (**B**) Comparison of NR and fastNR in terms of computational requirements and efficiency. We simulated real-world complex networks by randomly generated matrix with geometrically distributed singular values. We repeated the experiment 10 times to obtain the comparison between the two methods in terms of memory usage and running time.

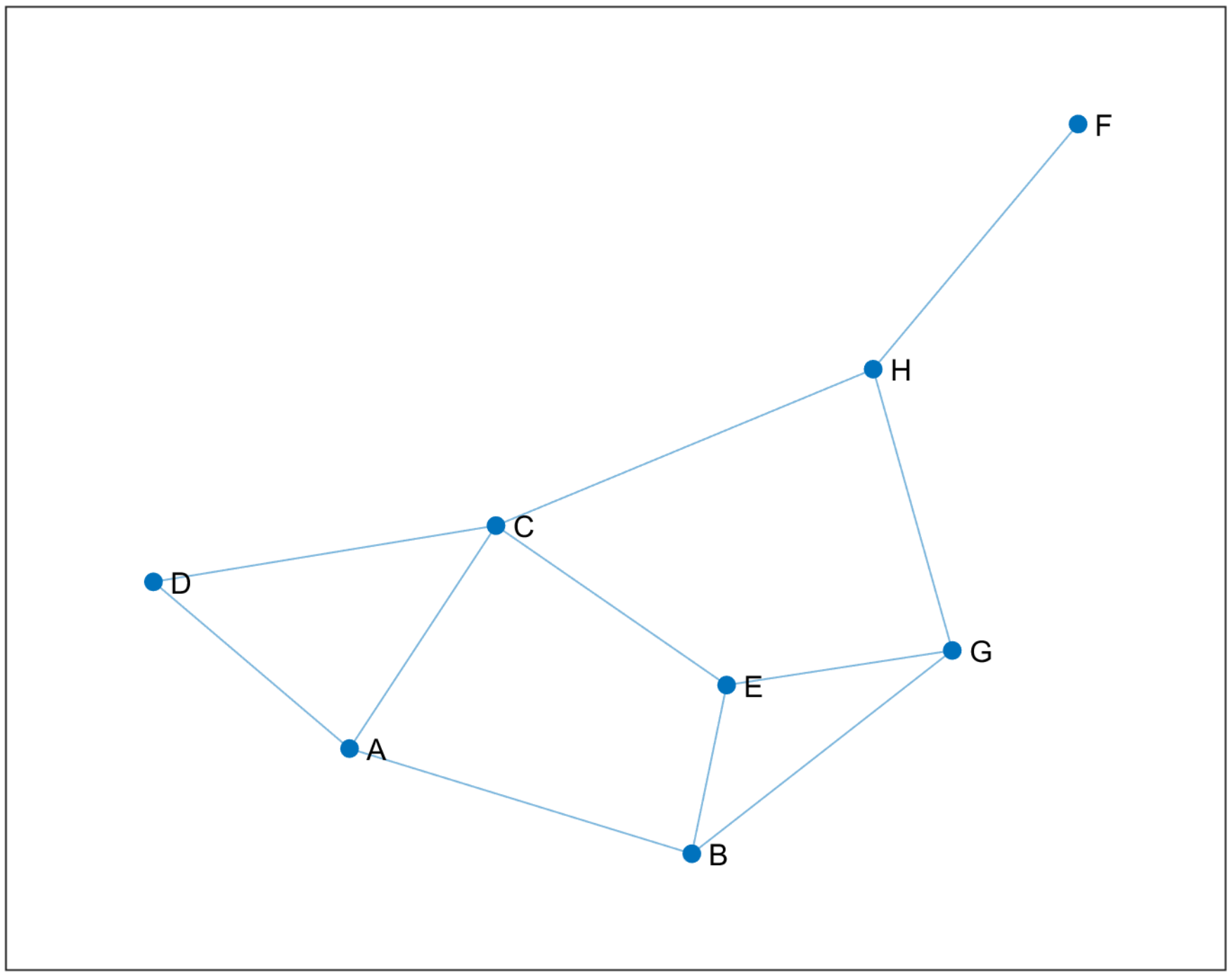

**Fig.S9 Network example illustrating the analysis of $n$-hop path types and their statistics**. Between nodes D and E, the counts for 2-hop, 4-hop, and 6-hop paths are 1, 10, and 89, while the respective quantities of $|P_3|$, $|P_5|$, and $|P_7|$ are 1, 3, and 2. The degrees of nodes B and D are 3 and 2, with the value of $\kappa_{ij}$ being 4.

**Table S1.** Thirty scRNA-seq datasets generated from different sequencing platforms.

| Dataset | Cells | Features | Types | Accession | Description |
|---|---|---|---|---|---|
| Baron_human1 | 1937 | 20124 | 14 | GSE84133 | human pancreatic cells |
| Baron_human2 | 1724 | 20124 | 14 | GSE84133 | human pancreatic cells |
| Baron_human3 | 3605 | 20124 | 14 | GSE84133 | human pancreatic cells |
| Baron_human4 | 1303 | 20124 | 14 | GSE84133 | human pancreatic cells |
| Baron_mouse | 822 | 14877 | 13 | GSE84133 | mouse pancreatic cells |
| Biase | 49 | 25736 | 3 | GSE57249 | mouse embryonic stem cells |
| Buettner | 288 | 38293 | 3 | E-MTAB-2805 | mouse embryonic stem cells |
| Chu_celltime | 758 | 19176 | 6 | GSE75748 | human pluripotent stem cells |
| Chu_celltype | 1018 | 19095 | 7 | GSE75748 | human pluripotent stem cells |
| Chung | 515 | 20345 | 5 | GSE75688 | human tumor and immune cells |
| Darmanis | 466 | 22085 | 9 | GSE67835 | human brain cells |
| Deng | 259 | 22958 | 10 | GSE45719 | mouse cells from different stages |
| Engel | 203 | 23337 | 4 | GSE74596 | mouse Natural killer T cells |
| Goolam | 124 | 41388 | 8 | E-MTAB-3321 | mouse cells from different stages |
| Grover | 135 | 15180 | 2 | GSE70657 | human hematopoietic stem cells |
| Karlsson | 94 | 59745 | 3 | E-MTAB-6142 | human myxoid liposarcoma cells |
| Kolodz | 704 | 38653 | 3 | E-MTAB-2600 | mouse embryonic stem cells |
| Kumar | 361 | 22394 | 4 | GSE60749 | mouse embryonic stem cells |
| Leng | 247 | 19084 | 3 | GSE64016 | human embryonic stem cells |
| Maria | 759 | 33694 | 7 | GSE124731 | human innate T cells |
| Pollen | 249 | 6981 | 11 | GSM1832359 | human brain cells |
| Puram | 3363 | 23686 | 8 | GSE103322 | non-malignant cells in Head and Neck Cancer |
| Robert | 194 | 23418 | 2 | GSE74923 | mouse leukemia cell line andprimary CD8+ T-cells |
| Ting | 187 | 21583 | 7 | GSE51372 | mouse circulating tumor cells |
| Treutlein | 80 | 23271 | 5 | GSE52583 | mouse lung epithelial cells |
| Yan | 90 | 20214 | 6 | GSE36552 | human embryonic stem cells |
| Yeo | 206 | 20345 | 3 | GSE85908 | human induced pluripotentstem cells |
| Zelsel | 3005 | 19486 | 9 | GSE60361 | mouse cerebral cortex cells |
| Zhou | 181 | 23937 | 8 | GSE67120 | mouse haematopoietic stem cells |
| Jurkat | 1580 | 32738 | 2 | 10X PBMC | Jurkat cells |

**Table S2.** Condition numbers of 88 real-world networks.

| Dataset | Cond | Dataset | Cond | Dataset | Cond | Dataset | Cond |
|---|---|---|---|---|---|---|---|
| UPG | 1.01E+34 | PB | 1.05e+36 | email-dnc-corecipient | inf | web-EPA | inf |
| HPD | 1.72E+71 | NS | inf | email-enron-only | 1.92E+17 | web-indochina-2004 | 1.49E+19 |
| ERD | 5.70E+126 | Ecoli | 4.01E+19 | email-univ | 4.84E+33 | web-polblogs | 3.13E+49 |
| YST | 3.07E+91 | bn-fly-drosophil | inf | en_adj | 1.38E+51 | web-spam | 2.79E+34 |
| EML | 4.83E+33 | bn-macaque-rhesus | 5.13E+65 | frb30-15-1 | 6.83E+03 | DDA_C | 1.32E+35 |
| ADV | 1.41E+20 | bn-mouse_retina_1 | inf | frb35-17-1 | 2.15E+04 | DDA_F | 5.47E+32 |
| KHN | 5.21E+49 | ca-CSphd | inf | frb40-19-5 | 4.97E+05 | DDI | 5.65E+04 |
| PGP | 4.33E+77 | ca-Erdos992 | 1.06E+139 | ia-crime-moreno | inf | DTI_en | 1.38E+51 |
| CEG | 1.53E+67 | ca-GrQc | 5.79E+18 | ia-fb-messages | 1.08E+19 | DTI_gpcr | 1.11E+18 |
| LDG | 6.17E+20 | ENZYMES_g295 | 4.54E+16 | ia-infect-dublin | 4.70E+03 | DTI_ic | 7.33E+49 |
| SMG | 1.06E+34 | ENZYMES_g296 | 3.19E+2 | movielens | inf | DTI_nr | inf |
| ZWL | 5.40E+33 | ENZYMES_g297 | 1.75E+16 | road-chesapeake | 5.10E+18 | LDA | 3.83E+16 |
| INF | 4.69E+3 | DIMACS-C4000-5 | 6.15E+06 | road-euroroad | 8.88E+18 | MDA_HMDD | inf |
| BUP | 7.20E+2 | DIMACS-MANN-a45 | 9.36E+19 | road-minnesota | 6.22E+17 | CircR2Disease | 3.11E+17 |
| HTC | 6.40E+19 | DIMACS-p-hat1500-1 | 2.91E+04 | rt_bahrain | 1.06E+178 | DisGeNET | 1.79E+34 |
| CGS | 2.21E+19 | eco-florida | inf | rt_http | inf | HMDAD | 2.43E+16 |
| GRQ | 8.86E+49 | eco-foodweb-baydry | inf | soc-advogato | inf | LncRNA2Target | 1.05E+17 |
| HMT | inf | eco-mangwet | inf | soc-hamsterster | inf | MiREnvironment | 8.27E+16 |
| FBK | 3.06E+21 | econ-beaflw | inf | soc-wiki-elec | inf | DrugBank | 2.16E+85 |
| UAL | 2.17E+35 | econ-mahindas | 9.76E+17 | tech-pgp | inf | PharmGKB | 1.12E+49 |
| CDM | 4.70E+18 | econ-mbeacxc | inf | tech-routers-rf | 1.99E+35 | SM2miR | 1.35E+17 |
| NSC | inf | econ-psmigr1 | 1.71E+05 | tech-WHOIS | 1.68E+67 | STRING | 1.36E+19 |

**Table S3.** Condition numbers of 44 Hi-C networks.

| Dataset | Cond | Dataset | Cond | Dataset | Cond | Dataset | Cond |
|---|---|---|---|---|---|---|---|
| Class1_1000 | 1.22E+17 | Class12_1000 | 1.25E+18 | Class1_5000 | 1.23E+01 | Class12_5000 | 1.29E+03 |
| Class2_1000 | 2.78E+18 | Class13_1000 | 6.05E+17 | Class2_5000 | 2.90E+17 | Class13_5000 | 1.11E+03 |
| Class3_1000 | 2.65E+17 | Class14_1000 | 1.38E+26 | Class3_5000 | 3.57E+01 | Class14_5000 | 1.41E+01 |
| Class4_1000 | 3.44E+18 | Class15_1000 | 1.90E+17 | Class4_5000 | 1.34E+17 | Class15_5000 | 4.39E+01 |
| Class5_1000 | 6.17E+17 | Class16_1000 | 4.19E+17 | Class5_5000 | 8.56E+00 | Class16_5000 | 1.80E+01 |
| Class6_1000 | inf | Class17_1000 | 1.34E+17 | Class6_5000 | 9.16E+00 | Class17_5000 | 1.80E+01 |
| Class7_1000 | 9.98E+16 | Class18_1000 | 4.19E+17 | Class7_5000 | 8.71E+00 | Class18_5000 | 2.56E+02 |
| Class8_1000 | 9.63E+17 | Class19_1000 | 4.03E+04 | Class8_5000 | 3.57E+16 | Class19_5000 | 2.59E+01 |
| Class9_1000 | 7.80E+16 | Class20_1000 | 1.13E+17 | Class9_5000 | 2.94E+01 | Class20_5000 | 6.44E+00 |
| Class10_1000 | 3.43E+17 | Class21_1000 | 9.53E+17 | Class10_5000 | 1.79E+01 | Class21_5000 | 7.36E+00 |
| Class11_1000 | 3.20E+17 | Class22_1000 | 9.49E+04 | Class11_5000 | 2.75E+03 | Class22_5000 | 7.80E+00 |

**Table S4.** Using NR to complement network enhancement methods for link prediction.

| Dataset | Methods | | | | | |
|---|---|---|---|---|---|---|
| | DSD | DSD+NR | ND | ND+NR | NE | NE+NR |
| UAL | 0.1523 | 0.1522 | 0.2638 | 0.6571 | 0.6136 | 0.5873 |
| CGS | 0.0037 | 0.0037 | 0.6238 | 0.6474 | 0.8347 | 0.7415 |
| UPG | 0.1251 | 0.1251 | 0.2304 | 0.4979 | 0.3269 | 0.3567 |
| PB | 0.0235 | 0.0235 | 0.3916 | 0.7073 | 0.5459 | 0.5833 |
| NS | 0.1290 | 0.1294 | 0.1809 | 0.3835 | 0.3790 | 0.4272 |
| YST | 0.0010 | 0.0010 | 0.1491 | 0.1314 | 0.1468 | 0.1152 |
| Ecoli | 0.1758 | 0.1758 | 0.4069 | 0.6785 | 0.4112 | 0.5599 |
| $DDA_F$ | 0.0213 | 0.0214 | 0.0376 | 0.5282 | 0.2086 | 0.2848 |
| $DDA_C$ | 0.0176 | 0.0176 | 0.0537 | 0.6173 | 0.2565 | 0.3298 |
| Average | **0.0721** | **0.0722** | **0.2597** | **0.5387** | **0.4137** | **0.4429** |

**Table S5.** Using NR to complement network enhancement methods for single-cell analysis.

| Dataset | Methods | | | | | |
|---|---|---|---|---|---|---|
| | DSD | DSD+NR | ND | ND+NR | NE | NE+NR |
| Baron_human1 | 0.7333 | 0.7163 | 0.7525 | 0.6952 | 0.6652 | 0.7197 |
| Baron_human2 | 0.6794 | 0.6818 | 0.7031 | 0.6755 | 0.7043 | 0.6906 |
| Baron_human3 | 0.6924 | 0.7294 | 0.7182 | 0.7199 | 0.6455 | 0.6984 |
| Baron_human4 | 0.7006 | 0.6721 | 0.6769 | 0.6583 | 0.6511 | 0.6901 |
| Baron_mouse | 0.5360 | 0.5813 | 0.5501 | 0.6374 | 0.6024 | 0.5873 |
| Biase | 0.4673 | 0.3639 | 0.5641 | 0.6337 | 0.6156 | 0.7975 |
| Buettner | 0.0099 | 0.0278 | 0.0951 | 0.1438 | 0.1337 | 0.0769 |
| Chu_celltime | 0.2811 | 0.3231 | 0.5392 | 0.5751 | 0.5171 | 0.5367 |
| Chu_celltype | 0.4908 | 0.5624 | 0.7490 | 0.7652 | 0.7230 | 0.7554 |
| Chung | 0.1165 | 0.2793 | 0.3892 | 0.4241 | 0.3961 | 0.4140 |
| Darmanis | 0.4414 | 0.5298 | 0.5423 | 0.5904 | 0.5762 | 0.6129 |
| Deng | 0.4186 | 0.5005 | 0.5821 | 0.5381 | 0.5714 | 0.5436 |
| Engel | 0.1750 | 0.0979 | 0.4216 | 0.4885 | 0.4247 | 0.4534 |
| Goolam | 0.2635 | 0.3044 | 0.3217 | 0.3517 | 0.3570 | 0.3645 |
| Grover | 0.0004 | 0.0078 | 0.0653 | 0.1405 | 0.1659 | 0.3259 |
| Karlsson | 0.0647 | 0.0214 | 0.0188 | 0.0166 | 0.0216 | 0.0153 |
| Kolodz | 0.0546 | 0.1487 | 0.2078 | 0.3114 | 0.0500 | 0.1890 |
| Kumar | 0.6362 | 0.7701 | 0.8088 | 0.8369 | 0.8198 | 0.8259 |
| Leng | 0.0640 | 0.1279 | 0.3864 | 0.4764 | 0.4562 | 0.4519 |
| Maria | 0.1047 | 0.1463 | 0.2017 | 0.2093 | 0.1780 | 0.2133 |
| Pollen | 0.7219 | 0.7437 | 0.8968 | 0.9033 | 0.8910 | 0.9499 |
| Puram | 0.8019 | 0.8418 | 0.7260 | 0.8234 | 0.8382 | 0.8356 |
| Robert | 0.0622 | 0.0585 | 0.1156 | 0.2300 | 0.1156 | 0.1156 |
| Ting | 0.4068 | 0.5331 | 0.5924 | 0.5134 | 0.5980 | 0.5755 |
| Treutlein | 0.3049 | 0.3181 | 0.6432 | 0.5881 | 0.5532 | 0.5560 |
| Yan | 0.5865 | 0.4808 | 0.7652 | 0.7652 | 0.7652 | 0.8301 |
| Yeo | 0.1602 | 0.6281 | 0.8064 | 0.8064 | 0.7922 | 0.8064 |
| Zelsel | 0.6513 | 0.6535 | 0.7504 | 0.7349 | 0.7163 | 0.7527 |
| Zhou | 0.3661 | 0.4480 | 0.6254 | 0.6232 | 0.6357 | 0.6357 |
| Jurkat | 0.0552 | 0.0467 | 0.0467 | 0.0588 | 0.0498 | 0.0502 |
| Average | **0.3682** | **0.4115** | **0.5087** | **0.5312** | **0.5076** | **0.5357** |